\documentclass[twocolumn,runningheads]{svjour3}

\usepackage[latin1]{inputenc}
\usepackage{times}

\usepackage{psfrag,amsmath, amssymb, amscd, amsfonts, latexsym, epsf, graphicx,amscd}
\usepackage{natbib}
\usepackage[switch]{lineno}

\def\diag{\operatorname{diag}}

\def\simple{\scriptsize\mbox{simple}}
\def\mono{\scriptsize\mbox{mono}}
\def\bino{\scriptsize\mbox{bino}}
\def\norm{\scriptsize\mbox{norm}}

\newcommand{\atantwo}{\operatorname{atan2}}
\newcommand{\sign}{\operatorname{sign}}
\newcommand{\orth}{\bot}

\journalname{arXiv preprint}

\begin{document}

\titlerunning{Do the receptive fields in V1 span the variabilities under spatial affine transformations?}

\title{\bf Relationships between the degrees of freedom in the affine
  Gaussian derivative model for visual receptive fields
  and 2-D affine image transformations, with
  application to covariance properties of simple cells in the primary visual cortex
  %Do the receptive fields in the primary visual cortex span the
  %variabilities in image structures generated by the group of spatial
  %affine image transformations?
\thanks{The support from the Swedish Research Council 
              (contract 2022-02969) is gratefully acknowledged. }}
\author{Tony Lindeberg}

\institute{Computational Brain Science Lab,
        Division of Computational Science and Technology,
        KTH Royal Institute of Technology,
        SE-100 44 Stockholm, Sweden.
        \email{tony@kth.se}
      ORCID: 0000-0002-9081-2170.}

\date{}

\maketitle

\begin{abstract}
  \noindent
  When observing the surface patterns of objects delimited by smooth surfaces, the projections of the surface patterns to the image domain will be subject to substantial variabilities, as induced by variabilities in the geometric viewing conditions, and as generated by either monocular or binocular imaging conditions, or by relative motions between the object and the observer over time. To first order of approximation, the image deformations of such projected surface patterns can be modelled as local linearizations in terms of local 2-D spatial affine transformations.
  
This paper presents a theoretical analysis of relationships between the degrees of freedom in 2-D spatial affine image transformations and the degrees of freedom in the affine Gaussian derivative model for visual receptive fields. For this purpose, we first describe a canonical decomposition of 2-D affine transformations on a product form, closely related to a singular value decomposition, while in closed form, and which reveals the degrees of freedom in terms of (i)~uniform scaling transformations, (ii)~an overall amount of global rotation, (iii)~a complementary non-uniform scaling transformation and (iv)~a relative normalization to a preferred symmetry orientation in the image domain.
Then, we show how these degrees of freedom relate to the degrees of freedom in the affine Gaussian derivative model.

Finally, we use these theoretical results to consider whether we could regard the biological receptive fields in the primary visual cortex of higher mammals as being able to span the degrees of freedom of 2-D spatial affine transformations, based on interpretations of existing neurophysiological experimental results.

% The presented results are aimed at providing a theoretical
%  understanding of basic covariance properties for the earliest layers
%  of receptive fields in
%  computational vision, as well as for generating hypotheses about the
%  computational functions in the earliest levels of visual processing
%  in biological vision.

 %\and Vision \and Theoretical neuroscience}

\end{abstract}

\section{Introduction}
\label{sec-intro}

\begin{figure*}[hbt]
  \begin{center}
    {\em\small Uniform scaling transformations caused by varying the distance between the object and the observer\/}

    \smallskip
    
    \begin{tabular}{ccc}
      \includegraphics[width=0.31\textwidth]{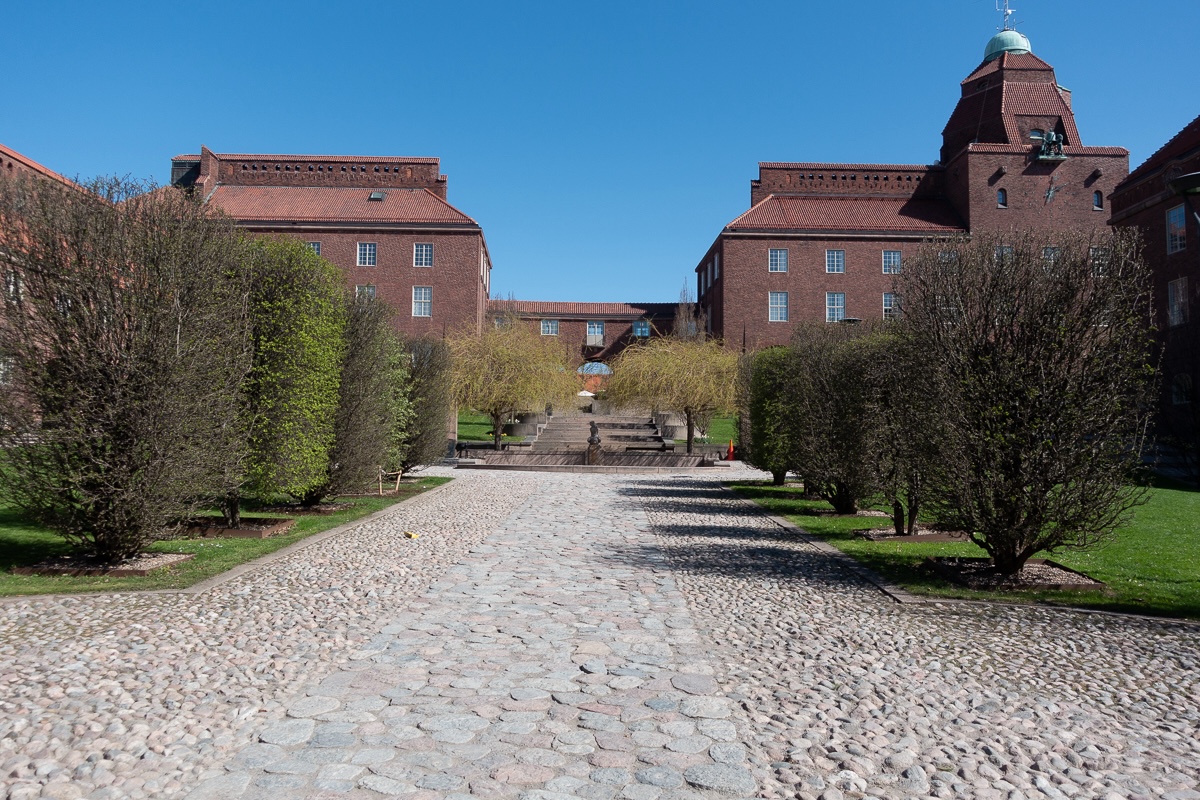}
      & \includegraphics[width=0.31\textwidth]{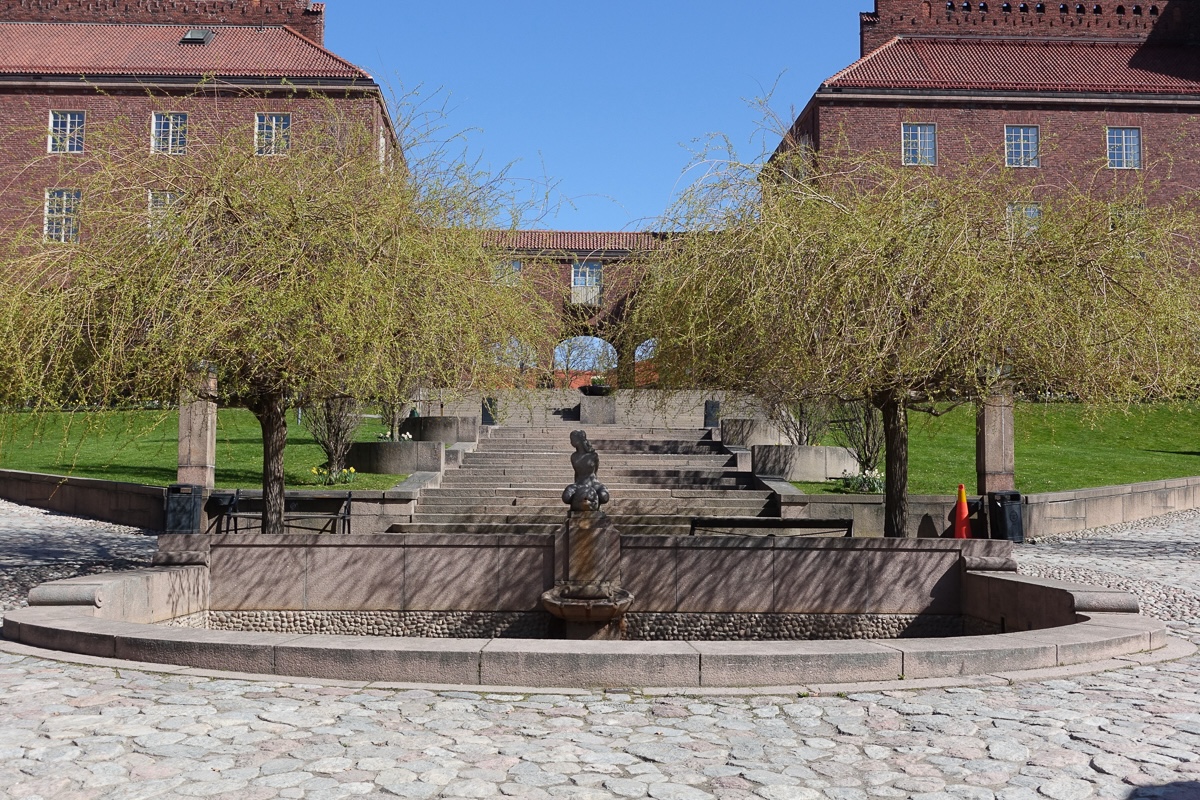}
      & \includegraphics[width=0.31\textwidth]{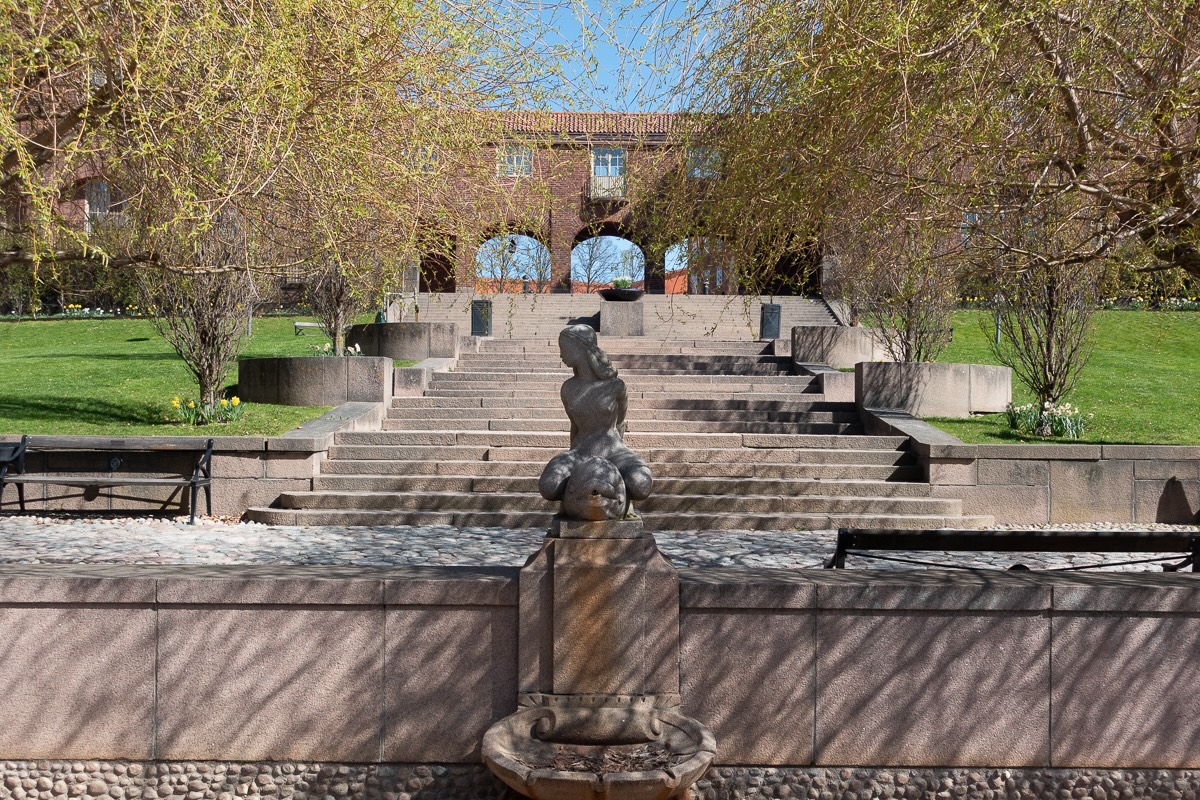}        
    \end{tabular}

    \medskip

        {\em\small Rotations in the image plane caused by relative rotations around the optical axis between the object and the observer\/}

    \smallskip
    
    \begin{tabular}{ccc}
      \includegraphics[width=0.31\textwidth]{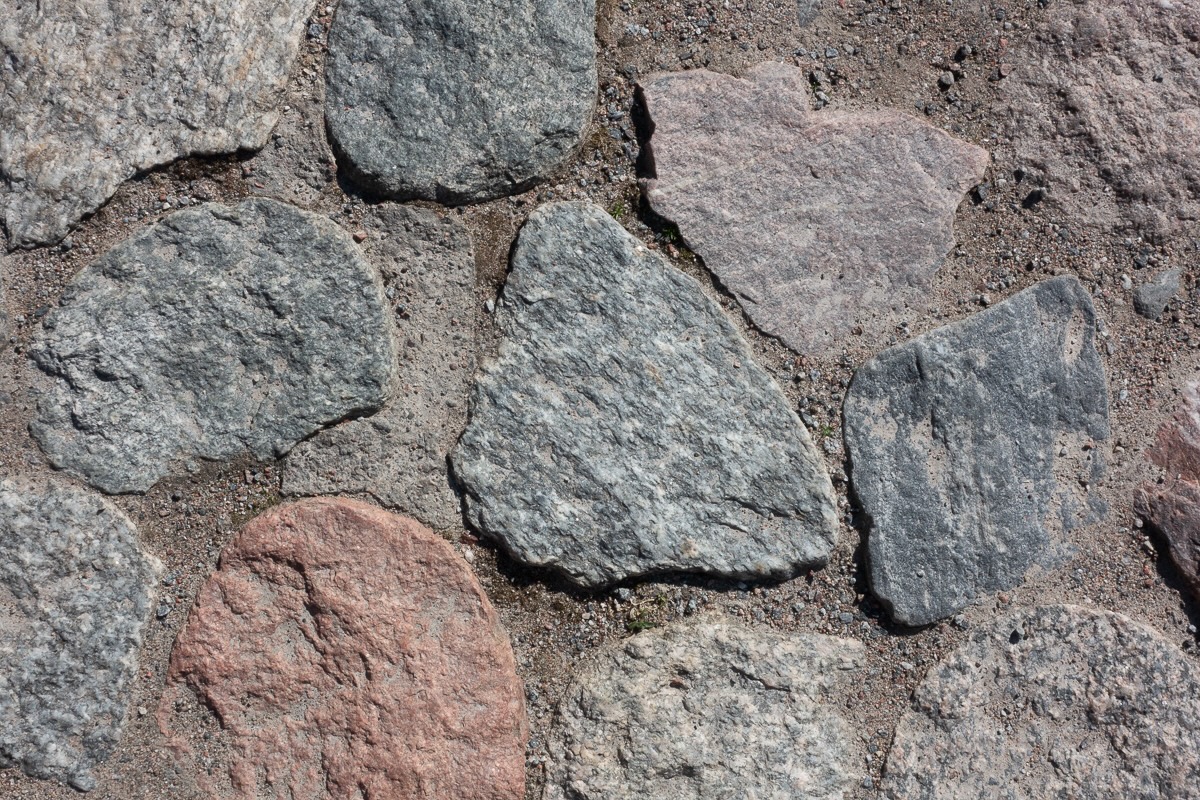}
      & \includegraphics[width=0.31\textwidth]{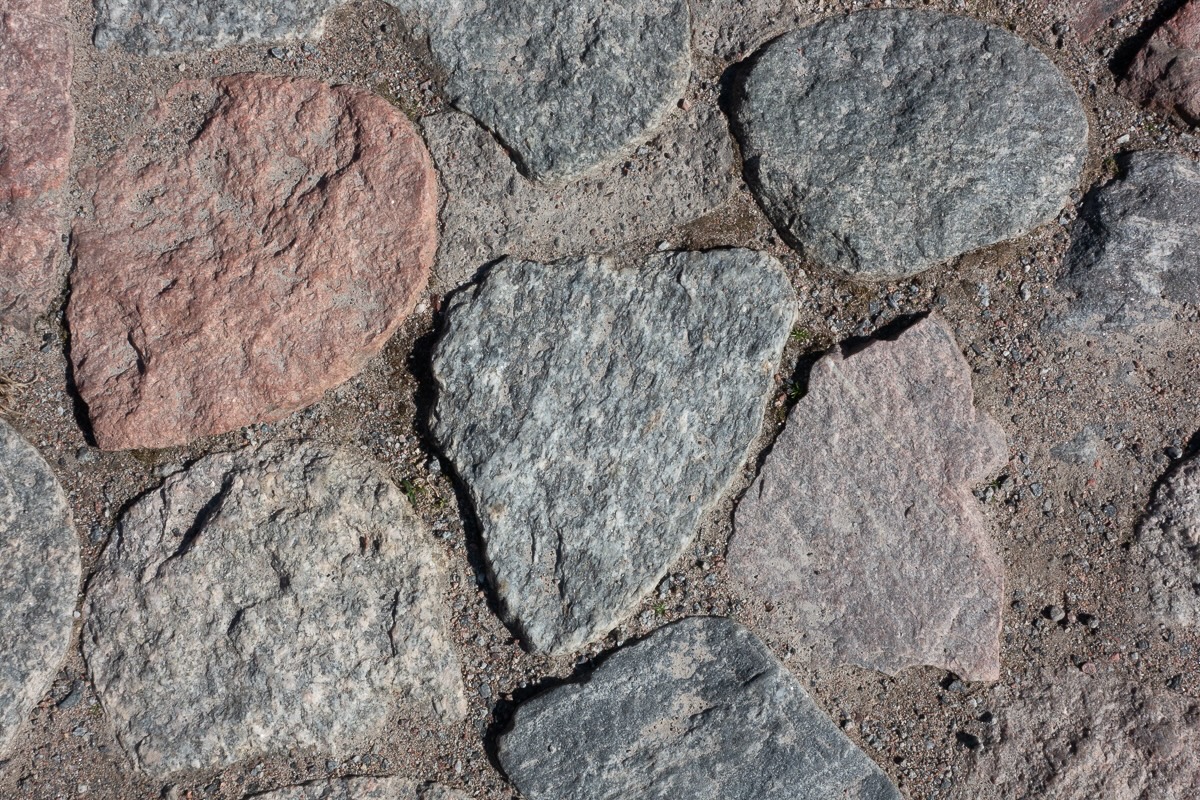}
      & \includegraphics[width=0.31\textwidth]{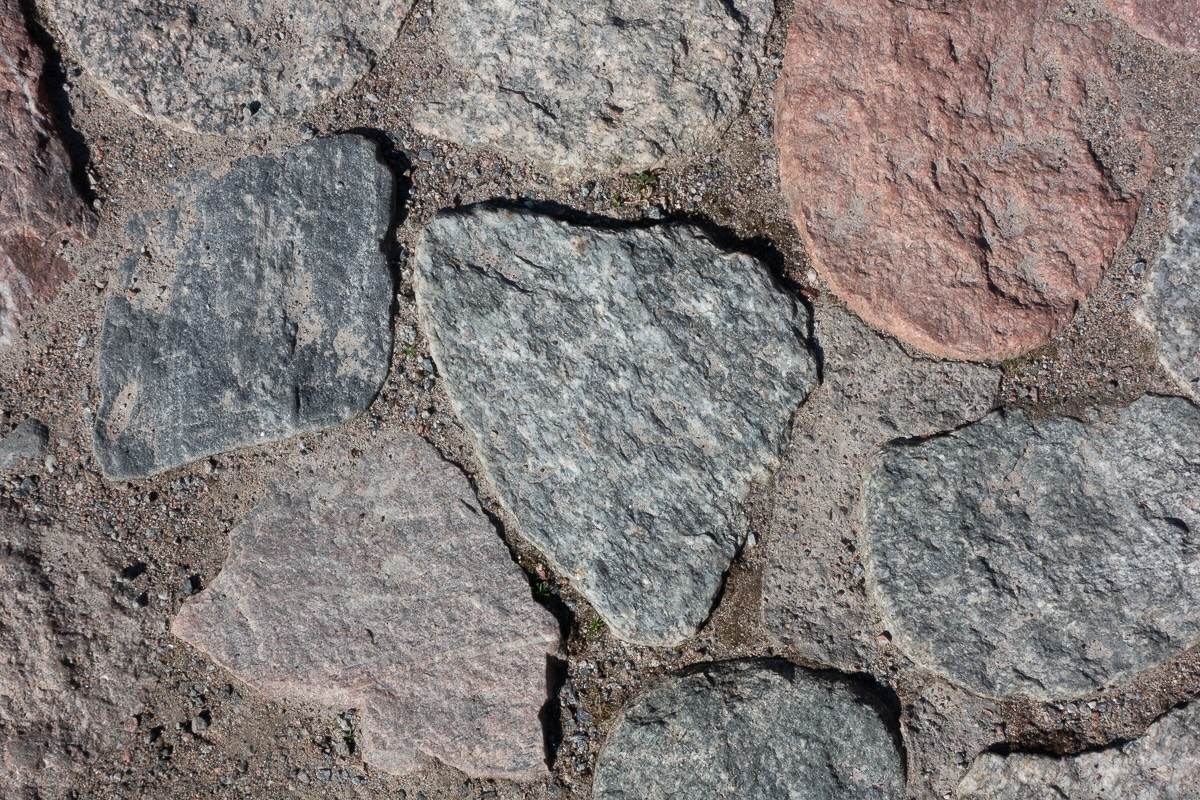}        
    \end{tabular}

    \medskip

     {\em\small Foreshortening transformations caused by varying the viewing direction relative to the object\/}

    \smallskip
    
    \begin{tabular}{ccc}
      \includegraphics[width=0.31\textwidth]{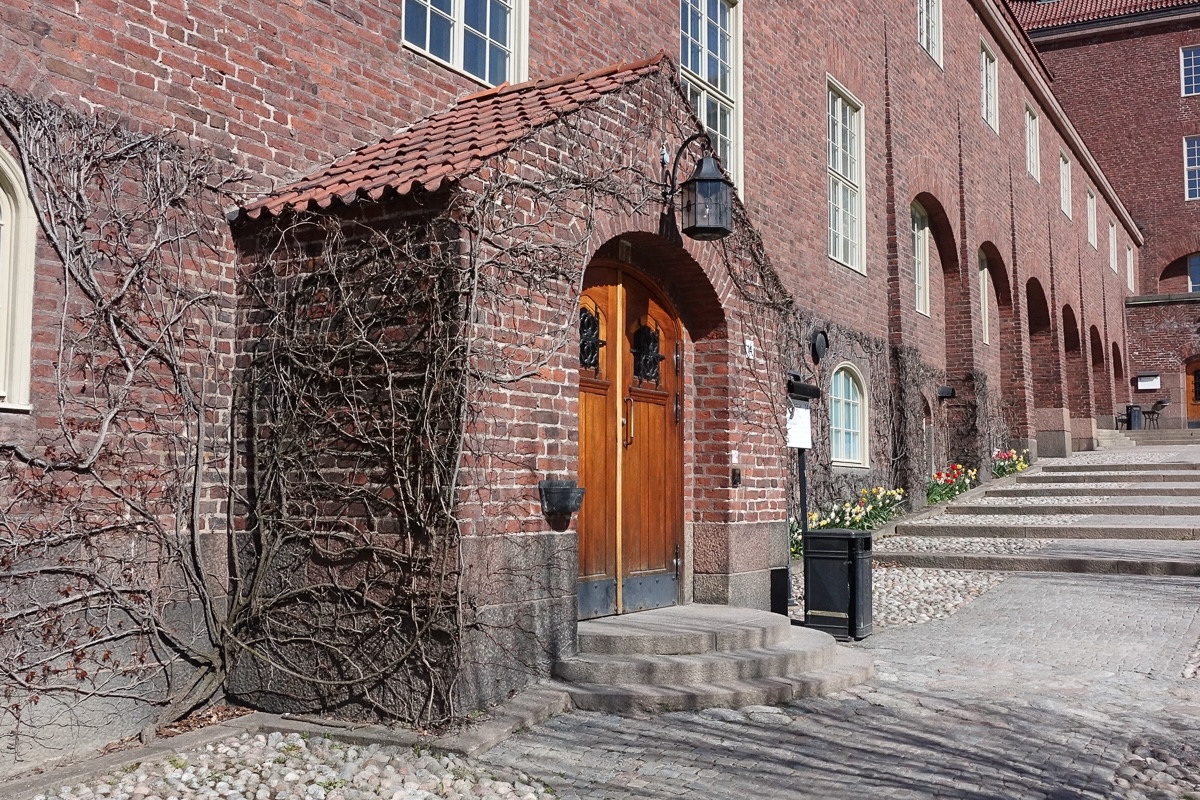}
      & \includegraphics[width=0.31\textwidth]{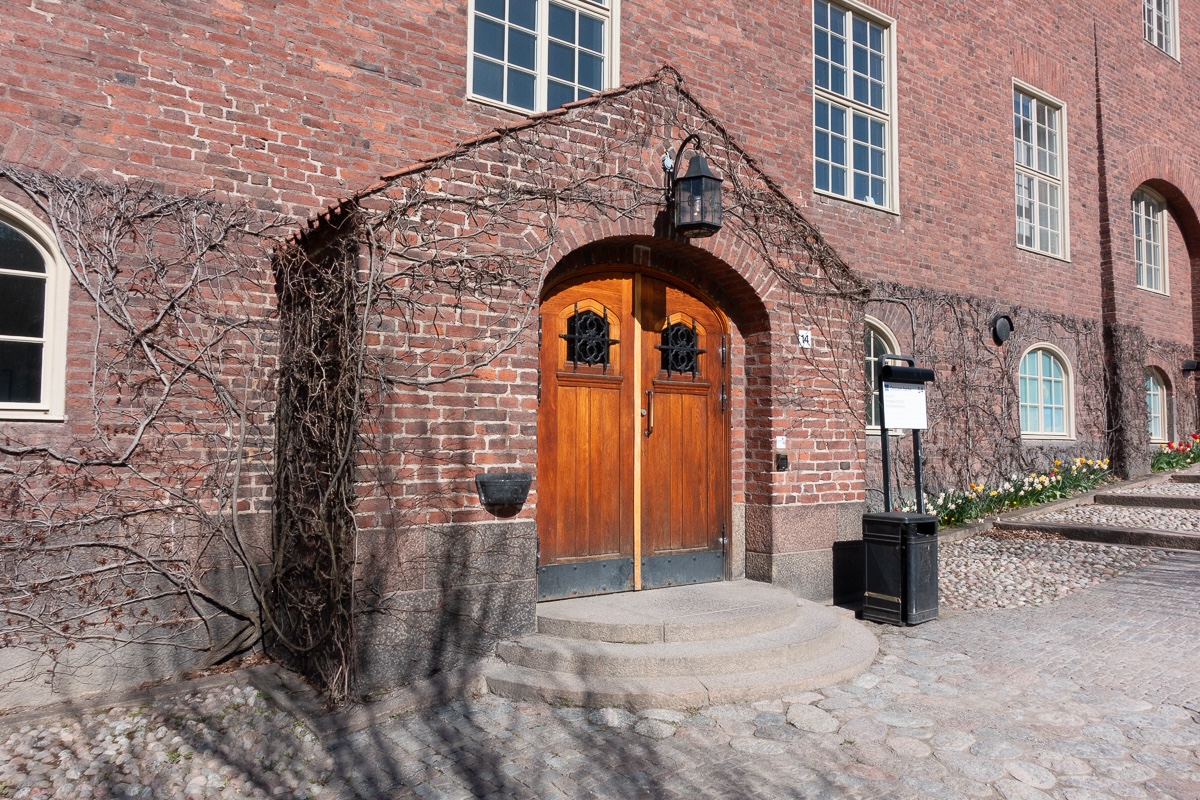}
      & \includegraphics[width=0.31\textwidth]{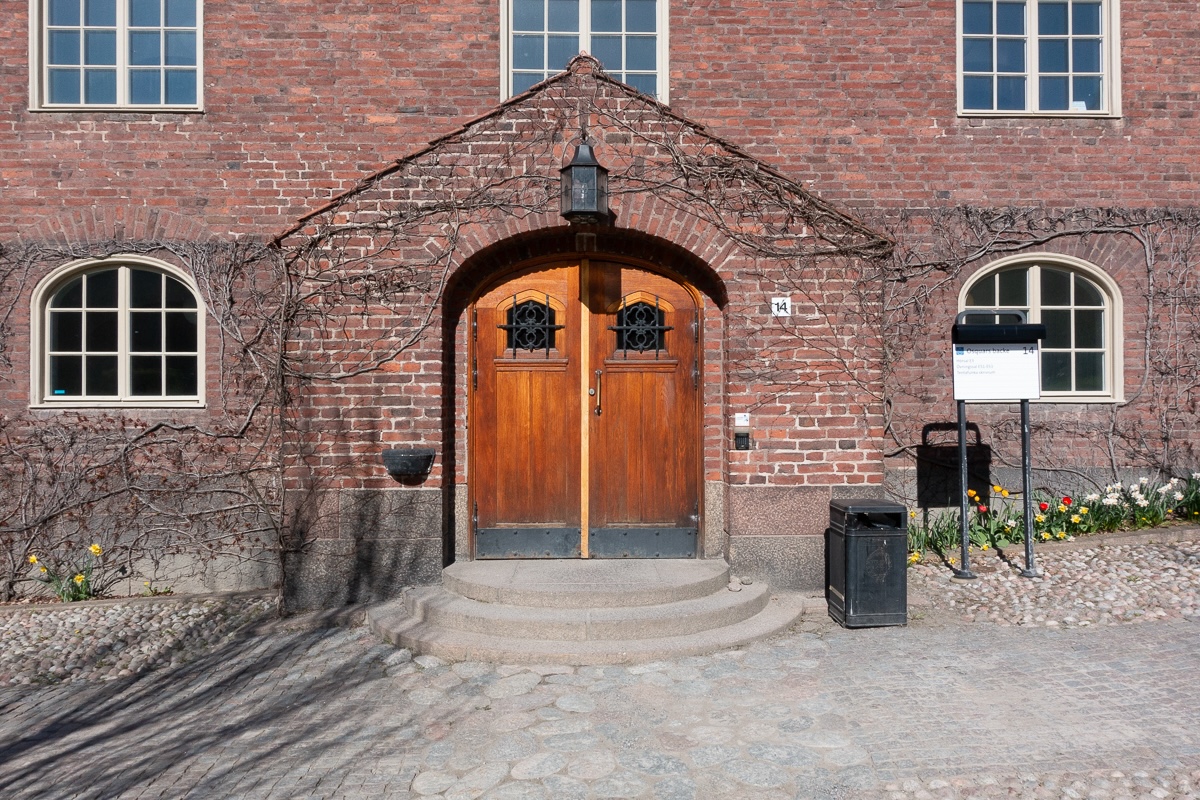}        
    \end{tabular}
    
  \end{center}
  \caption{{\em Basic types of variabilities in natural image data as caused by geometric image transformations:\/}
  {\bf (top row)} Uniform scaling transformations caused by varying
  the distance between the object and the observer. {\bf (middle row)}
  Rotations in the image domain caused by relative rotations between
  the object and the observer around the optical axis. {\bf (bottom row)} Foreshortening transformations that lead to non-uniform scaling transformations with the tilt direction (the projection of the surface normal to the image plane) as a preferred orientation in the image plane. Notably, despite these variabilities in the image domain, our perceptual vision system is able to perceive the objects in the world as stable, although the receptive fields, which operate on the set of image values over local regions in image space, will then be exposed to input information that may differ significantly between different views of the same object. In this work, we explore the consequences of using covariance properties of the visual receptive fields as a fundamental constraint, to make it possible to compare and relate the receptive field responses from different views, with {\em a priori\/} possible unknown geometric image transformations between such views.}
  \label{fig-natural-img-transf}
\end{figure*}

When viewing a 3-D object in the environment from different distances
and different viewing directions, the projected 2-D images on either
the retina or the camera sensor will be subject to a substantial
variability, as caused by the variability in the viewing conditions
(see Figure~\ref{fig-natural-img-transf} for illustrations).

To first order of approximation, by approximating the perspective
mappings from a smooth local surface patch on the object to any two
different perspective images by local linearizations (first-order
derivatives), the resulting variability of resulting image data from
such multi-view observations of the same surface patch can be modelled
in terms of local affine transformations of the form
\begin{equation}
  \label{eq-aff-transf}
  x' = {\cal A} \, x.
\end{equation}
For simplicity, we have here discarded a
possible complementary variability with respect to an added
translation vector $b$ in the image domain of the
form $x' = {\cal A} \, x + b$, by without essential loss
of generality assuming that all the multi-view observations of the
same surface patch are focused on the same viewing point $P$ on the
3-D object. The origins of the coordinate systems in the two
image domains are also assumed to correspond to the projections of that same fixation point.

A non-trivial, while sometimes overlooked, aspect of visual perception is that we perceive an object in the world as the same, although the 2-D perspective projections of such an object can differ substantially, depending on the viewing distance and the viewing direction of the observer.

A conceptual question that one could then ask is if the ability of
perceptual system to derive seemingly stable representations of
external objects in the 3-D environment would lead to constraints
regarding the variabilities in the shapes of the receptive fields at
the earliest levels in the visual processing hierarchy. In a
theoretically motivated and axiomatically determined 
normative theory of visual receptive fields (Lindeberg
\citeyear{Lin13-BICY,Lin21-Heliyon}), idealized models for the
receptive fields of simple cells in the primary visual cortex (V1) have
specifically been derived. These receptive field models lead to provable covariance properties
under spatial affine transformations, and regarding extensions to
time-dependent spatio-temporal image data also provable covariance
properties under Galilean transformations, to handle the variabilities in
the relative motion between the objects
in the world and the observer.
See Lindeberg (\citeyear{Lin23-FrontCompNeuroSci,Lin25-JMIV})
for more in-depth treatments regarding the importance of
covariance properties for visual receptive fields under
geometric image transformations.

\begin{figure*}[hbt]
    \begin{center}
    \begin{tabular}{cc}
      {\em\small (a) Non-covariant receptive fields\/}
      $\quad$ & $\quad$
     {\em\small (b) Covariant receptive fields\/} \\
      \includegraphics[width=0.35\textwidth]{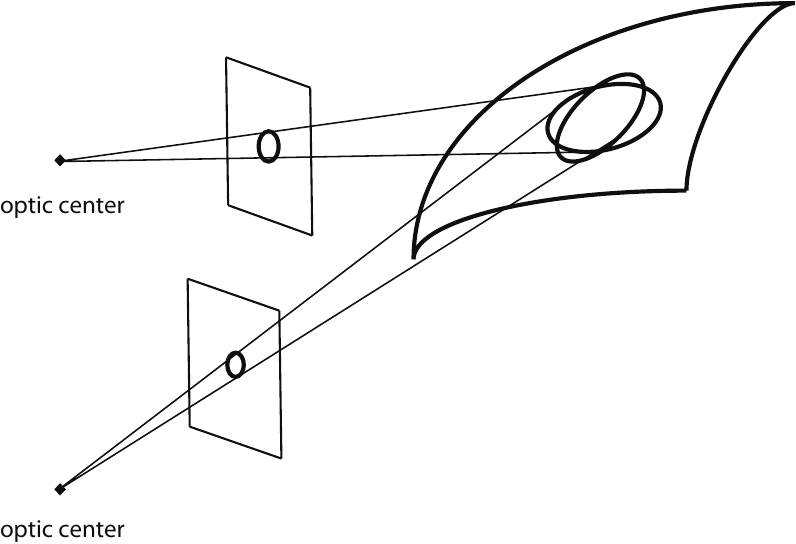}
      $\quad$ & $\quad$
      \includegraphics[width=0.35\textwidth]{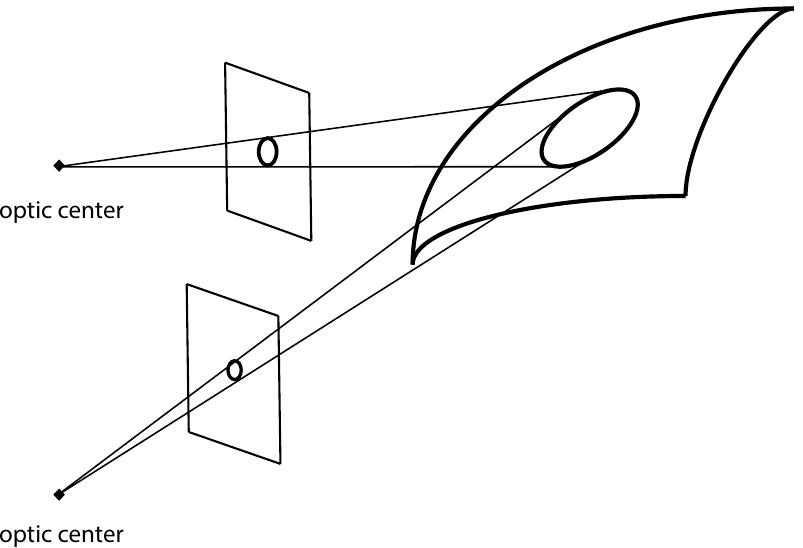} \\
    \end{tabular}
  \end{center}
  \caption{{\em Geometric illustration of the motivation for the underlying hypothesis
    concerning affine covariant receptive fields in biological vision\/}.
    The importance of the notion of covariance for the receptive fields
    under geometric image transformations originates from the
    difference in the backprojected receptive fields that will result,
    if the receptive fields are not covariance.
    (a)~In the left figure, which shows backprojections of non-covariant
    receptive fields to the tangent plane of a local surface patch, it is clear
    that largely different support regions in the tangent plane to the surface
    will affect the receptive field responses for observations of the same
    surface patch from different viewing directions.
    (b) In the right figure, which shows backprojections of covariant receptive
    fields, for which the parameters of the receptive fields have been additionally
    matched, such that the backprojections of the receptive fields for the two
    different observations are equal, the contributions from the different
    image points within the spatial support regions of the visual receptive
    fields will contribute in a similar manner to the receptive field responses,
    which implies that the mismatch source of error to cues about properties
    of the environment will, to first order of approximation, be completely eliminated.
    (Figures adapted from Lindeberg  (\citeyear{Lin23-FrontCompNeuroSci})
  with permission (OpenAccess).)}
  \label{fig-ill-cov-rfs}
\end{figure*}

From a computational viewpoint, if we regard the earliest layers in
the visual perception system as a computational engine, that
infers successively more complex cues about the structure of the
environment from the image measurements that reach the retina, then
covariance properties of visual receptive fields do from a theoretical
perspective constitute a highly useful notion. Specifically, if we
design a computational algorithm for computing the local surface
orientation of a smooth surface patch from binocular cues, then
compared to not basing the computations on covariant spatial receptive
fields, the use of provably affine-covariant visual receptive
fields can improve the accuracy by an order of magnitude in estimates
of the local surface orientation (see Tables~1--4 in Lindeberg and
G{\aa}rding (\citeyear{LG96-IVC})). This is achieved by eliminating a source of error
corresponding to a mismatch between the backprojected receptive fields
to the tangent plane of the surface, from the two different
perspective views, see Figure~\ref{fig-ill-cov-rfs} for an
illustration.

Given these theoretically motivated considerations, which have also
been empirically tested in computer vision algorithms, one could then
raise the question whether the biological vision systems in higher
mammals could be regarded as obeying covariance properties under
affine spatial transformations, and thus spanning a variability over
the degrees of freedom of 2-D spatial affine transformations. 

The subject of this paper is to address this issue, by:
\begin{itemize}
\item
  First theoretically describing the different degrees of freedom in
  spatial affine transformations, in terms of a canonical
  parameterization, based on a closed-form factorization
  of the affine transformation matrix ${\cal A}$ very closely
  related to a singular value decomposition.
\item
  Then, relating
  these degrees of freedom to the degrees of freedom in the parameters
  of the idealized receptive models of simple cells in the primary
  visual cortex, in terms of the generalized Gaussian derivative theory
  for visual receptive fields.
\item
  Finally, relating these results
  to existing evidence regarding variabilities in the shapes of
  biological receptive fields established from neurophysiological
  measurements.
\end{itemize}
Specifically, we will
after an initial overview of related work in Section~\ref{sec-rel-work},
in Section~\ref{sec-dofs-2d-affine} describe a decomposition of
2-D spatial affine image transformations, based on a closed-form decomposition of
the affine transformation matrix in a way very closely related to a singular
value decomposition. The added value of this parameterization is that
it reveals the degrees of freedom of 2-D
spatial affine image transformations in a very geometric manner.

Then, after reviewing the covariance properties of the affine
Gaussian derivative model for visual receptive fields in
Section~\ref{sec-cov-aff-rfs}, we will in
Section~\ref{sec-dofs-aff-gauss-der-rfs} explicitly describe the
degrees of freedom in the affine Gaussian derivative model for visual
receptive fields.

Based on this theoretical basis, we will then in
Section~\ref{sec-rels-var-aff-transf-aff-gauss-ders} relate
the degrees of freedom in the affine Gaussian derivative model to the degrees
of freedom revealed by our proposed decomposition of
2-D affine transformations. This is achieved in terms of explicit mappings between the
parameters in our decomposition of 2-D spatial affine transformations
and the parameters in the affine Gaussian derivative model. These
connections will then be used in
Section~\ref{sec-dofs-spanned-by-V1-RFs} for addressing the question of
whether we can regard the receptive fields in the primary visual
cortex as being able to span the degrees of freedom of 2-D affine
transformations.

Finally, Section~\ref{sec-summ-disc} gives a summary of the main
results, with suggestions to further neurophysiological studies to
firmly determine to what extent the presented biological hypotheses
hold in the primary visual cortex, and then also for what species, as
well as outlooks concerning further implications of the presented
theoretical results.

The presented results are aimed at providing a theoretical
understanding of basic covariance properties for the earliest layers
of receptive fields in
computational vision, as well as for generating hypotheses about the
computational functions in the earliest levels of visual processing
in biological vision.

A main significance of this study is that it
provides a theoretical framework for how the variability in the spatial
receptive field shapes in higher mammals can be accounted for
based on symmetry properties of the environment.
In particular, the theoretical relations to be derived reveal the practical usefulness for the visual
system to explicitly handle geometric image transformations
from the 3-D environment to the 2-D visual patterns registered
in the earliest layers of the visual hierarchy.

More specifically,
by expanding the receptive field shapes over the degrees of freedom
of the geometric image transformations, the visual system will have
the possibility to
compute explicitly readable signals to the higher cortices, that are
suitable for a variety of different imaging conditions, for extracting
invariant features from the objects that are projected to the retina.

In this way, the visual system can meet the huge variability in
spatial or spatio-temporal image
structures, that are generated from different possible imaging
conditions of any object in the environment
by a corresponding variability in the receptive field shapes
in the primary visual cortex. Thereby, it is maintaining a notion of identity
between measurements of the same object performed under different
geometric configurations between the object and the observer.

An important consequence of the covariance property of receptive
fields, used as a theoretical foundation in following treatment,
is that it makes it possible for receptive field responses to be matched
between different viewing conditions. By the notion of geometric
covariance, the visual system will be able to meet the variability in
image structures generated by geometric image transformations by a
corresponding variability in receptive field shapes. This implies that the
receptive field responses will, to first order of approximation,
be the same under the geometric image
transformations, and thus achieving a constancy of the representation,
given that the parameters of receptive fields can be matched
to the geometric image transformations.

Thus, in summary, the proposed theory predicts a set of variabilities
in receptive field shapes as induced by corresponding variabilities in
image structures, and as generated by the variabilities of geometric image transformations.

\section{Relations to previous work}
\label{sec-rel-work}

Characterizing the functions of the receptive fields in
the early visual pathway in terms of computational models constitutes
a main topic in the task of understanding the functionalities in the
primary visual cortex.

With regard to the task of characterizing the functionality of simple
cells, neurophysiological recordings of receptive field profiles of
simple cells in the primary visual cortex have
been performed by, among others, 
DeAngelis {\em et al.\/}\
(\citeyear{DeAngOhzFre95-TINS,deAngAnz04-VisNeuroSci}),
Ringach (\citeyear{Rin01-JNeuroPhys,Rin04-JPhys}),
Conway and Livingstone (\citeyear{ConLiv06-JNeurSci}),
Johnson {\em et al.\/}\ (\citeyear{JohHawSha08-JNeuroSci}),
Ghodrati {\em et al.\/} (\citeyear{GhoKhaLeh17-ProNeurobiol}),
Walker {\em et al.\/} (\citeyear{WalSinCobMuhFroFahEckReiPitTol19-NatNeurSci})
and De and Horwitz (\citeyear{DeHor21-JNPhys}).

The receptive field shapes of simple cells have, in turn, been modelled
mathematically in terms of, most commonly, either Gabor filters by
Marcelja (\citeyear{Mar80-JOSA}),
Jones and Palmer (\citeyear{JonPal87a,JonPal87b}),
Porat and Zeevi (\citeyear{PorZee88-PAMI}),
Ringach (\citeyear{Rin01-JNeuroPhys,Rin04-JPhys}),
or Gaussian derivatives by
Koenderink and van Doorn (\citeyear{Koe84,KoeDoo87-BC,KoeDoo92-PAMI})
Young and his co-workers (\citeyear{You87-SV,YouLesMey01-SV,YouLes01-SV}) and
Lindeberg (\citeyear{Lin13-BICY,Lin21-Heliyon}).
See specifically Figures~16 and~17 in Lindeberg
(\citeyear{Lin21-Heliyon}) for examples of how spatial receptive
fields of simple cells in the primary visual cortex can be modelled in
terms of affine Gaussian derivatives.

Functional modelling of different processes in biological vision has
also specifically been performed in terms of Gaussian derivatives by
Lowe (\citeyear{Low00-BIO}),
May and Georgeson (\citeyear{MayGeo05-VisRes})
Hesse and Georgeson (\citeyear{HesGeo05-VisRes}),
Georgeson  {\em et al.\/}\ (\citeyear{GeoMayFreHes07-JVis}),
Wallis and Georgeson (\citeyear{WalGeo09-VisRes}),
Hansen and Neumann (\citeyear{HanNeu09-JVis}),
Wang and Spratling (\citeyear{WanSpra16-CognComp}) and
Pei {\em et al.\/}\ (\citeyear{PeiGaoHaoQiaAi16-NeurRegen}).

With regard to spatial scaling transformations of the image data that
is registered by a visual observer, 
at a higher level of abstraction, evidence for processing over
multiple scales, with scale invariance constituting a basic functional
property in biological vision,
have been presented by
Biederman and Cooper (\citeyear{BieCoo92-ExpPhys}),
Logothetis {\em et   al.\/} (\citeyear{LogPauPog95-CurrBiol}),
Ito {\em et al.\/} (\citeyear{ItoTamFujTan95-JNeuroPhys}),
Furmanski and Engel (\citeyear{FurEng00-VisRes}),
Hung {\em et al.\/} (\citeyear{HunKrePogDiC05-Science}),
Wiskott (\citeyear{Wis06-ProblSystNeuroSci}),
Isik {\em et al.\/} (\citeyear{IsiMeyLeiPog13-JNPhys}),
Murty and Arun (\citeyear{MurAru17-eNeuro}),
Benvenuti {\em et al.\/} (\citeyear{BenCheRamDeiGeiSei18-Neuron})
and
Han  {\em et al.\/}  (\citeyear{HanRoiGeiPog20-SciRep}).

Studies of the geometric relationships between pairwise binocular
views have been presented by
Koenderink and van Doorn (\citeyear{KoeDoo76-BICY}),
Jones and Malik (\citeyear{JM92-ECCVa}),
G{\aa}rding and Lindeberg (\citeyear{GL94-ECCV}),
Mitiche and L{\'e}tang (\citeyear{MitLet98-RobAutSyst}),
Acz{\'e}l  {\em et al.\/}\ (\citeyear{AczBorHelNg00-MathPsych}),
Uka and DeAngelis (\citeyear{UkaDeAng02-CurrBio}),
Hansard and Horaud (\citeyear{HanHor09-JOSA}) and
Turski (\citeyear{Tur16-VisRes,Tur24-SciRep}), and more general
studies of the image geometry between multiple views by
Hartley and Zisserman (\citeyear{HarZis04-Book}) and
Faugeras (\citeyear{Faug-book}).

Concerning invariances of image features to geometric image transformations, invariant
representations have been modelled in terms of group theory by
Poggio and Anselmi  (\citeyear{PogAns16-book}).
An overview of computational methods to
achieve invariance to scaling transformations and locally linearized
perspective transformations for monocular cues, as well as locally
linearized projections between projective pairwise views, within a
larger framework that also involves invariances to Galilean
transformations in joint space-time, has been given in
(Lindeberg \citeyear{Lin13-PONE}). Specifically, it is proposed in
(Lindeberg \citeyear{Lin13-BICY,Lin13-PONE,Lin21-Heliyon})
how covariance properties of the
receptive fields in the earliest layers of the visual hierarchy make
it possible to compute invariant image features at the higher levels,
to, for example, support invariant object recognition.

In view of such a computational view to geometric invariance
and covariance properties, one may therefore ask if the receptive fields in
the primary visual cortex would be covariant under the basic classes
of geometric image transformations, as proposed in
(Lindeberg \citeyear{Lin23-FrontCompNeuroSci}).

The subject of this paper is to address this issue with regard to the
special case of spatial affine image transformations, and specifically
based on a canonical decomposition of such 2-D spatial affine
transformations. This decomposition is in terms of a product form of
(i)~a uniform scaling transformation,
(ii)~pure rotations, (iii)~a non-uniform scaling transformation, and (iv)~a normalization of
the non-uniform scaling transformation to a preferred symmetry
orientation in the image domain.

An earlier study of parameterizations of affine transformations, which
we will build upon in this work, was presented in
(Lindeberg \citeyear{Lin95-ICCV}). In this paper, we extend that
parameterization by first of all providing an explicit derivation
in Appendix~\ref{sec-app-deriv-2d-decomp}, which was not presented
in the original paper because of space limitations. Then, we analyse
the properties of this decomposition for four basic classes of
primitive affine transformations in Appendix~\ref{app-aff-decomp-spec-cases}.
We do additionally give a more general treatment regarding
special cases for the parameters in the affine transformation model,
which thereby increases the domain of
applicability of the proposed parameterization to more general families of
affine transformation matrices, that are reasonably near the unit matrix
multiplied by a scalar scaling factor.

Then, we will relate the degrees of freedom obtained from this
decomposition arising parameterization of 2-D spatial affine
transformations to the variabilities in the shapes of the receptive
fields in the primary visual cortex.

\begin{figure*}[hbtp]
  \begin{center}
    \begin{tabular}{cc}
      {\footnotesize\em Pure scaling transformation for $S_x = 1.2$}
      & {\footnotesize\em Pure rotation for $\gamma = \frac{\pi}{16}$} \\
      {\footnotesize ($\rho_1 = S_x$, $\rho_2 = S_x$, $\varphi = 0$, $\psi = 0$)}
      & {\footnotesize ($\rho_1 = 1$, $\rho_2 = 1$, $\varphi = \gamma$, $\psi = 0$)} \\
      \includegraphics[width=0.32\textwidth]{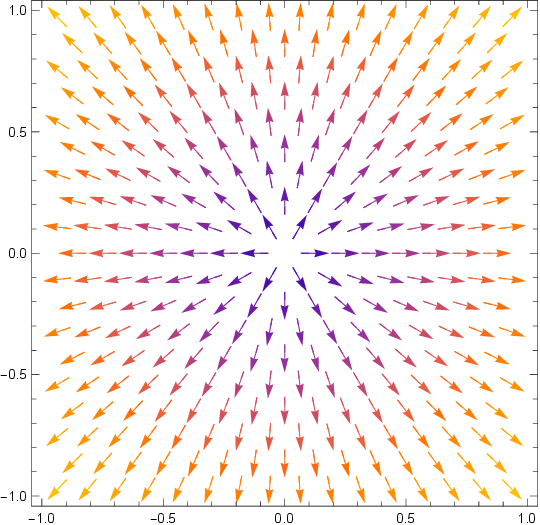}
      &
        \includegraphics[width=0.32\textwidth]{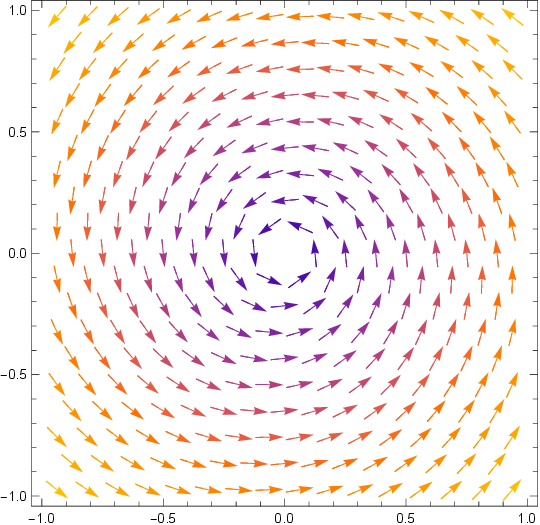}
      \\
      $\,$ \\
     {\footnotesize\em Non-uniform scaling for $S_1 = 1.2$,
      $S_2 = 0.8$ and $\gamma = \frac{\pi}{4}$}
      & {\footnotesize\em Pure skewing transformation for $\gamma = \frac{\pi}{8}$} \\
      {\footnotesize ($\rho_1 = S_1$, $\rho_2 = S_2$, $\varphi = 0$, $\psi = \gamma$)}
      & {\footnotesize ($\rho_{1,2} = \sqrt{1 + \frac{\tan^2 \gamma}{4}} \pm
        \frac{| \tan \gamma |}{2}$, $\varphi = -\arctan
        \left(\frac{\tan \gamma}{2} \right)$,
        $\psi = - \frac{\pi}{2} \, \sign \gamma$)} \\
\includegraphics[width=0.32\textwidth]{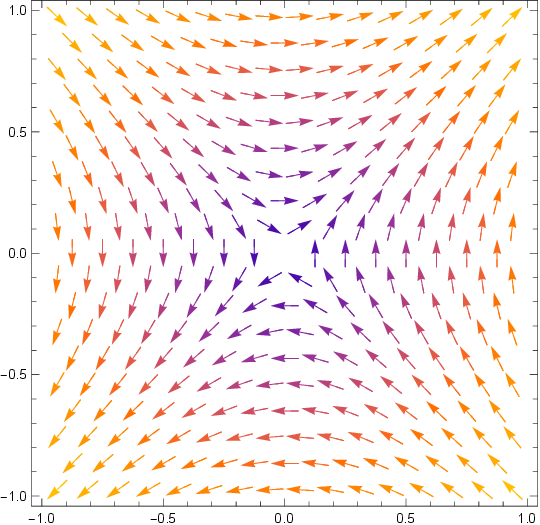}
      &
        \includegraphics[width=0.32\textwidth]{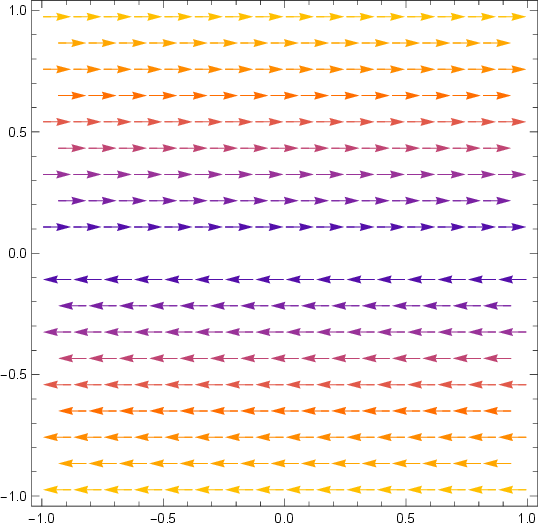}
      \\
      $\,$ \\
       {\footnotesize\em Composed affine transformation}
      & {\footnotesize\em  Composed affine transformation} \\
      {\footnotesize ($\rho_1 = 1.2$, $\rho_2 = 0.8$,
      $\varphi = \frac{\pi}{32}$, $\psi = \frac{\pi}{4}$)}
      & {\footnotesize ($\rho_1 = 1.3$, $\rho_2 = 1.1$, $\varphi = -\frac{\pi}{32}$, $\psi = \frac{2\pi}{3}$)} \\ \includegraphics[width=0.32\textwidth]{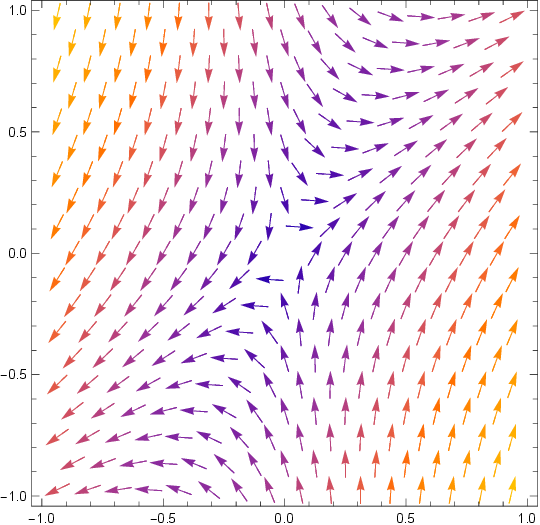}
      & \includegraphics[width=0.32\textwidth]{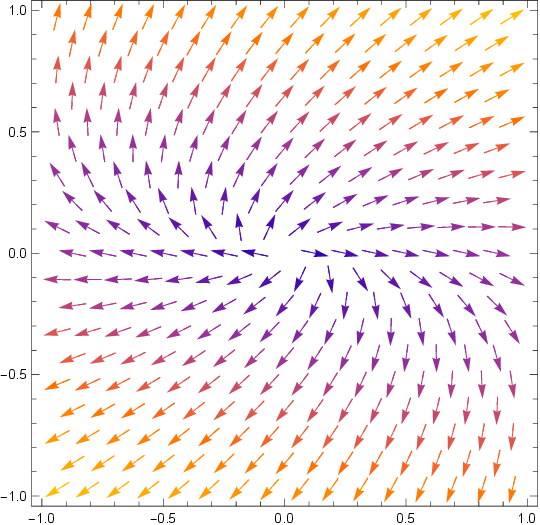}
      \\
    \end{tabular}
  \end{center}
  \caption{{\em Visualizations of deformation fields\/}
    $\Delta x = x' - x = {\cal A} \, x - x = ({\cal A} - {\cal I}) \, x$
    {\bf (top and middle rows)} for four basic types of
    primitive affine transformations and {\bf (bottom row)} two
    composed affine image transformations, with the directions of the
    deformation vectors $\Delta x$ shown as arrows of unit length, and
  with the magnitude of the deformation coded in colour, with blue
  corresponding to low magnitudes and yellow to higher magnitudes.
  For illustration, also the parameters ($\rho_1$, $\rho_2$,
  $\varphi$, $\psi$) in the proposed decomposition of the affine
  transformation matrix according to (\ref{eq-aff-decomp}) are also
  shown, based on the detailed analysis of these four special cases of
  primitive affine image transformations in
  Appendix~\ref{app-aff-decomp-spec-cases}, with specifically the
  pure scaling transformation matrix according to (\ref{eq-def-scale-mat-app}), the
  pure rotation matrix according to (\ref{eq-def-rot-mat-app}), the non-uniform scaling
  transformation according to (\ref{eq-non-uni-scale-transf-app}) and
  the skewing transformation according to (\ref{eq-def-skew-mat-app}).
  ({\bf Horizontal axes:} spatial coordinate $x_1$.
  {\bf Vertical axes:} spatial coordinate $x_2$.)}
  \label{fig-basic-def-fields}
\end{figure*}

\section{The degrees of freedom in 2-D spatial affine transformations}
\label{sec-dofs-2d-affine}

Let us parameterize the affine transformation matrix ${\cal A}$ in the
affine transformation (\ref{eq-aff-transf}) as
\begin{equation}
  {\cal A}
  =
  \left(
    \begin{array}{cc}
      a_{11} & a_{12} \\
      a_{21} & a_{22}
    \end{array}
  \right)
\end{equation}
and parameterize the image coordinates as $x = (x_1, x_2)^T$ and $x' = (x'_1, x'_2)^T$.
Here, we are specifically interested in the special case when the
affine transformation matrix ${\cal A}$ is reasonably close to the
unit matrix ${\cal I}$
multiplied by some scalar scaling factor.

Following Section~3 in (Lindeberg \citeyear{Lin95-ICCV}),
let us next define the following descriptors from the elements $a_{ij}$ of ${\cal A}$:
\begin{align}
  \begin{split}
     T & = \frac{a_{11} + a_{22}}{2},
  \end{split}\\
  \begin{split}
    A & = \frac{a_{21} - a_{12}}{2},
  \end{split}\\
  \begin{split}
    C & = \frac{a_{11} - a_{22}}{2},
  \end{split}\\
  \begin{split}
    S & = \frac{a_{12} + a_{21}}{2}.
  \end{split}
\end{align}
Let us also define the following derived descriptors
\begin{align}
 \begin{split}
    P  & = \sqrt{T^2 + A^2},
  \end{split}\\
  \begin{split}
    Q & = \sqrt{C^2 + S^2}.
  \end{split}
\end{align}
Then, according to the treatment in
Appendix~\ref{sec-app-deriv-2d-decomp},
specifically
Equations~(\ref{eq-expr-rho1})--(\ref{eq-expr-rho2}),
it follows that, if we decompose the affine transformation matrix ${\cal A}$
into a modified singular value decomposition of the form
\begin{equation}
  \label{eq-decomp-A}
   {\cal A} = {\cal R}_{\alpha} \, \diag(\rho_1, \rho_2) \, {\cal R}_{\beta}^T,
\end{equation}
where
\begin{equation}
  {\cal R}_{\alpha}
  =
  \left(
    \begin{array}{cc}
       \cos \alpha & - \sin \alpha \\
       \sin \alpha & \cos \alpha 
    \end{array}
  \right)
\end{equation}
and 
\begin{equation}
  {\cal R}_{\beta}
 =
 \left(
    \begin{array}{cc}
       \cos \beta & - \sin \beta \\
       \sin \beta & \cos \beta
    \end{array}
  \right)
\end{equation}
are enforced to be pure rotation matrices,
then
the diagonal entries $\rho_1$ and $\rho_2$
in the diagonal matrix $\diag(\rho_1, \rho_2)$
in the decomposition (\ref{eq-decomp-A}) are given by
\begin{align}
 \begin{split}
    \rho_1 & = P + Q,
  \end{split}\\
  \begin{split}
    \rho_2 & = P - Q.
  \end{split}
\end{align}
Due to our assumption of the affine transformation being reasonably close to the
unit matrix ${\cal I}$ multiplied by some scalar scaling factor,
we will for the purpose of this treatment
specifically assume that $\rho_1 > \rho_2 > 0$,
such that the decomposition (\ref{eq-decomp-A}) also constitutes a
genuine singular value decomposition%
\footnote{Note that in a general singular value decomposition of an
  arbitrary affine transformation matrix ${\cal A}$ of the form
  ${\cal A} = U \, \Sigma \, V^T$, the matrices $U$ and $V$ are only restricted to
  be unitary matrices, not necessarily rotation matrices.}
of the affine transformation matrix ${\cal A}$, with the notable
distinction that the matrices ${\cal R}_{\alpha}$ and ${\cal R}_{\beta}$
are here guaranteed to be actual rotation matrices.
    
If we furthermore define the following angles 
\begin{align}
 \begin{split}
    \tan \varphi & = \frac{A}{T},
  \end{split}\\
  \begin{split}
    \tan \psi & = \frac{S}{C},
  \end{split}
\end{align}
as well as the corresponding rotation matrices
\begin{equation}
  {\cal R}_{\frac{\varphi}{2}}
  =
  \left(
    \begin{array}{cc}
       \cos \frac{\varphi}{2} & - \sin \frac{\varphi}{2} \\
       \sin \frac{\varphi}{2} & \cos \frac{\varphi}{2}
    \end{array}
  \right)
\end{equation}
and
\begin{equation}
  {\cal R}_{\frac{\psi}{2}}
  =
  \left(
    \begin{array}{cc}
       \cos \frac{\psi}{2} & - \sin \frac{\psi}{2} \\
       \sin \frac{\psi}{2} & \cos \frac{\psi}{2}
    \end{array}
  \right),
\end{equation}
then it can be shown (see Appendix~\ref{sec-app-deriv-2d-decomp}
specifically Equations~(\ref{eq-aff-decomp-raw-app})
and~(\ref{eq-aff-decomp-app})) that
the singular value decomposition of the affine transformation matrix ${\cal A}$,
can be factorized on the form
\begin{align}
  \begin{split}
    {\cal A}
    & = {\cal R}_{\frac{\psi}{2}} \, {\cal R}_{\frac{\varphi}{2}}
           \diag(\rho_1, \rho_2) \, 
           {\cal R}_{\frac{\varphi}{2}} \, {\cal R}_{-\frac{\psi}{2}}
  \end{split}\nonumber\\
  \begin{split}
    \label{eq-aff-decomp}
    & = \sqrt{\rho_1 \, \rho_2} \,\,
           {\cal R}_{\frac{\psi}{2}} \, {\cal R}_{\frac{\varphi}{2}}
           \diag(\sqrt{\frac{\rho_1}{\rho_2}}, \sqrt{\frac{\rho_2}{\rho_1}}) \, 
           {\cal R}_{\frac{\varphi}{2}} \, {\cal R}_{-\frac{\psi}{2}},
  \end{split}
\end{align}
where
\begin{itemize}
\item
  $\rho_1$ and $\rho_2$ are the singular values of ${\cal A}$,
\item
  the factor
  \begin{equation}
    \label{eq-def-overall-sc-factor-aff}
    S = \sqrt{\rho_1 \, \rho_2}
  \end{equation}
  corresponds to an overall
  spatial scaling transformation,
\item
  the diagonal matrix
  \begin{equation}
    \label{eq-def-noniso-diag-aff}
    {\cal D} = \operatorname{diag}(\sqrt{\frac{\rho_1}{\rho_2}},
                                             \sqrt{\frac{\rho_2}{\rho_1}})
  \end{equation}
   describes a non-uniform stretching transformation, corresponding
   to a relative spatial stretching factor of
   \begin{equation}
     \lambda = \frac{\rho_1}{\rho_2},
   \end{equation}
\item
    ${\cal R}_{\frac{\psi}{2}}$ and ${\cal R}_{\frac{\varphi}{2}}$ are rotation
    matrices with rotation angles $\frac{\psi}{2}$ and $\frac{\varphi}{2}$,
    respectively.
\end{itemize}
In terms of overall effects of the affine transformation, the
interpretations of the two rotation angles $\varphi$ and $\psi$
are then, specifically, that:
\begin{itemize}
\item
  $\varphi = \frac{\varphi}{2} + \frac{\varphi}{2}$
  describes the overall total amount of spatial rotation,
\item
  $\frac{\psi}{2}$ represents the orientation of a local symmetry axis for the
  non-isotropic part of the affine transformation, 
  {\em e.g.\/} the pure stretching transformation.
\end{itemize}
The top and the middle rows in Figure~\ref{fig-basic-def-fields} show
examples of spatial deformation fields
for four basic classes of primitive affine transformations, with the parameters for each one of the different special cases
determined according to the analysis of these four special cases in
Appendix~\ref{app-aff-decomp-spec-cases}.
The bottom row in Figure~\ref{fig-basic-def-fields} does additionally
show examples of composed affine deformation fields, as specified from the
parameters $(\rho_1, \rho_2, \varphi, \psi)$ in the proposed 
decomposition of affine transformation matrices according to (\ref{eq-aff-decomp}).

\begin{figure*}[hbtp]
  \begin{center}
    \begin{tabular}{cc}
      {\footnotesize\em Monocular deformation field}
      & {\footnotesize\em Binocular deformation field}  \\
      {\footnotesize ($\Lambda = 1.4$, $\nu = \frac{\pi}{4}$)}
      & {\footnotesize ($Z_X = -0.5$, $Z_Y = 1$,
        $\gamma= \frac{\pi}{4}$, $\mu = \frac{\pi}{16}$)} \\
      \includegraphics[width=0.32\textwidth]{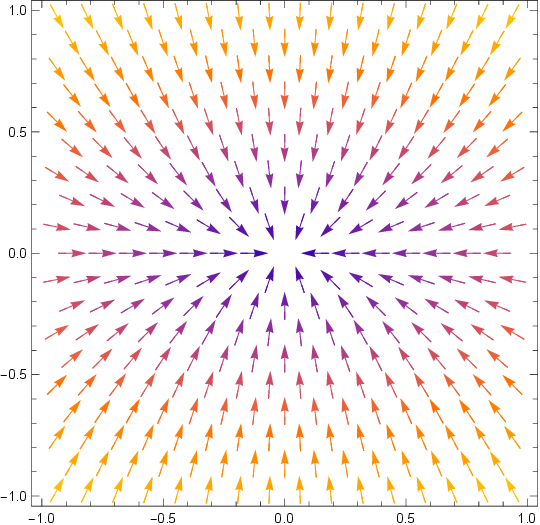}
      &
        \includegraphics[width=0.32\textwidth]{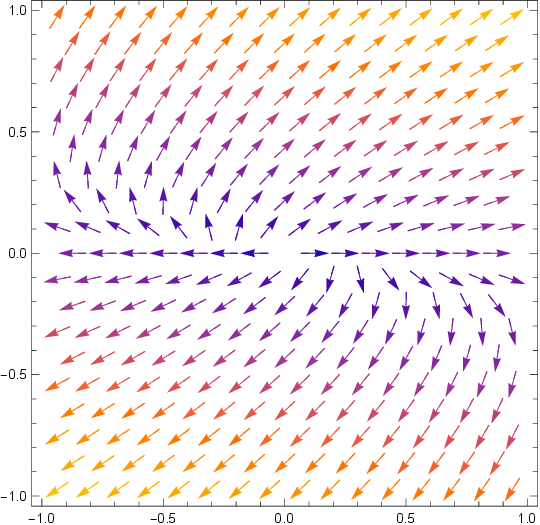}
      \\
   \end{tabular}
  \end{center}
  \caption{{\em Visualizations of deformation fields\/}
    $\Delta x = x' - x = {\cal A} \, x - x = ({\cal A} - {\cal I}) \, x$ from
    locally linearized monocular and binocular perspective or projective
    projection models. {\bf (left)} A monocular deformation field as arising
    from a locally linearized perspective projection according to
    (\ref{eq-monocular-proj-model-aff}) of a slanted surface with
    slant angle $\nu$ under variations of
    the distance $\Lambda$ between the object and the observer.
    {\bf (right)} A binocular deformation field as arising from a
    locally linearized projective mapping between the tangent planes of
    two smooth surface according to
    (\ref{eq-binocular-proj-model-aff}), for specific values of the
    depth gradient $\nabla Z = (Z_X, Z_Y)$, the gaze angle $\gamma$
    between the frontal direction and direction to the fixation point as
    well as the opening angle $2\mu$ between the viewing directions of
    the two observers. ({\bf Horizontal axes:} spatial coordinate $x_1$.
    {\bf Vertical axes:} spatial coordinate $x_2$.)}
  \label{fig-geom-def-fields-mono-bino}
\end{figure*}

\subsection{Geometric interpretations of affine image deformations}

As previously mentioned, let us consider surface patterns of possibly moving
smooth objects in the world are projected to
the image plane, by either (i)~the perspective mapping to a monocular
image,
(ii)~the projective mapping between two binocular views, or
(iii)~the temporally integrated optic flow fields obtained by observing moving objects
between adjacent time moments. Then, the resulting image
transformations or deformation fields can to first order of
approximation be modelled as local affine transformations.

For this purpose, the decomposition in (\ref{eq-aff-decomp}) constitutes a
geometrically very natural way to parameterize the degrees of freedom
in such local affine transformations of the perspective/projective
mappings.
Specifically, these geometric image transformations
determine the variabilities of the image structures,
that a visual system is exposed to,
when observing sufficiently smooth surface patterns in the environment
from different viewing directions relative to the objects in the
environment.

\subsubsection{Monocular locally linearized perspective projection}

To make such relations more explicit, consider, for example, the monocular
perspective projection of a smooth local surface patch onto a
spherical camera. Let us specifically consider two local coordinate systems, with
\begin{itemize}
\item
  the first local coordinate system being in the tangent
  plane of the surface patch, with the second local coordinate
  direction aligned with the tilt%
\footnote{The tilt direction is the direction where the depth along a
    smooth surface varies most strongly.}
  direction of the surface and
\item
  the second local coordinate system
  being in the tangent plane of the
  spherical camera, with the second local coordinate direction aligned with the
  backprojected tilt direction onto the resulting image plane parallel
  to the tangent plane of the surface.
\end{itemize}
Then, the first-order locally
linearized mapping from the tangent plane of the surface to the
tangent plane on the spherical camera is given by
(see G{\aa}rding and Lindeberg (\citeyear{GL94-IJCV}) Equation~(28),
although here reformulated with somewhat different conventions and notation)
\begin{equation}
  \label{eq-monocular-proj-model-aff}
  {\cal A}_{\mono}
  =
  \frac{1}{\Lambda}
  \left(
    \begin{array}{cc}
      1 & 0\\
      0 & \cos \nu
    \end{array}
  \right),
\end{equation}
where
\begin{itemize}
\item
  $\Lambda$ is the distance between the observed point on the
  surface of the object and the observer, and
\item
  $\nu$ is the slant angle, that is the angle between the surface
  normal to the surface of the object and the viewing direction.
\end{itemize}
Figure~\ref{fig-geom-def-fields-mono-bino}(left) shows a schematic illustration of
such a deformation field, as arising from the appearance of
deformations of a local surface pattern on a planar surface to the
image plane, as we by the local linearization disregard higher-order
non-linear components in the perspective mapping.

\subsubsection{Binocular locally linearized projective projection}

In the binocular case, consider two eyes or cameras that observe a smooth
local surface patch from a cyclopean observer.
This means that locus of the cyclopean observer is
on the midpoint between the two optical centers
of the cameras on a circle through the two optical centers 
and the observed point on the surface, see Figure~1 in
G{\aa}rding and Lindeberg (\citeyear{GL94-ECCV}) for an illustration.

Let $2 \mu$ denote the angle between the viewing directions from the two eyes
or cameras to the fixation point, and let $\gamma$ denote the gaze
angle between the frontal direction and the direction to the fixation point.
Furthermore, let $Z_X$ and $Z_Y$ represent
the components of the depth gradient,
with $Z$ denoting the depth, and $X$ and $Y$ being
the world coordinates parallel to the directions of the image
coordinates $x_1$ and $x_2$, respectively. Then, the disparity gradient from the
left image to the right image for a calibrated visual observer is
given by
(see G{\aa}rding and Lindeberg (\citeyear{GL94-ECCV})
Equation~(3))
\begin{align}
  \begin{split}
    {\cal A}_{\bino}
    & =
    \left(
      \begin{array}{cc}
        1 + h_1 & h_2 \\
        v_1 & 1 + v_2
      \end{array}
    \right)
  \end{split}\nonumber\\
  \begin{split}
  \label{eq-binocular-proj-model-aff}    
    & =
    \frac{\cos(\gamma - \mu)}{\cos(\gamma + \mu)}
    \left(
      \begin{array}{cc}
        \frac{\cos \mu + Z_X \, \sin \mu}{\cos \mu - Z_X \, \sin \mu}
        &  \frac{Z_Y \, \sin 2 \mu}{\cos \mu - Z_X \, \sin \mu} \\
        0 & 1
      \end{array}
    \right),
  \end{split}
\end{align}
where $\nabla h = (h_1, h_2)^T$ denotes the horizontal disparity
gradient and $\nabla v = (v_1, v_2)^T$ denotes the vertical disparity gradient.

Figure~\ref{fig-geom-def-fields-mono-bino}(right) shows a schematic example of such a disparity field, as
arising from a local linearization between corresponding points
under the projective mapping between the two image domains
(seen from the two views) for specific
values of the geometric parameters $\mu$, $\gamma$ and $(Z_X, Z_Y)$.

\section{Covariance properties of affine Gaussian derivative based
  receptive fields under spatial affine transformations}
\label{sec-cov-aff-rfs}

To characterize the effects that affine image transformations have on
receptive field responses, let us next consider receptive fields according to
the generalized Gaussian derivative model for visual receptive fields
(Lindeberg \citeyear{Lin21-Heliyon}). With this regard, let us specifically
only consider the effects in relation to the pure spatial and
spatio-temporal smoothing transformations applied to static image data or
video data, respectively.

\subsection{Affine covariance for purely spatial receptive fields}
\label{sec-aff-cov-pure-spat-rfs}

For a purely spatial receptive field applied on pure spatial image
data, the influence of a spatial affine transformation on a
receptive field response can be described as the convolution of any input
image $f(x)$ with an affine Gaussian kernel $g(x;\; \Sigma)$
(see Lindeberg (\citeyear{Lin93-Dis}) Equation~(15.17))
\begin{equation}
  \label{sec-aff-scsp-repr}
  L(x;\; \Sigma) = (g(\cdot;\; \Sigma) * f(\cdot))(x;\; \Sigma)
\end{equation}
where
\begin{equation}
  \label{eq-aff-gauss-def}
  g(x;\; \Sigma)
  = \frac{1}{2 \pi \sqrt{\det \Sigma}} \, e^{-x^T \Sigma^{-1} x/2}.
\end{equation}
Then, it can be shown that, under an affine transformation of the image
domain $f'(x') = f(x)$ for $x' = {\cal A} \, x$ according to (\ref{eq-aff-transf}), the
corresponding purely spatial scale-space representation according to
(\ref{sec-aff-scsp-repr}) over the transformed domain
\begin{equation}
  \label{sec-aff-scsp-repr-prim}
  L'(x';\; \Sigma') = (g(\cdot;\; \Sigma') * f'(\cdot))(x';\; \Sigma')
\end{equation}
is related to the scale-space representation (\ref{sec-aff-scsp-repr})
over the original domain according to
\begin{equation}
  L'(x';\; \Sigma') = L(x;\; \Sigma),
\end{equation}
provided that the spatial covariance matrices $\Sigma'$ and $\Sigma$
in the two images domains (before and after the geometric image
transformation) are related according to
(see Equations~(29) and (30) in Lindeberg and G{\aa}rding (\citeyear{LG96-IVC}))
\begin{equation}
  \label{eq-transf-prop-Sigma}
  \Sigma' = {\cal A} \, \Sigma \, {\cal A}^T.
\end{equation}
Given these raw purely smoothed components of the receptive field
responses, corresponding receptive field responses in terms of spatial derivatives
over the two image domains (before and after the geometric image
transformation),
smoothed by affine Gaussian kernels with
matching spatial covariance matrices $\Sigma$ and $\Sigma'$,
can, in turn, be related according to
\begin{equation}
  \nabla_{x'} = {\cal A}^{-T} \, \nabla_{x}.
\end{equation}
These relationships, thus, show that, provided that spatial covariance
matrices are related in an appropriate way, according to an actual
affine image transformation, it is then possible to match the receptive
field responses perfectly between the two image domains. For such matching
to be possible in an actual situation, the actual value of the spatial
covariance matrices, and hence the shapes of the corresponding
affine Gaussian derivative based receptive fields, must thereby be adapted to
the actual affine transformation. Specifically, if the input stimuli
involve a variability over a certain subspace or subspace of the
degrees of freedom of the affine transformations, then the family of receptive field
shapes must also comprise a variability over a that corresponding subspace
or subdomain of the parameter space of the
affine Gaussian derivative based receptive fields, in order for it to
be possible to perfectly match the outputs from the affine Gaussian derivative
kernels between the two image domains.

\subsection{Affine covariance for joint spatio-temporal receptive fields}
\label{sec-aff-cov-spat-temp-rfs}

A similar analysis can also be performed for joint
spatio-temporal receptive fields according to the idealized model
(see Lindeberg (\citeyear{Lin21-Heliyon}) Equation~(24))
\begin{equation}
  \label{eq-def-spat-temp-rf-model}
  T(x, t;\; \Sigma, \tau, v) = g(x - v t;\; \Sigma) \, h(t;\; \tau)
\end{equation}
where $\tau$ denotes the temporal scale in units of the temporal variance
$\tau = \sigma_t^2$ of the temporal smoothing kernel $h(t;\; \tau)$
and $v = (v_1, v_2)$ denotes an image velocity parameter.

Given two video sequences $f(x, t)$ and $f'(x', t')$,
that are related according to a pure spatial affine transformation
\begin{equation}
  x' = {\cal A} \, x, \quad\quad \mbox{and} \quad\quad t' = t,
\end{equation}
let
\begin{multline}
    L(x, t;\; \Sigma, \tau, v) = \\
    = (T(\cdot, \cdot;\; \Sigma, \tau, v)
       * f(\cdot, \cdot))(x, t;\; \Sigma, \tau, v),
\end{multline}
\begin{multline}  
    L'(x', t';\; \Sigma', \tau', v') = \\
    = (T(\cdot, \cdot;\; \Sigma', \tau', v')
    * f(\cdot, \cdot))(x', t';\; \Sigma', \tau', v'),
\end{multline}
denote the spatio-temporal smoothed scale-space representations in the
two spatio-temporal image domains. Then, these joint spatio-temporal scale-space
representations are related according to
(see Lindeberg (\citeyear{Lin23-FrontCompNeuroSci}) Equation~(54))
\begin{equation}
   L'(x', t';\; \Sigma', \tau', v') = L(x, t;\; \Sigma, \tau, v),
\end{equation}
provided that the spatial covariance matrices $\Sigma$ and $\Sigma'$
are related according to
\begin{equation}
  \Sigma' = {\cal A} \, \Sigma \, {\cal A}^T,
\end{equation}
as well as provided that the other receptive field parameters are equal, that is
$\tau' = \tau$ and $v' = v'$.
Thus, this result shows that also joint spatio-temporal receptive field responses
can be made to perfectly match each other under spatial affine
transformations, provided that the spatial shapes of the joint
spatio-temporal receptive fields are appropriately adapted with
respect to the actual spatial affine transformation.

\begin{figure*}[hbtp]
  \begin{center}
    \begin{tabular}{cccccc}
      $\bar{\sigma} = 1/\sqrt{2}$
      & $\bar{\sigma} = 1$
      & $\bar{\sigma} = \sqrt{2}$
      & $\bar{\sigma} = 2$
      & $\bar{\sigma} = 2\sqrt{2}$
      & $\bar{\sigma} = 4$ \\
      \includegraphics[width=0.14\textwidth]{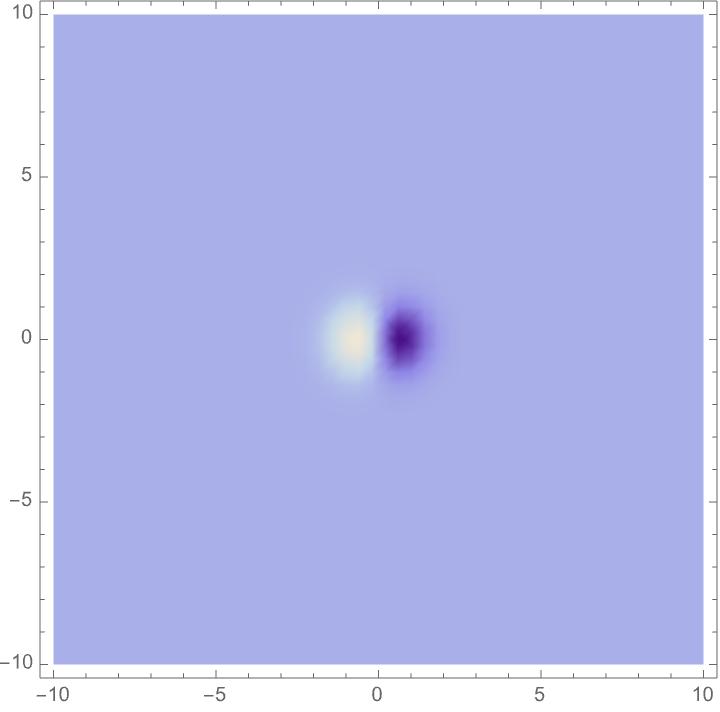}
      & \includegraphics[width=0.14\textwidth]{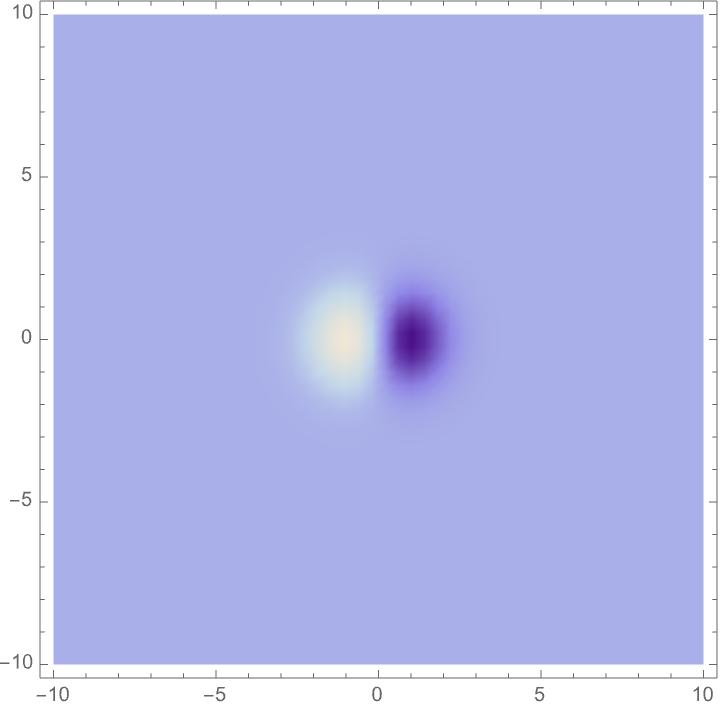}
      & \includegraphics[width=0.14\textwidth]{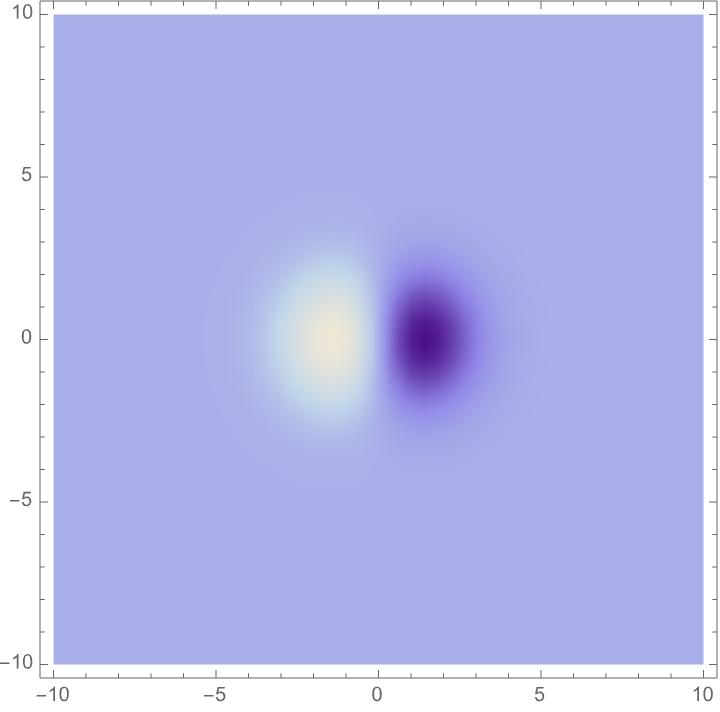}
      & \includegraphics[width=0.14\textwidth]{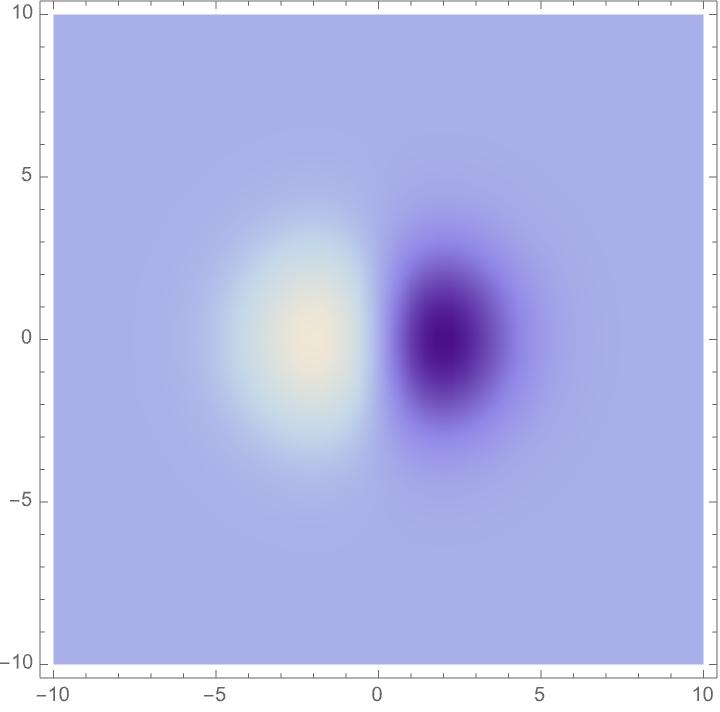}
      & \includegraphics[width=0.14\textwidth]{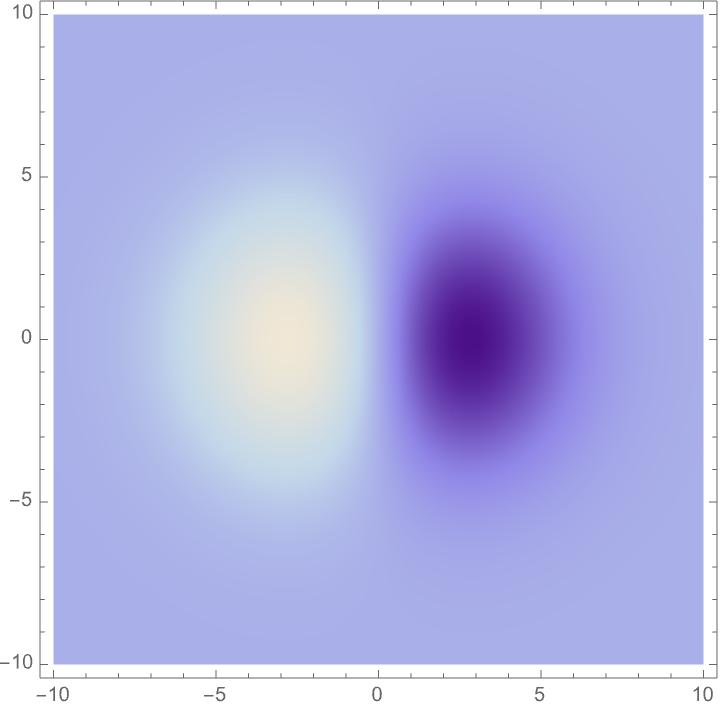}
      & \includegraphics[width=0.14\textwidth]{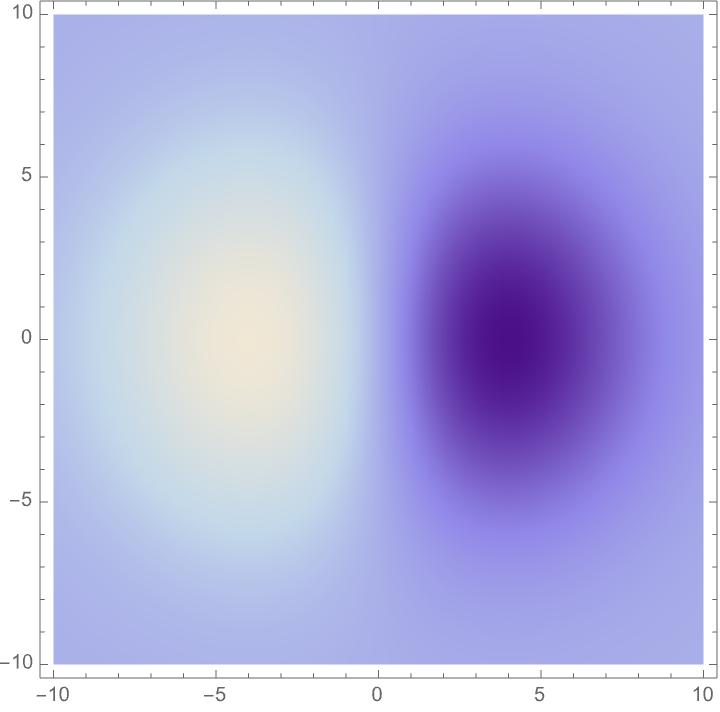} \\
      \includegraphics[width=0.14\textwidth]{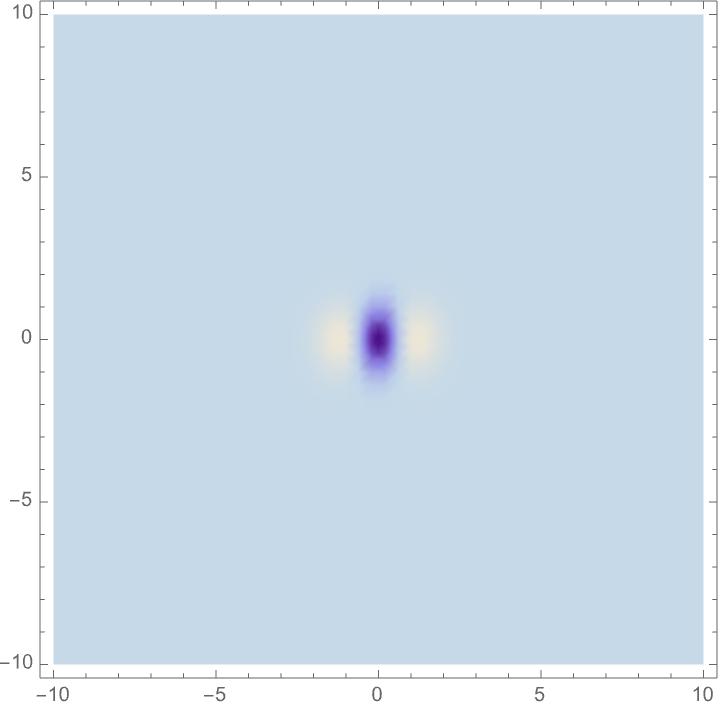}
      & \includegraphics[width=0.14\textwidth]{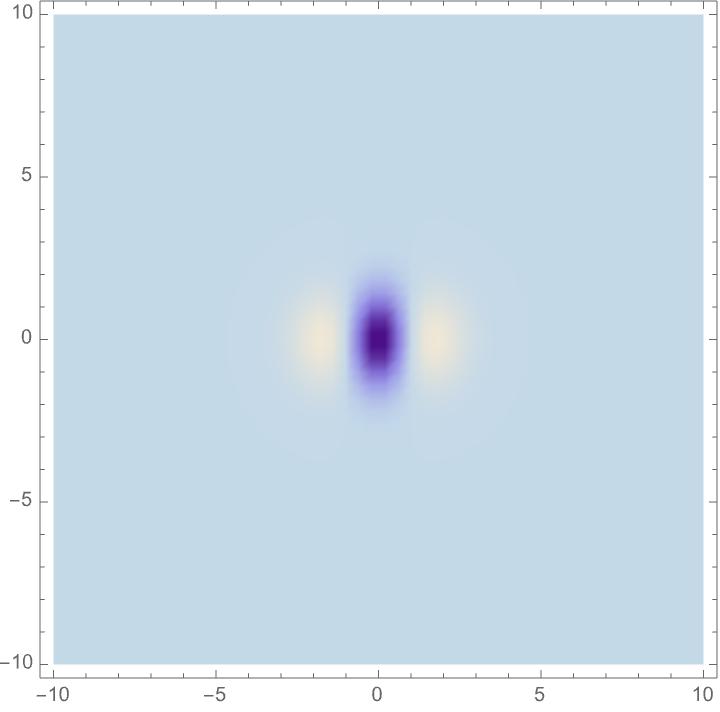}
      & \includegraphics[width=0.14\textwidth]{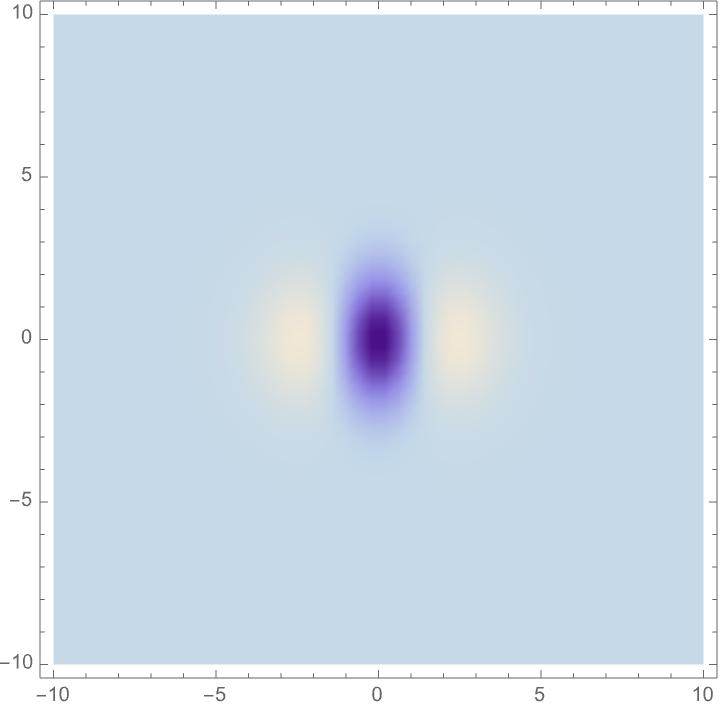}
      & \includegraphics[width=0.14\textwidth]{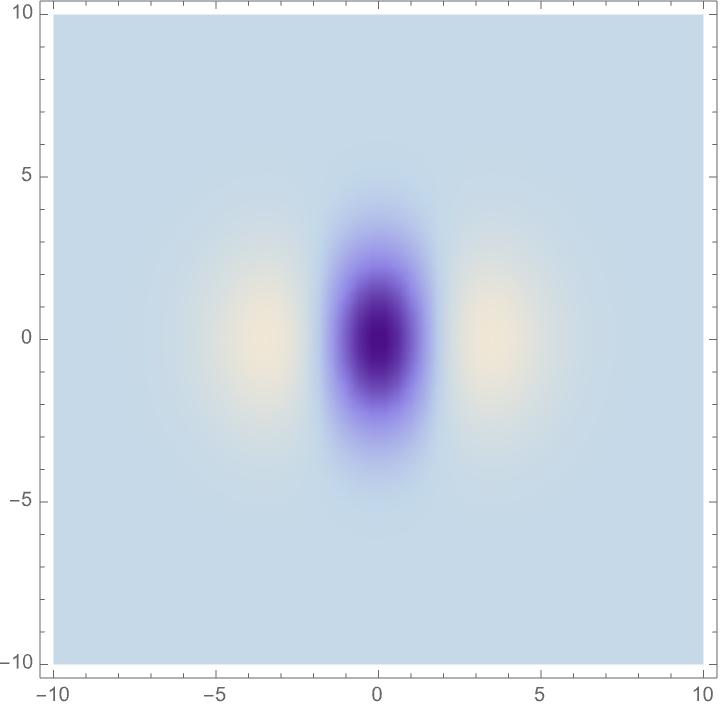}
      & \includegraphics[width=0.14\textwidth]{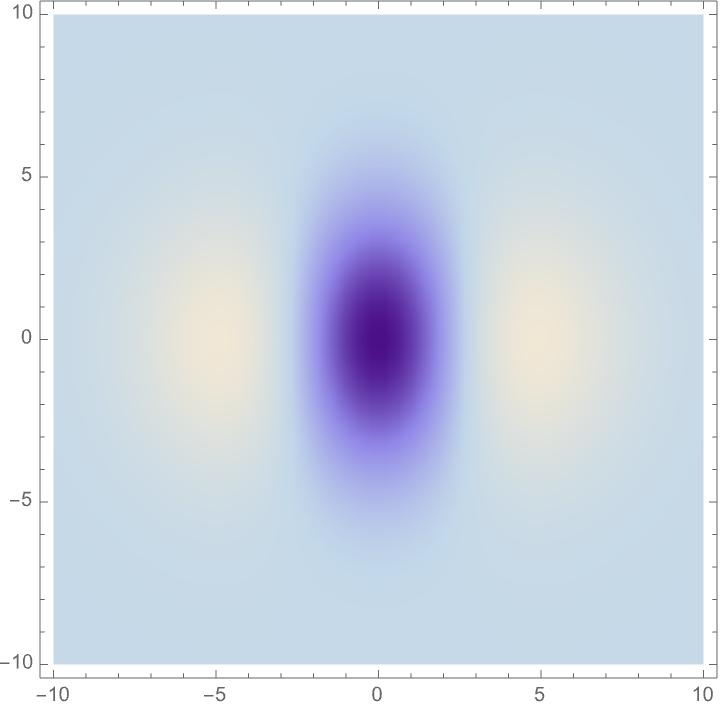}
      &
        \includegraphics[width=0.14\textwidth]{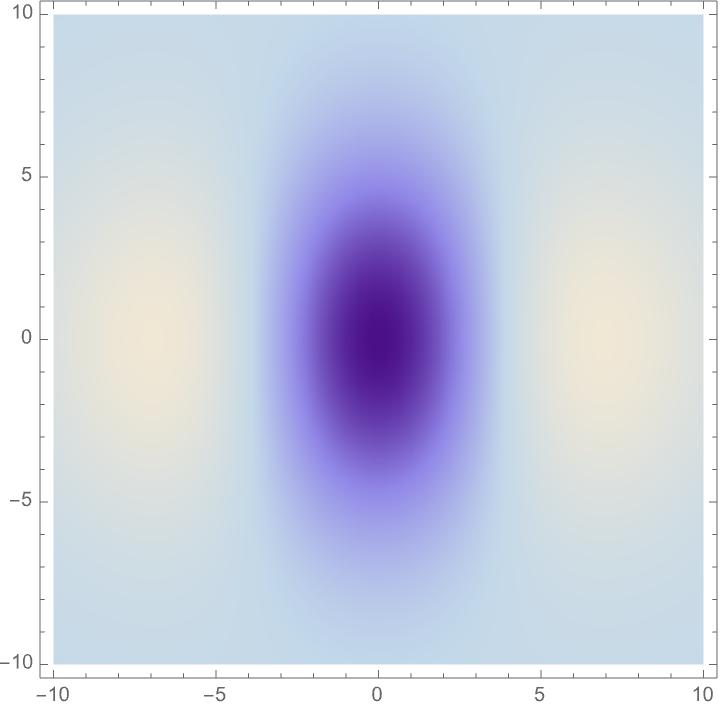} \\
    \end{tabular}
  \end{center}
  \caption{{\em Variability in the size \/}of affine Gaussian derivative
    receptive fields (for $\sigma_1 = \sigma_2$ and
    image orientation $\varphi = 0$),
    with the overall size $\bar{\sigma} =
    \sqrt{\sigma_1 \sigma_2}$ increasing from $1/\sqrt{2}$ to $4$
    according to a logarithmic distribution, from left to right.
    (top row) First-order directional derivatives of affine Gaussian
    kernels according to (\ref{eq-1dirder-affgauss}).
    (bottom row) Second-order directional derivatives of affine
    Gaussian kernels according to (\ref{eq-2dirder-affgauss}).
    ({\bf Horizontal axes:} image coordinate $x_1 \in [-10, 10]$.
     {\bf Vertical axes:} image coordinate $x_2 \in [-10, 10]$.)}
  \label{fig-size-variability}

  \bigskip

  \begin{center}
    \begin{tabular}{cccccc}
        $\varphi = 0$
      & $\varphi= \frac{\pi}{6}$
      & $\varphi= \frac{\pi}{3}$
      & $\varphi= \frac{\pi}{2}$
      & $\varphi= \frac{2\pi}{3}$
      & $\varphi= \frac{5\pi}{6}$ \\
      \includegraphics[width=0.14\textwidth]{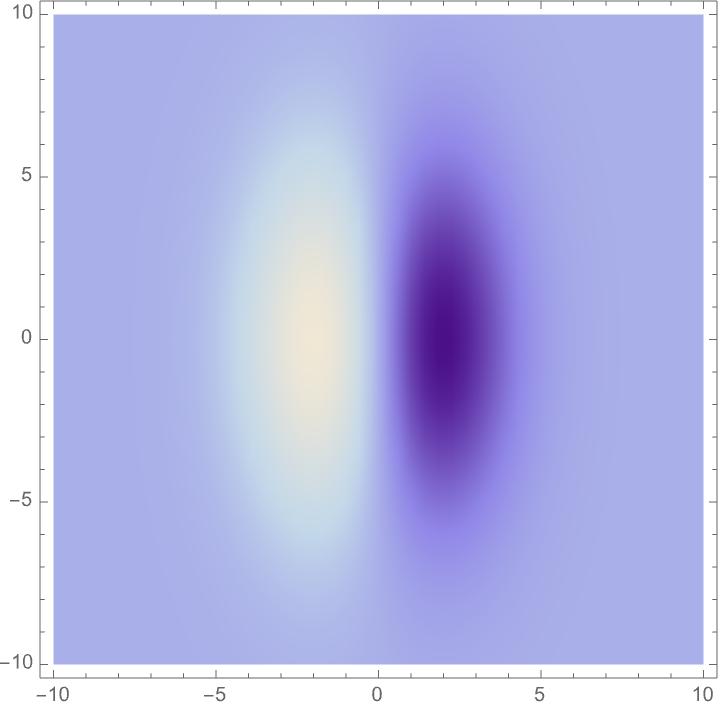}
      & \includegraphics[width=0.14\textwidth]{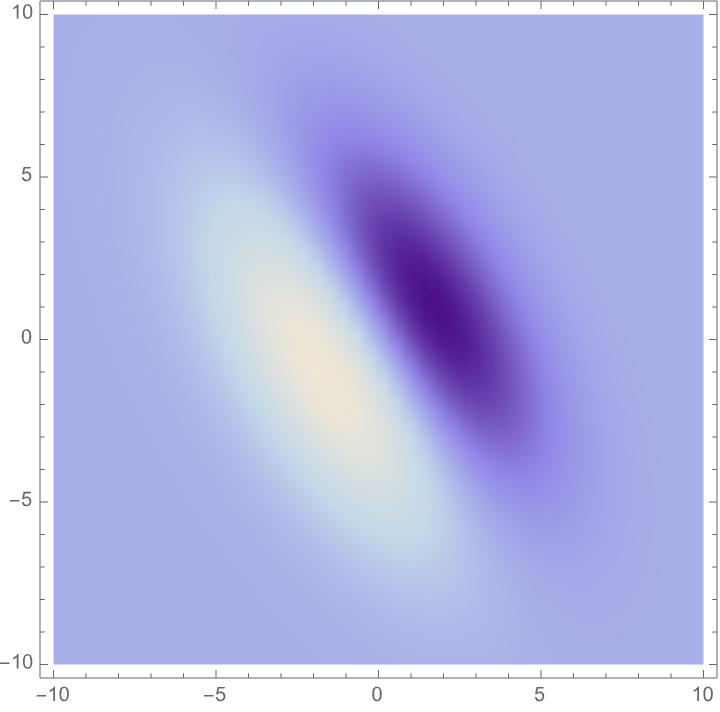}
      & \includegraphics[width=0.14\textwidth]{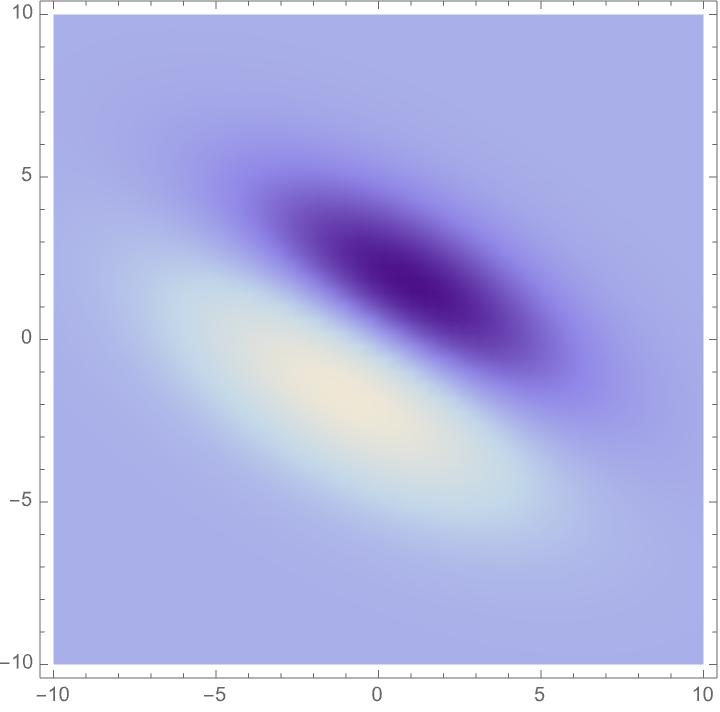}
      & \includegraphics[width=0.14\textwidth]{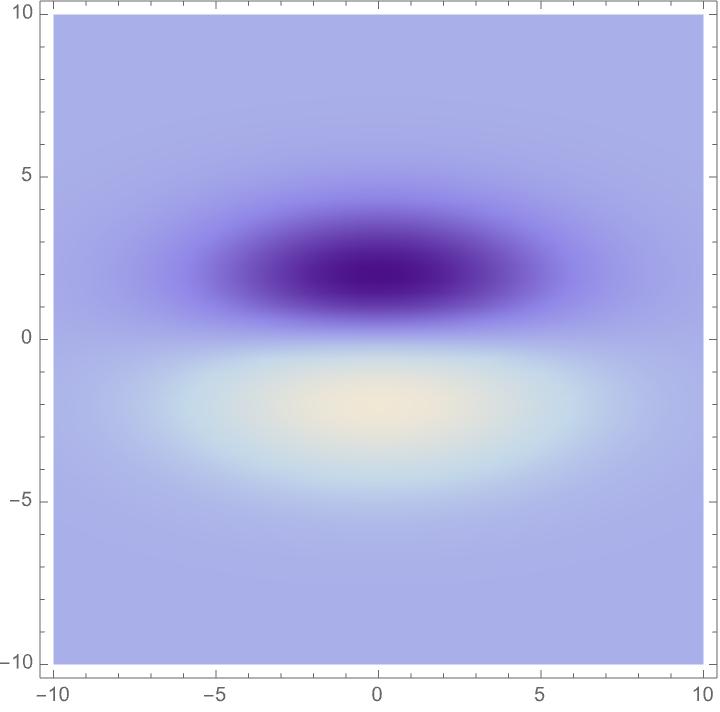}
      & \includegraphics[width=0.14\textwidth]{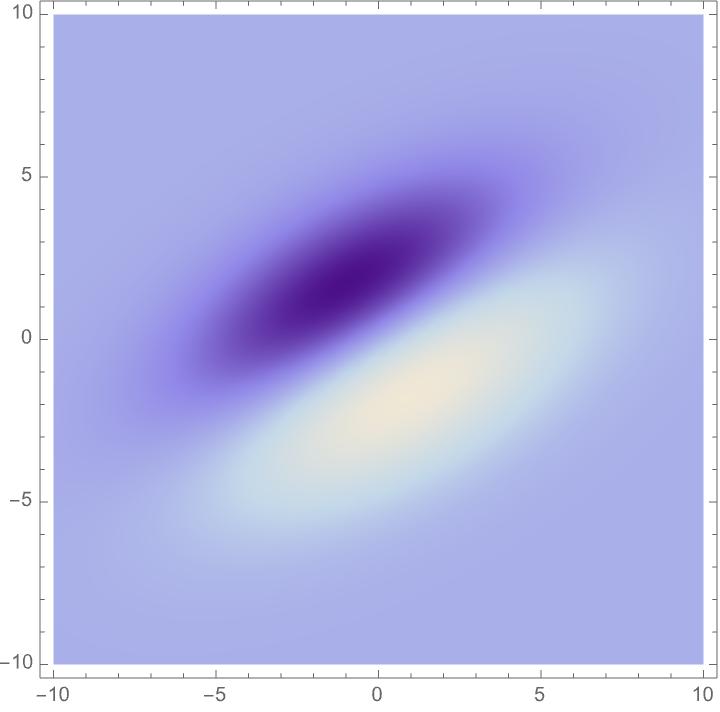}
      & \includegraphics[width=0.14\textwidth]{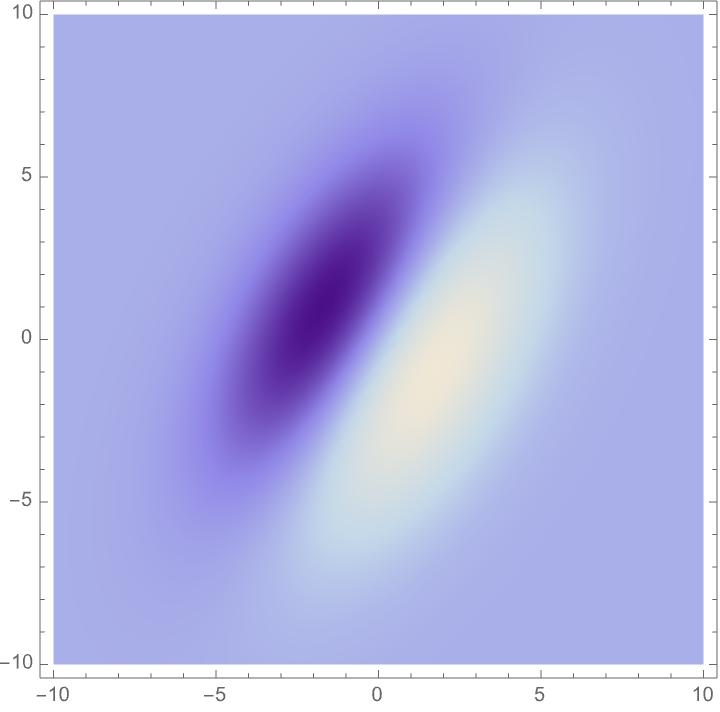} \\
      \includegraphics[width=0.14\textwidth]{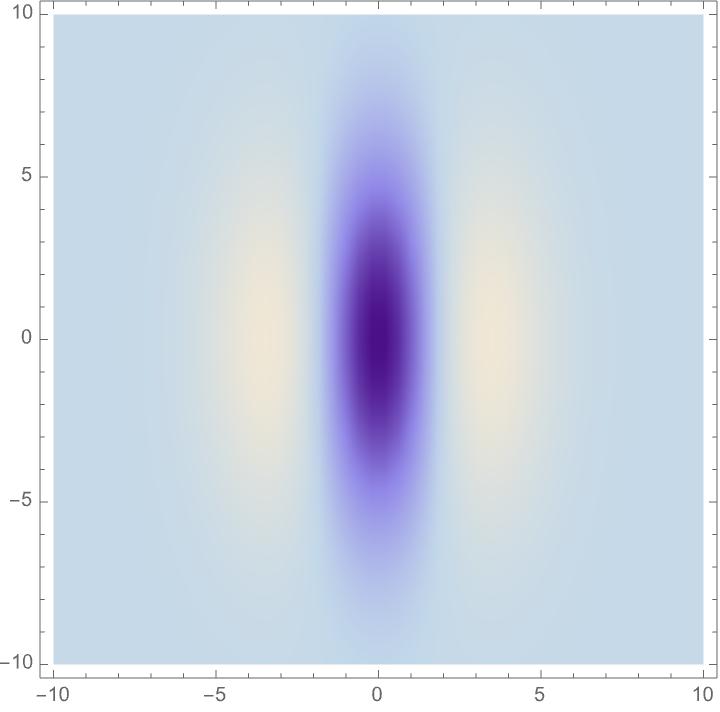}
      & \includegraphics[width=0.14\textwidth]{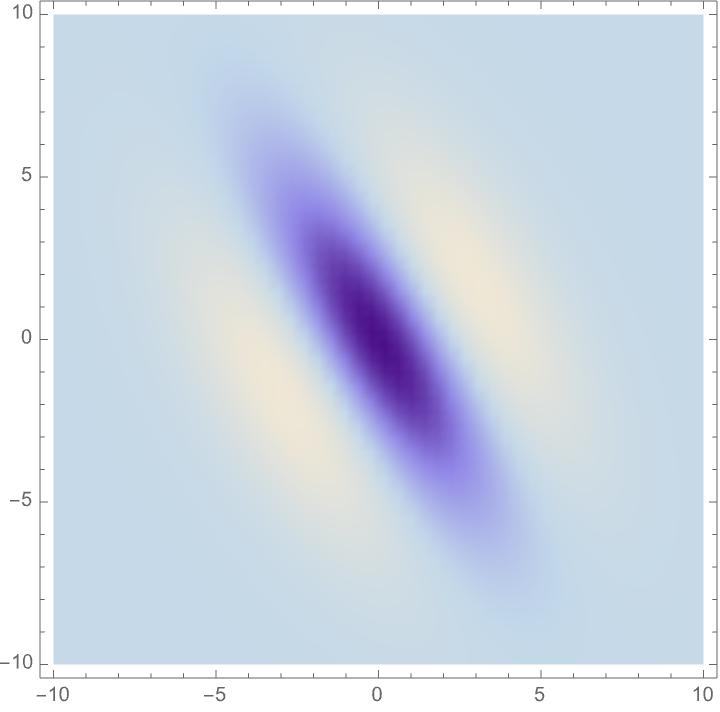}
      & \includegraphics[width=0.14\textwidth]{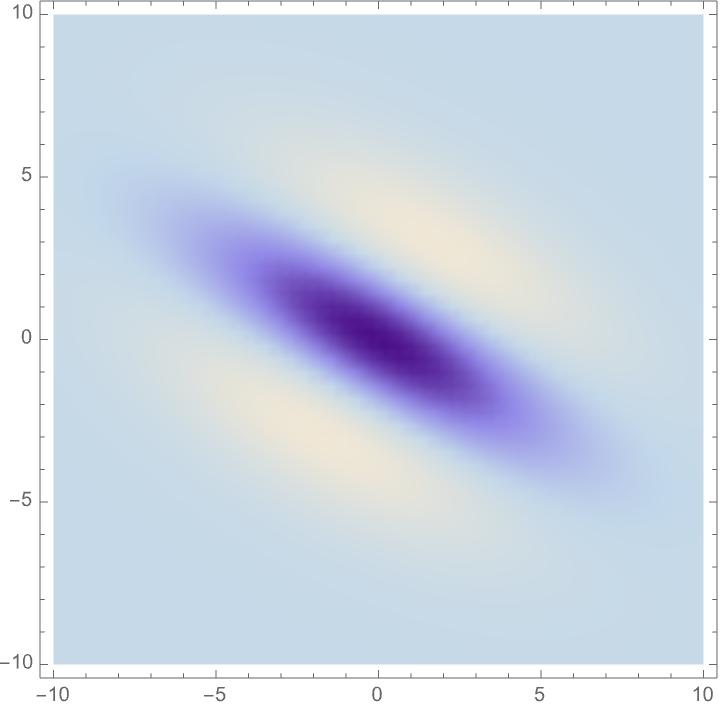}
      & \includegraphics[width=0.14\textwidth]{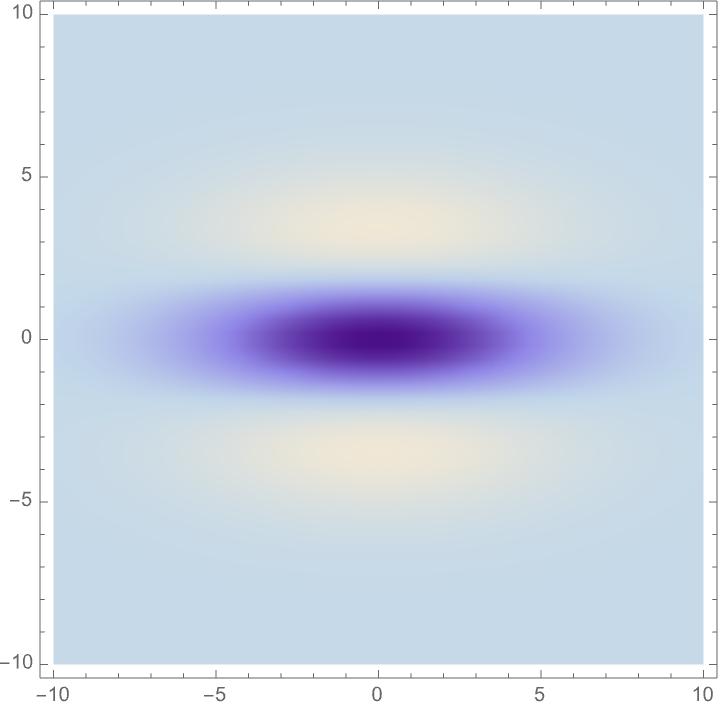}
      & \includegraphics[width=0.14\textwidth]{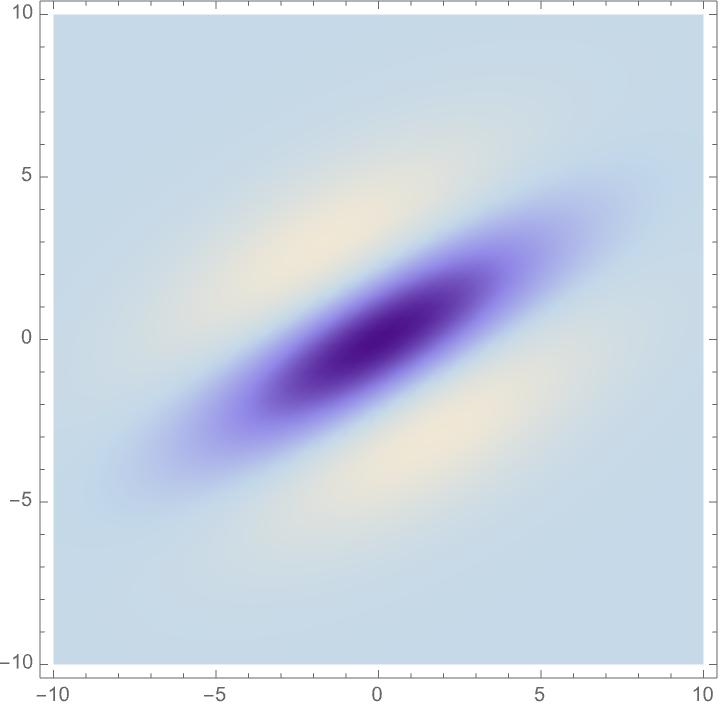}
      & \includegraphics[width=0.14\textwidth]{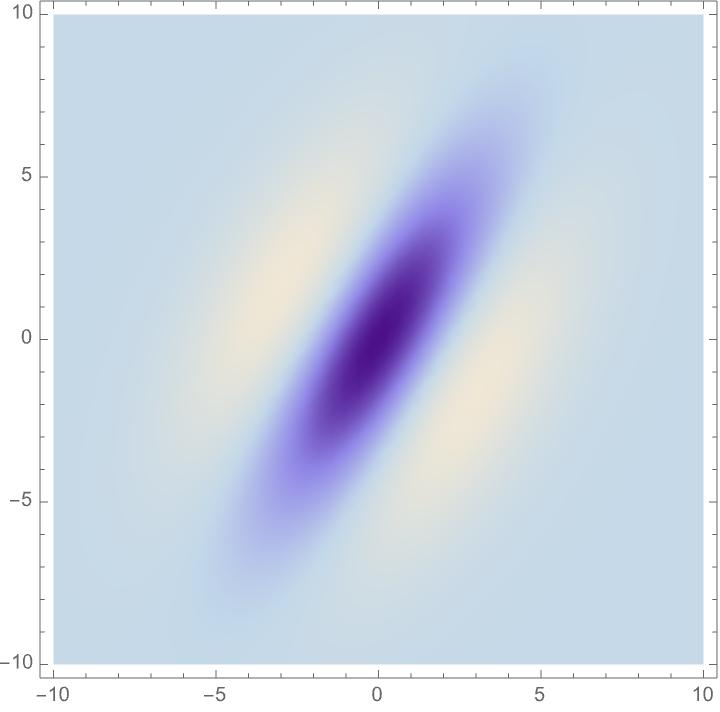} \\
   \end{tabular}
  \end{center}
  \caption{{\em Variability in the orientation\/} of affine Gaussian derivative
    receptive fields (for $\sigma_1 = 2$ and $\sigma_2 = 4$),
    with the orientation angle $\varphi$ increasing from left to right.
    (top row) First-order directional derivatives of affine Gaussian
    kernels according to (\ref{eq-1dirder-affgauss}).
    (bottom row) Second-order directional derivatives of affine
    Gaussian kernels according to (\ref{eq-2dirder-affgauss}).
    ({\bf Horizontal axes:} image coordinate $x_1 \in [-10, 10]$.
     {\bf Vertical axes:} image coordinate $x_2 \in [-10, 10]$.)}
  \label{fig-ori-variability}

  \bigskip
  
 \begin{center}
    \begin{tabular}{cccccc}
      $\epsilon= 1$
      & $\epsilon= \frac{1}{\sqrt{2}}$
      & $\epsilon= \frac{1}{2}$
      & $\epsilon= \frac{1}{2\sqrt{2}}$
      & $\epsilon= \frac{1}{4}$        
      & $\epsilon= \frac{1}{4\sqrt{2}}$ \\
     \includegraphics[width=0.14\textwidth]{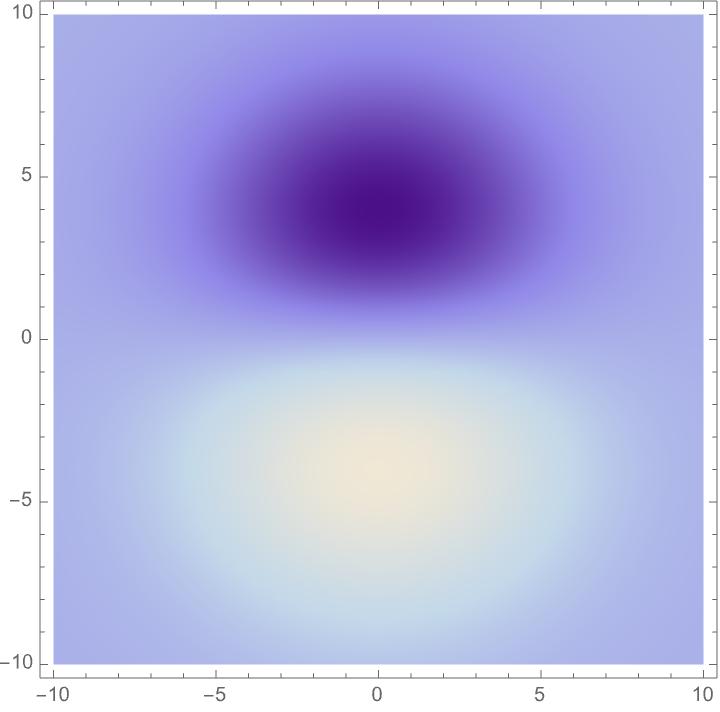}
      & \includegraphics[width=0.14\textwidth]{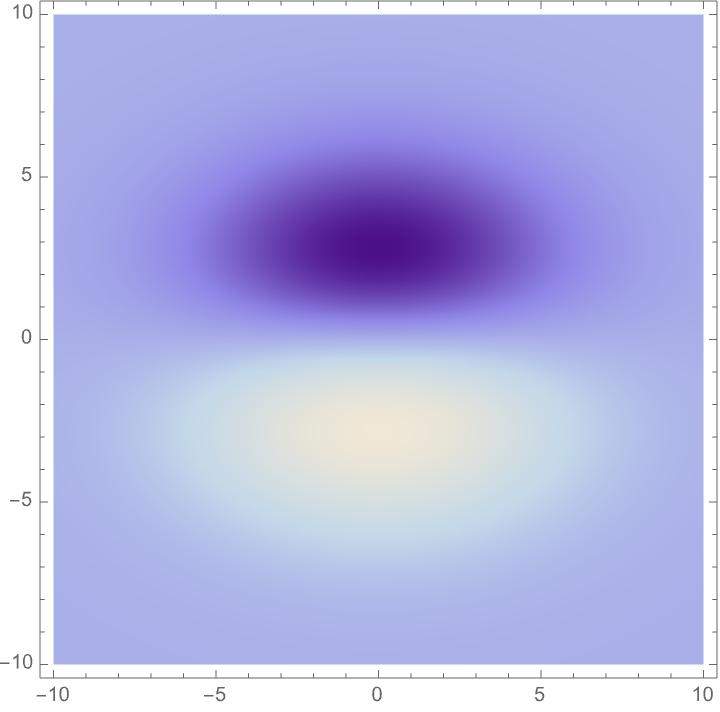}
      & \includegraphics[width=0.14\textwidth]{affgauss-1dirder-sigma-2-4-phi-piover2-mono.jpg}
      & \includegraphics[width=0.14\textwidth]{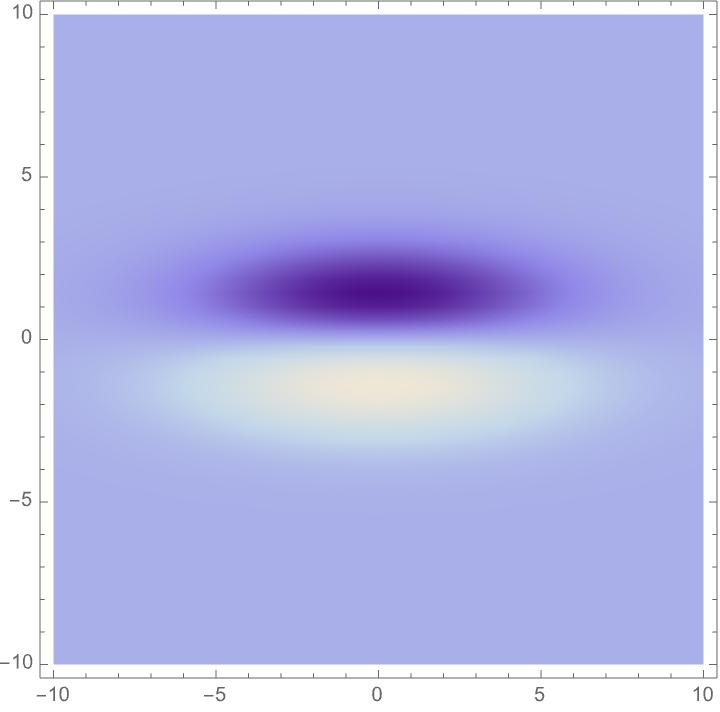}
      & \includegraphics[width=0.14\textwidth]{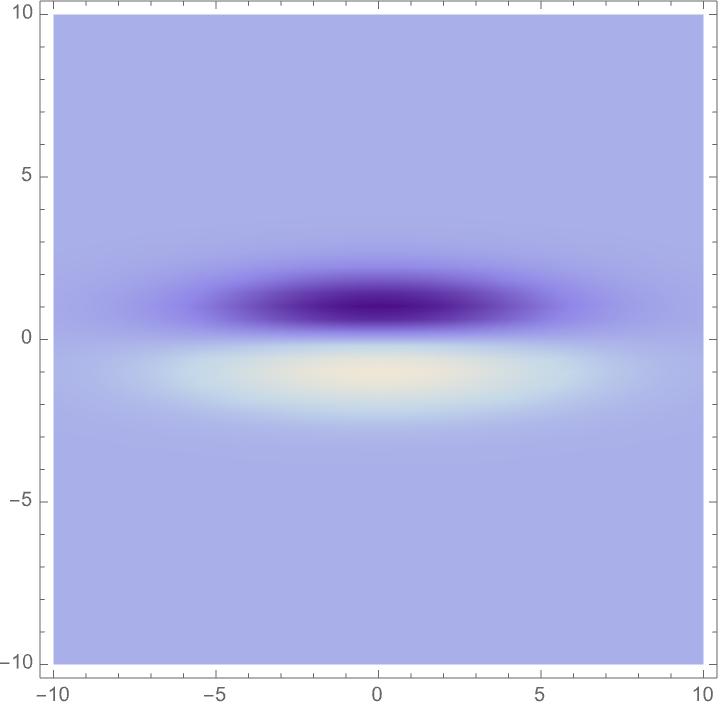}
      & \includegraphics[width=0.14\textwidth]{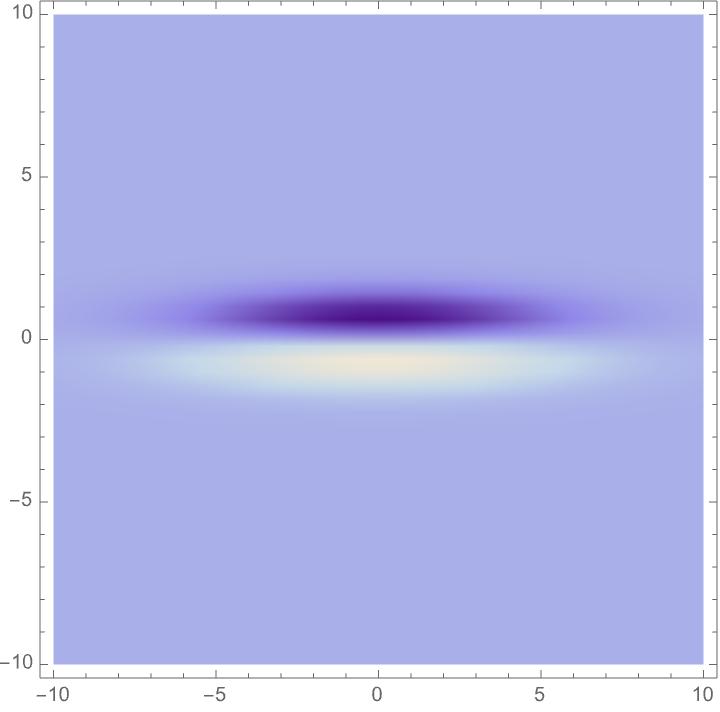} \\
     \includegraphics[width=0.14\textwidth]{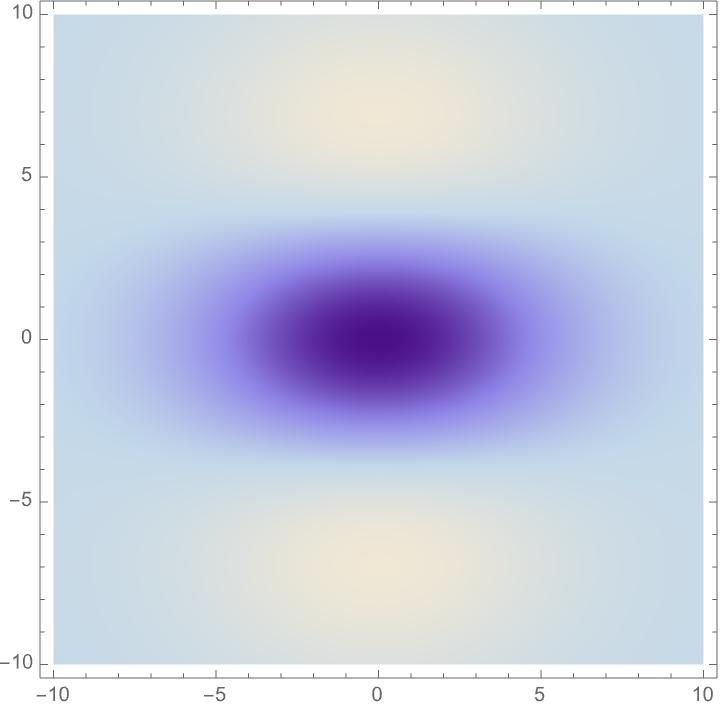}
      & \includegraphics[width=0.14\textwidth]{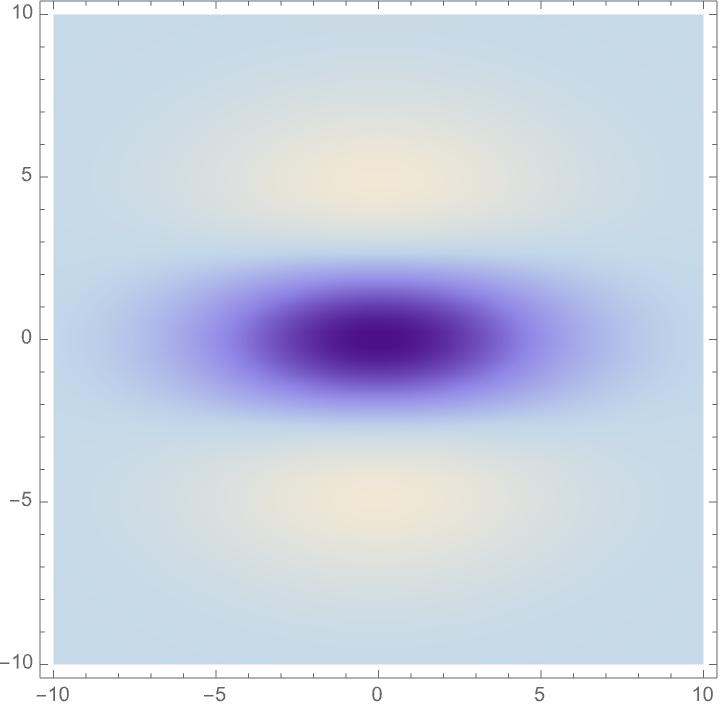}
      & \includegraphics[width=0.14\textwidth]{affgauss-2dirder-sigma-2-4-phi-piover2-mono.jpg}
      & \includegraphics[width=0.14\textwidth]{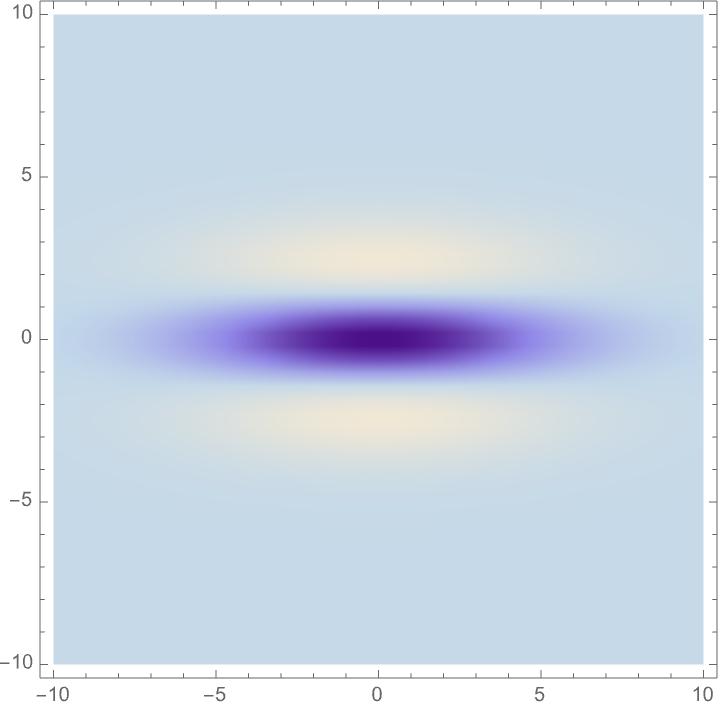}
      & \includegraphics[width=0.14\textwidth]{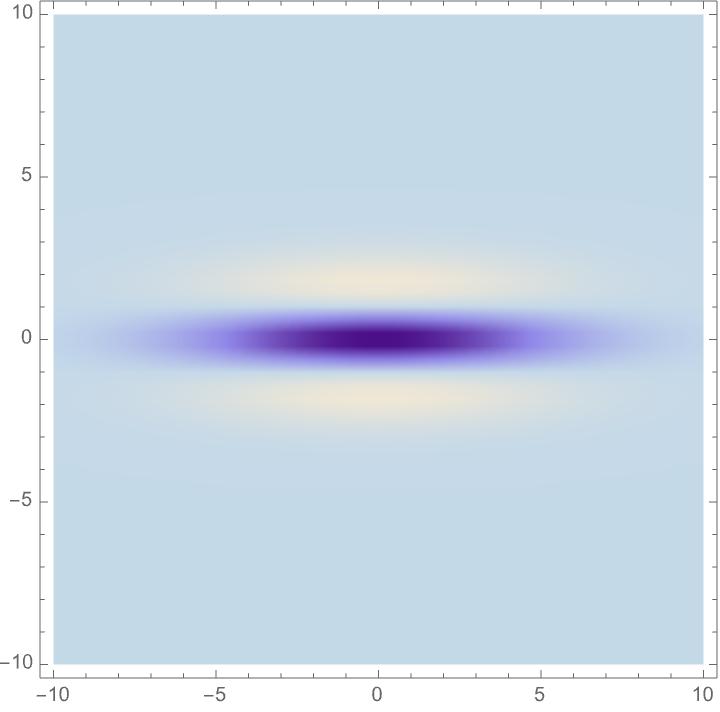}
      & \includegraphics[width=0.14\textwidth]{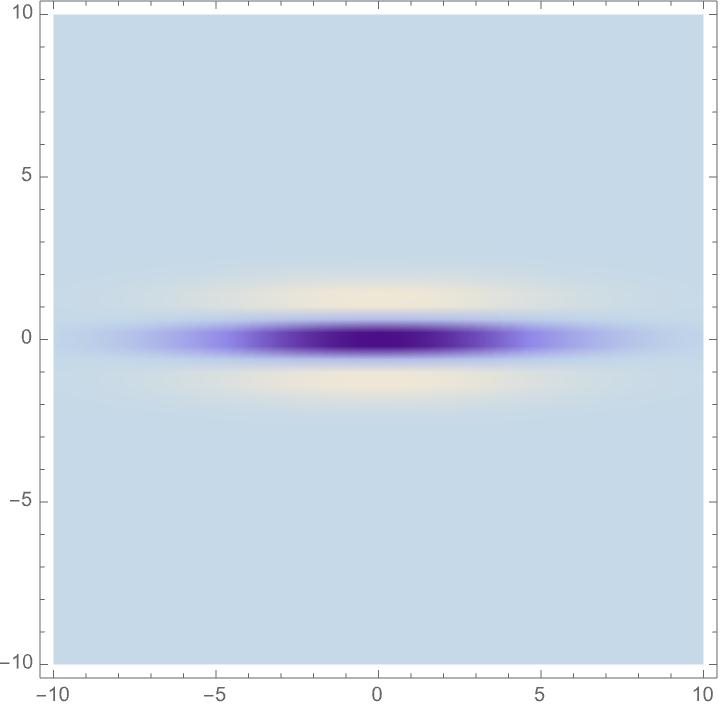}     
    \end{tabular}
  \end{center}
  \caption{{\em Variability in the eccentricity\/} of affine Gaussian derivative
    receptive fields (for image orientation $\varphi = \frac{\pi}{2}$),
    with the eccentricity 
    $\epsilon = \sigma_1/\sigma_2$ decreasing from $1$ to
    according to a logarithmic distribution, from left to right,
    with $\sigma_2$ kept constant.
    (top row) First-order directional derivatives of affine Gaussian
    kernels according to (\ref{eq-1dirder-affgauss}).
    (bottom row) Second-order directional derivatives of affine
    Gaussian kernels according to (\ref{eq-2dirder-affgauss}).
    ({\bf Horizontal axes:} image coordinate $x_1 \in [-10, 10]$.
     {\bf Vertical axes:} image coordinate $x_2 \in [-10, 10]$.)}
  \label{fig-ecc-variability}
\end{figure*}

\section{Degrees of freedom in the affine Gaussian derivative model
  for visual receptive fields}
\label{sec-dofs-aff-gauss-der-rfs}

The spatial covariance matrix
\begin{equation}
  \label{def-spat-cov-mat}
  \Sigma
  = \left(
        \begin{array}{cc}
          C_{11} & C_{12} \\
          C_{12} & C_{22}
        \end{array}
      \right)
\end{equation}
in the affine Gaussian derivative model
for visual receptive fields can be parameterized on the form
\begin{align}
    \begin{split}
       \label{eq-expl-par-Cxx}
       C_{11} & = \sigma_1^2 \, \cos^2 \varphi + \sigma_2^2 \, \sin^2 \varphi,
    \end{split}\\
    \begin{split}
          \label{eq-expl-par-Cxy}
        C_{12} & = (\sigma_1^2 - \sigma_2^2)  \cos \varphi  \, \sin \varphi,
    \end{split}\\
    \begin{split}
       \label{eq-expl-par-Cyy}
        C_{22} & = \sigma_1^2 \, \sin^2 \varphi + \sigma_2^2  \, \cos^2 \varphi,
   \end{split}
\end{align}
where $\lambda_1 = \sigma_1^2$ and $\lambda_2^2= \sigma_2^2$
constitute the eigenvalues of $\Sigma$.
Under variabilities of the spatial scale parameters $\sigma_1 > 0$ and
$\sigma_2 > 0$, and the spatial orientation $\varphi$, these variabilities span
the variability of the shapes of the receptive fields in the receptive
field model.

In relation to overall shape properties of the receptive fields, the
degrees of freedom in this parameterization have the following interpretation:
\begin{itemize}
\item
  The product
  \begin{equation}
    \label{eq-def-sigma-bar-aff-RF}
    \bar{\sigma} = \sqrt{\sigma_1 \, \sigma_2}
  \end{equation}
  describes the overall {\em spatial size\/} of the receptive field.
\item
  The ratio
  \begin{equation}
    \label{eq-def-eccentricity}
    \epsilon = \frac{\sigma_2}{\sigma_1}
  \end{equation}
  describes the
   eccentricity or the {\em degree of elongation\/} of the receptive field, in the sense
   that deviations in this ratio from the rotationally symmetric
   special case when $\epsilon = 1$ correspond to more
    elongated or anisotropic receptive fields. 
  \item
    The angle $\varphi$ in (\ref{eq-expl-par-Cxx})--(\ref{eq-expl-par-Cyy})
    represents the {\em spatial orientation\/} of the
    receptive field.
  \end{itemize}
Figures~\ref{fig-size-variability}--\ref{fig-ecc-variability}
show illustrations of these variabilities for purely
spatial receptive fields, in terms of first- and second-order
directional derivatives of affine Gaussian kernels,
according to the following idealized models for the receptive fields of simple cells in the primary visual cortex
(based on Equation~(31) in Lindeberg (\citeyear{Lin21-Heliyon})):
\begin{equation}
  \label{eq-dir-der-aff-gauss-kern}
  T_{\simple}(x_1, x_2;\; \sigma_1, \sigma_2, \varphi, m)
  = \sigma_1^m \, \partial_{\varphi}^m \, g(x_1, x_2;\; \Sigma),
\end{equation}
which for first and second orders of spatial differentiation $m$ assume the
following explicit forms
\begin{align}
  \begin{split}
     T_{\simple}(x_1, x_2;\; \sigma_1, \sigma_2, \varphi, 1) =
  \end{split}\nonumber\\
  \begin{split}
    = \sigma_1 \, (\cos (\varphi) \, \partial_{x_1} + \sin(\varphi) \, \partial_{x_2}) \, 
        g(x_1, x_2;\; \Sigma)
  \end{split}\nonumber\\
  \begin{split}
     =
     -\frac{(x_1 \cos (\varphi )+x_2 \sin (\varphi ))}
               {2 \pi  \, \sigma_1^2 \, \sigma_2} \times
   \end{split}\nonumber\\
  \begin{split}
    \label{eq-1dirder-affgauss}
    e^{
   -\frac{\left(\sigma_1^2+\sigma_2^2\right)
   \left(x_1^2+x_2^2\right)-(\sigma_1-\sigma_2) (\sigma_1+\sigma_2) (2 x_1
   x_2 \sin (2 \varphi )+\cos (2 \varphi ) (x_1-x_2) (x_1+x_2))}{4 \sigma_1^2
   \sigma_2^2}}
  \end{split}\\
  \begin{split}
     T_{\simple}(x_1, x_2;\; \sigma_1, \sigma_2, \varphi, 2) =
  \end{split}\nonumber\\
 \begin{split}
   = \sigma_1^2 \, (\cos^2 (\varphi) \, \partial_{x_1 x_1} +
   2 \cos(\varphi) \sin(\varphi) \, \partial_{x_1 x_2} +
    \sin^2(\varphi) \, \partial_{x_2 x_2}) \, 
 \end{split}\nonumber\\
  \begin{split}
         \quad\quad g(x_1, x_2;\; \Sigma)
  \end{split}\nonumber\\
  \begin{split}
     =
    \frac{\left(\cos (2 \varphi ) \left(x_1^2-x_2^2\right)+2 x_1 x_2 \sin (2 \varphi )-2
                     \sigma_1^2+x_1^2+x_2^2\right)}
            {4 \pi  \, \sigma_1^3 \, \sigma_2} \times
  \end{split}\nonumber\\
  \begin{split}
    \label{eq-2dirder-affgauss}    
      e^{
   -\frac{\left(\sigma_1^2+\sigma_2^2\right)
   \left(x_1^2+x_2^2\right)-(\sigma_1-\sigma_2) (\sigma_1+\sigma_2) (2 x_1
   x_2 \sin (2 \varphi )+\cos (2 \varphi ) (x_1-x_2) (x_1+x_2))}{4 \sigma_1^2
   \sigma_2^2}}.
  \end{split}
\end{align}
Figures~\ref{fig-size-variability}--\ref{fig-ecc-variability} 
illustrate the basic variabilities of the shapes of the receptive fields
arising in this way according to the affine Gaussian derivative model. A main question
addressed in this paper concerns what variabilities in receptive field
shapes may be spanned in the primary visual cortex.

\section{Relationships between the variabilities in affine image
  transformations and the variabilities in affine Gaussian receptive
  fields}
\label{sec-rels-var-aff-transf-aff-gauss-ders}

From a comparison between the degrees of freedom in 2-D spatial affine
transformations according to Section~\ref{sec-dofs-2d-affine}
with the degrees of freedom in the shapes of the
affine Gaussian receptive fields in Section~\ref{sec-dofs-aff-gauss-der-rfs},
it does specifically hold that:
\begin{itemize}
\item
  Variabilities in the {\em size\/} of the overall size receptive field,
  as represented by the product
  \begin{equation}
    \label{eq-def-sigma-bar}
    \bar{\sigma} = \sqrt{\sigma_1 \, \sigma_2}
  \end{equation}
  in (\ref{eq-def-sigma-bar-aff-RF}), span
  the variability over the overall {\em spatial scaling factor\/}
  \begin{equation}
    S = \sqrt{\rho_1 \, \rho_2}
  \end{equation}
  according to (\ref{eq-def-overall-sc-factor-aff})
  in the affine image transformations.
\item
  Variabilities in the {\em eccentricity\/} of the receptive field, as can be
  described by the ratio
  \begin{equation}
    \label{eq-def-eccentricity-rf}
    \epsilon = \frac{\sigma_2}{\sigma_1}
  \end{equation}
  in (\ref{eq-def-eccentricity}), do, for a fixed value of the image
  orientation angle $\varphi$ of the affine Gaussian receptive fields,
  span the variability in the {\em ratio between the singular values\/}
  \begin{equation}
    \label{eq-def-lambda-ratio-sing-vals}
    \lambda = \frac{\rho_1}{\rho_2}
  \end{equation}
  of the affine transformation, which, in
  turn, represents the essential variability in the amount of non-uniform spatial
  stretching in the pure stretching transformation ${\cal D}$ according to
  (\ref{eq-def-noniso-diag-aff}) in the affine image transformations.
\item
  Variabilities in the {\em orientation angle\/}
  \begin{equation}
    \varphi
  \end{equation}
  of the affine
  Gaussian receptive fields according to
  (\ref{eq-expl-par-Cxx})--(\ref{eq-expl-par-Cyy})
  span the variability in the amount of
  {\em overall rotation\/} $\varphi$ of the total rotation component
  \begin{equation}
    \label{eq-def-tot-amount-rot}
    {\cal R}_{\varphi} = {\cal R}_{\frac{\varphi}{2}} {\cal R}_{\frac{\varphi}{2}}
  \end{equation}
  according to
  (\ref{eq-aff-decomp}) in the affine image transformations.
\end{itemize}
In these respects, there is a one-to-one mapping between three of the
degrees of freedom of 2-D spatial affine transformations and the three
degrees of freedom in the regular formulation of idealized receptive
field models in terms of the regular formulation of directional
derivatives of affine Gaussian kernels.

Specifically, if the 
image data in an imaging situation would be subject to these degrees
of freedom in 2-D spatial affine transformations, we can then 
let the affine Gaussian derivative kernels exhibit corresponding
variabilities in the shapes of their receptive fields,
to be able to match the receptive field responses
between the image domains before {\em vs.\/}\ after the affine image deformation.

\begin{figure*}[hbtp]
  \begin{center}
    \begin{tabular}{cccccc}
        $\varphi = 0$
      & $\varphi= \frac{\pi}{6}$
      & $\varphi= \frac{\pi}{3}$
      & $\varphi= \frac{\pi}{2}$
      & $\varphi= \frac{2\pi}{3}$
      & $\varphi= \frac{5\pi}{6}$ \\
      \includegraphics[width=0.14\textwidth]{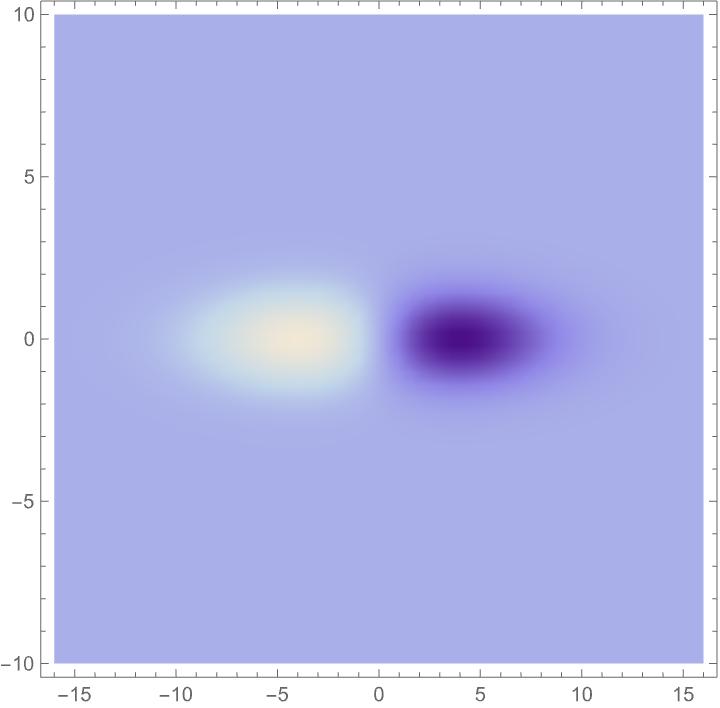}
      & \includegraphics[width=0.14\textwidth]{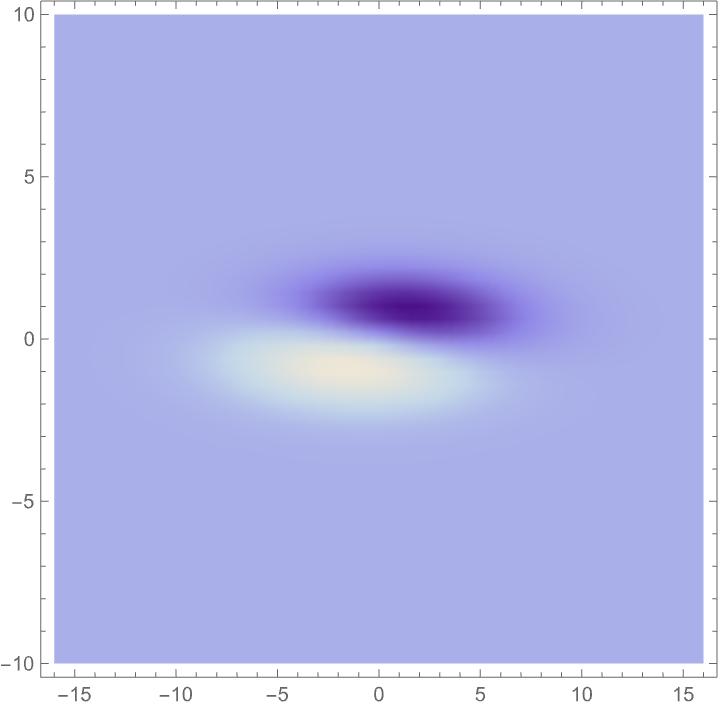}
      & \includegraphics[width=0.14\textwidth]{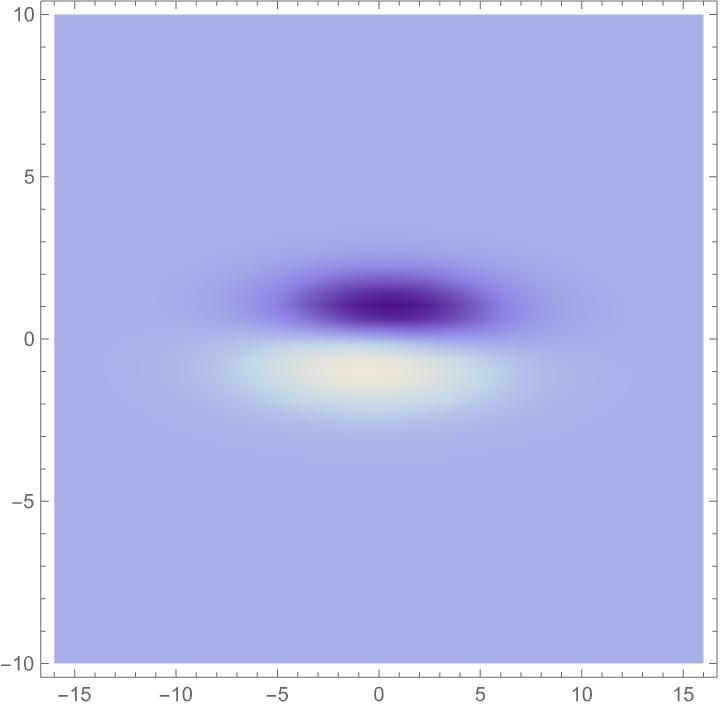}
      & \includegraphics[width=0.14\textwidth]{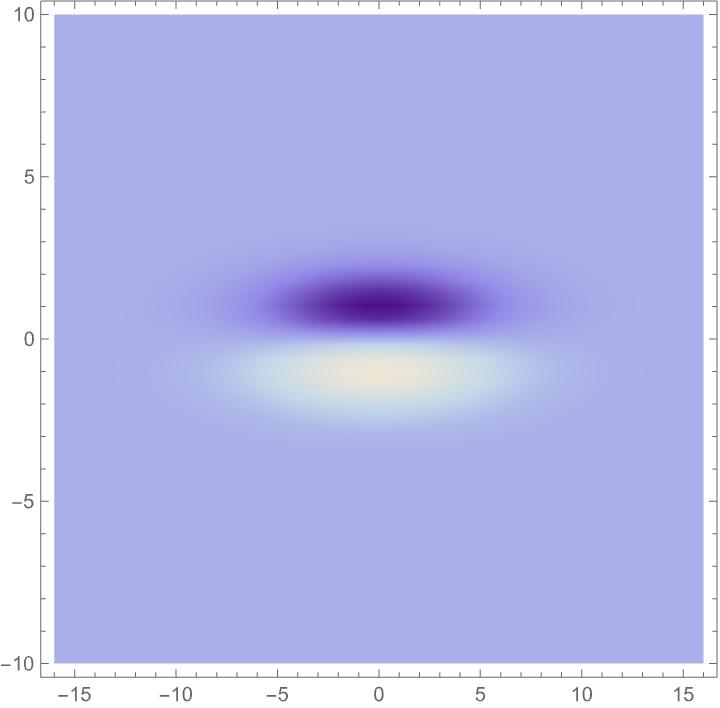}
      & \includegraphics[width=0.14\textwidth]{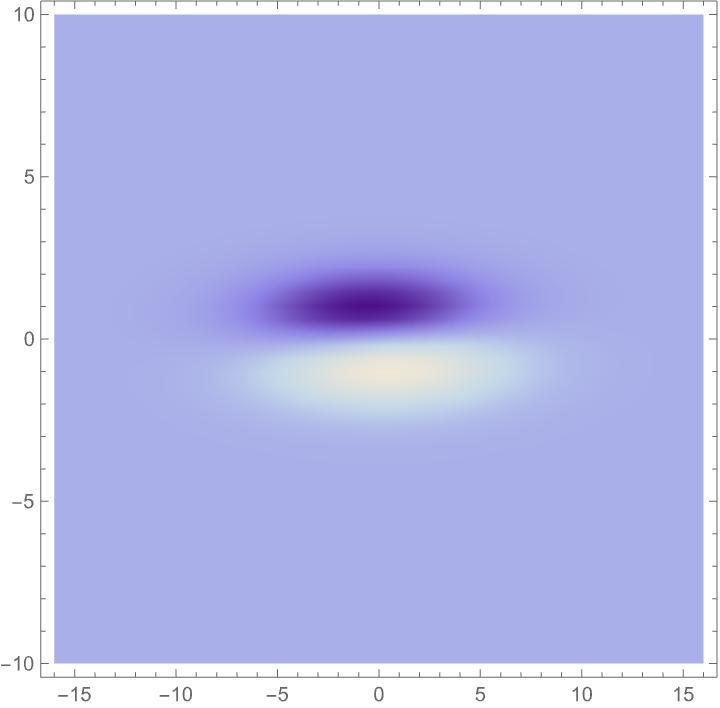}
      & \includegraphics[width=0.14\textwidth]{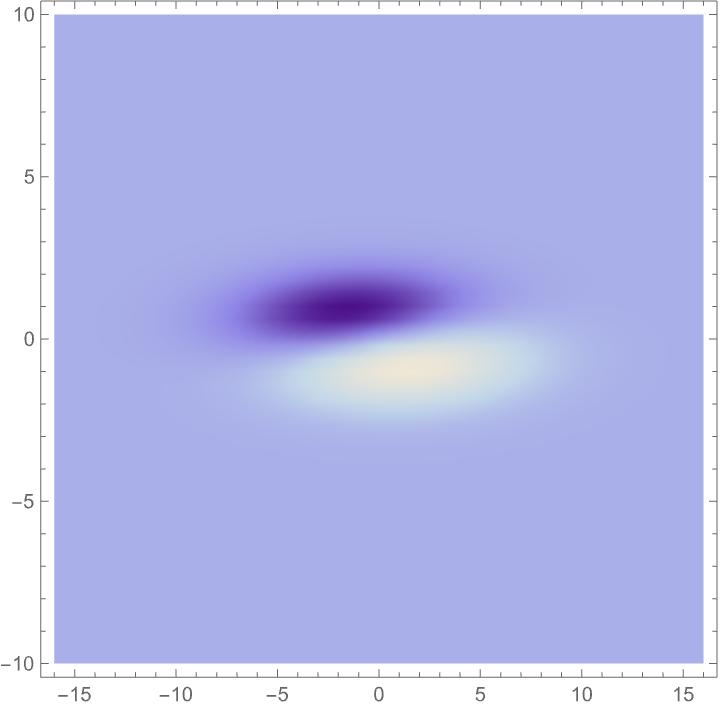} \\
      \includegraphics[width=0.14\textwidth]{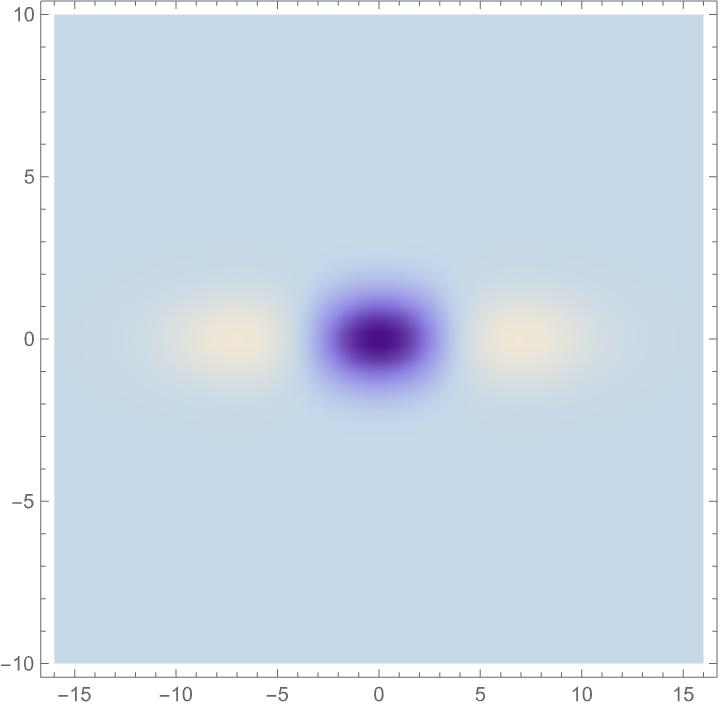}
      & \includegraphics[width=0.14\textwidth]{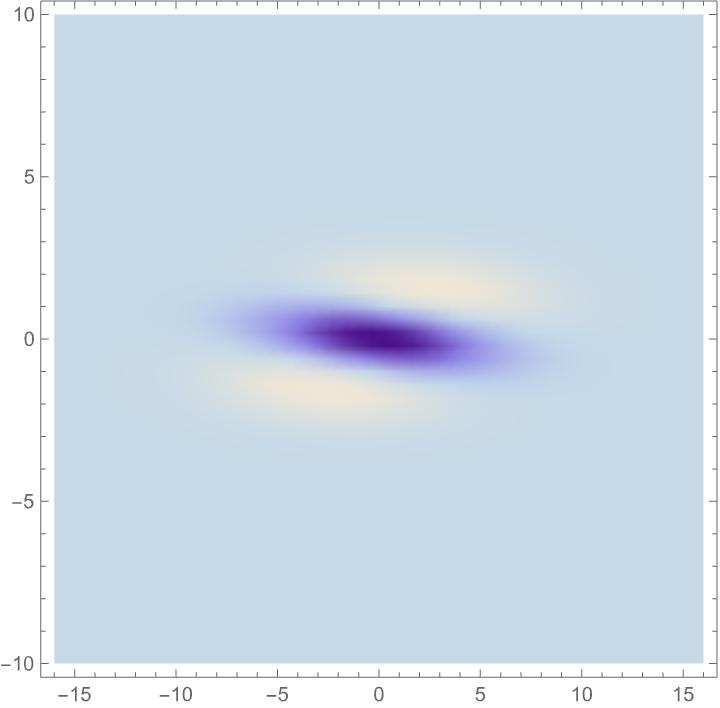}
      & \includegraphics[width=0.14\textwidth]{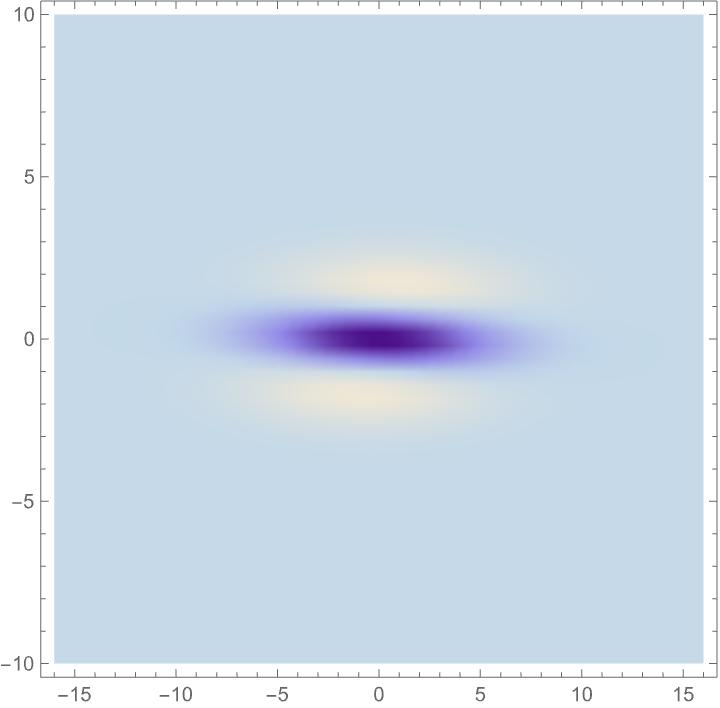}
      & \includegraphics[width=0.14\textwidth]{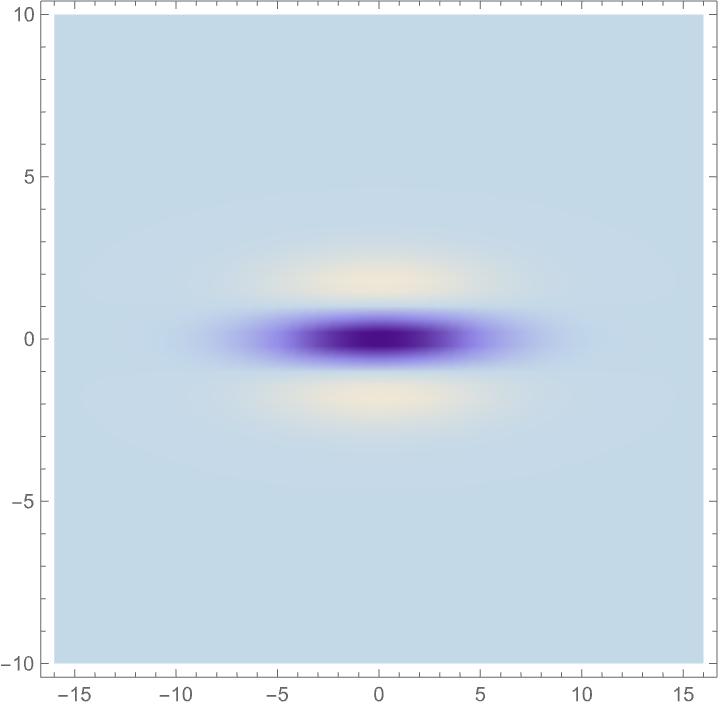}
      & \includegraphics[width=0.14\textwidth]{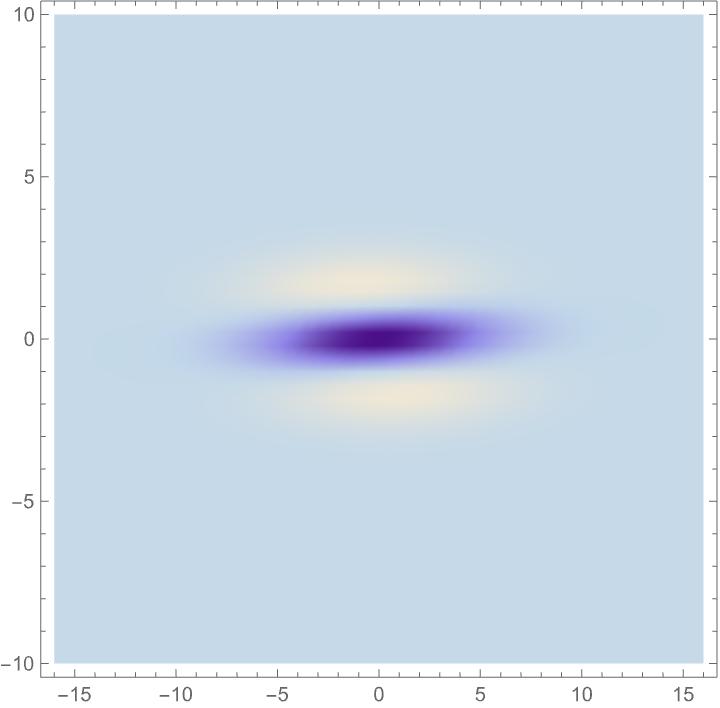}
      & \includegraphics[width=0.14\textwidth]{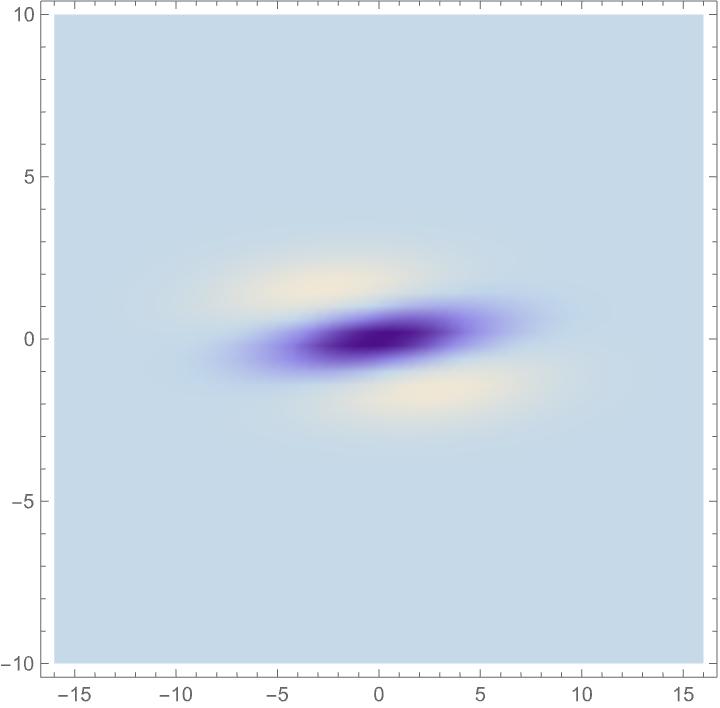} \\
   \end{tabular}
  \end{center}
  \caption{{\em Variability in the orientation of directional
      derivatives\/}
    of affine Gaussian derivative
    receptive fields (for $\sigma_1 = 4$ and $\sigma_2 = 1$ and
    preferred image orientation $\varphi = 0$),
    with the orientation angle $\varphi$ for the directional
    derivative operators increasing from left to right.
    (top row) First-order directional derivatives of affine Gaussian
    kernels. % according to (\ref{eq-1dirder-affgauss}).
    (bottom row) Second-order directional derivatives of affine
    Gaussian kernels. % according to (\ref{eq-2dirder-affgauss}).
    ({\bf Horizontal axes:} image coordinate $x_1 \in [-10, 10]$.
     {\bf Vertical axes:} image coordinate $x_2 \in [-10, 10]$.)}
  \label{fig-ori-variability-fixed}
\end{figure*}

\subsection{Restricted 3-D variability of regular affine Gaussian derivative kernels}

Consider the affine
Gaussian derivative model used for formulating the explicit
expressions for the idealized models for the receptive fields of
simple cells according to Equations~(\ref{eq-1dirder-affgauss}) and
(\ref{eq-2dirder-affgauss}),
with the directions for which the directional derivatives of the
affine Gaussian kernels are computed in directions parallel with the
eigendirections of the affine Gaussian kernels.
Because the variability of the resulting shapes of affine Gaussian
derivative-based receptive
fields in three-dimensional, while the variability of the affine 
image transformations is four-dimensional, the variability of the
shapes of the receptive fields according that regular affine Gaussian derivative model,
in its original formulation, cannot
span the full variability of the affine group.

\subsection{Possible extensions to a 4-D variability to enable matching
  of affine Gaussian derivative responses}
\label{sec-ext-4D-variabil}

An operational explanation, why affine covariance cannot be achieved
for the regular affine Gaussian derivative kernels,
with the orientation for computing the directional derivative of that
regular affine Gaussian derivative kernel being required to be aligned with a
principal direction of the covariance matrix of the affine Gaussian
kernel, can be stated as follows: If the affine image transformation
comprises a non-uniform scaling transformation in a direction that not
coincides with the principal directions of the spatial covariance
matrix in the affine Gaussian smoothing kernel, then the transformed
receptive field of a regular affine Gaussian derivative kernel, will be no longer within the definition of such a regular
affine Gaussian directional derivative kernel.

Instead, to enable full affine covariance, the image
orientation of the directional derivative operator in the
transformed affine Gaussian derivative kernel must
be allowed to be different from the restriction of the image direction for
the directional derivative being aligned to either of the principal
directions of the affine Gaussian smoothing kernel.

\subsubsection{Alternative 1: Explicit expansion over the directions of
  directional derivatives of affine Gaussian kernels}

The variability of the receptive fields of receptive fields
according to the affine Gaussian derivative model for visual receptive
fields can, however, be extended. Consider the orientation angle for which
the directional derivatives in the direction $\varphi$ of the
affine Gaussian derivative kernels are computed according to
(\ref{eq-dir-der-aff-gauss-kern})
\begin{equation}
  \label{eq-dir-der-aff-gauss-kern-again}
  T_{\simple}(x_1, x_2;\; \sigma_1, \sigma_2, \varphi, m)
  = \sigma_1^m \, \partial_{\varphi}^m \, g(x_1, x_2;\; \Sigma).
\end{equation}
Let us instead allow the directional derivatives to be computed
in directions $\varphi$ in the spatial domain that do
not necessarily have to be aligned
with either of the eigenvectors of the spatial covariance matrix
$\Sigma$, see Figure~\ref{fig-ori-variability-fixed} for examples of
such receptive fields for fixed values of the spatial scale parameters
$\sigma_1$ and $\sigma_2$.
Then, such extended models for the spatial receptive fields, which
will span a four-dimensional variability, will allow for expanded covariance
properties over the full group of non-singular spatial affine transformations,
according to the treatment in
(Lindeberg \citeyear{Lin25-JMIV} Section~3.5).

\subsubsection{Alternative 2: Expanded matching to linear combinations of
  affine Gaussian directional derivatives as opposed to restricted
  matching to just plain receptive field responses}

Concerning the above possible extension, it should, however, be noted
that due to the definition of directional derivatives, the directional
derivative of a 2-D affine Gaussian kernel can be parameterized to the form
\begin{align}
  \begin{split}
    T_{\varphi^{m_1}\orth\varphi^{m_2},\norm}(x_1, x_2;\; \sigma_1, \sigma_2, \varphi, m)
 \end{split}\nonumber\\
  \begin{split}    
    = \partial_{\varphi,\norm}^{m_1} \,
        \partial_{\orth\varphi, \norm}^{m_2} \, g(x_1, x_2;\; \Sigma)
  \end{split}\nonumber\\
  \begin{split}
    & = \sigma_1^m \, 
           (\cos \varphi \, \partial_{x_1} + \sin \varphi \, \partial_{x_2})^{m_1} 
  \end{split}\nonumber\\
  \begin{split}
    \label{eq-dir-der-aff-gauss-kern-expanded}
    & \phantom{=} \quad
           \, \sigma_2^{m_2} \,
           (-\sin \varphi \, \partial_{x_1} + \cos \varphi \, \partial_{x_2})^{m_2} \,
           g(x_1, x_2;\; \Sigma),
    \end{split}
\end{align}
where $\orth\varphi$ denotes the orthogonal direction to $\varphi$.

Therefore, given a fixed value of the spatial covariance matrix
$\Sigma$, it is sufficient to compute the directional derivatives in
$m + 1$ sufficiently different directions, in order to then be able to
span the space for
computing the directional derivative of order $m = m_1 + m_2$ in any other
direction, according to
\begin{multline}
    T_{\varphi^{m}}(x_1, x_2;\; \sigma_1, \sigma_2, \varphi, m) = \\
    = \sum_{k=1}^M
         p_k \, T_{\varphi_k^{m}}(x_1, x_2;\; \sigma_1, \sigma_2, \varphi_k, m),
\end{multline}
for some $M \geq m + 1$ and some constants $p_k$.
For this purpose, the actual values for the parameters $p_k$ can, in turn,
be determined by setting up and solving a
linear system of equations relations, based on relationships of the form
(\ref{eq-dir-der-aff-gauss-kern-expanded}), mapped down to
relationships in terms of the partial derivatives
$\partial_{x_1^i x_2^{m-i}} \, g(x_1, x_2;\; \Sigma)$ as the underlying
  basis for these relationships for all $j \in [0, m]$.

Thus, from a computational viewpoint, it is not necessary
to fully span the variability of this fourth dimension, while
nevertheless being able to match the receptive field responses between
the two images domains, before and after the geometric image
transformation. This can be achieved by
simultaneously expanding the matching
process to not just match plain receptive field responses between the
two image domains, but instead allowing for matching of the receptive field
response from the first image domain to a linear combination of receptive
field responses over the second image domain of the form
\begin{multline}
  T_{\varphi^{m}}(x_1, x_2;\; \sigma_1, \sigma_2, \varphi, m)
  * f(x_1, x_2) = \\
    = \sum_{k=1}^M
         q_k \, T_{{\varphi'}_k^{m}}(x'_1, x'_2;\; \sigma'_1, \sigma'_2, \varphi'_k, m)
 * f'(x'_1, x'_2)
\end{multline}
for some constants $q_k$ chosen specific to the relationships between
the orientation $\varphi$ in the first domain and the selection of sampled
orientations $\varphi'$ in the second domain, as well as also
depending on the
degrees of freedom of the affine transformation matrix ${\cal A}$.

Note, however, that for such a computational structure to
work, for directional derivatives of higher orders $m \geq 2$,
it is not sufficient to use the directional derivatives along the
eigendirections of the spatial covariance matrix $\Sigma$ as the
basis. Instead, for directional derivative orders $m \geq 2$,
directional derivatives would also have to be computed in directions
that are significantly different from the eigendirections of the
spatial covariance matrix $\Sigma$. Furthermore, that number of
additional required complementary directions for directional
derivative computations would be required to increase with the order
$m$ of spatial differentiation.

\section{Which degrees of freedom of 2-D spatial affine
  transformations are spanned by the receptive fields in the primary visual cortex?}
\label{sec-dofs-spanned-by-V1-RFs}

Given (i)~these theoretical results, and given that (ii)~the biological
receptive fields corresponding to simple cells in the primary visual
cortex, as measured by
(\citeyear{DeAngOhzFre95-TINS,deAngAnz04-VisNeuroSci}),
Conway and Livingstone (\citeyear{ConLiv06-JNeurSci}) and
Johnson {\em et al.\/}\ (\citeyear{JohHawSha08-JNeuroSci}),
can be qualitatively rather well modelled with spatial components in
terms of affine Gaussian derivatives according to the
axiomatically derived normative theory for visual receptive
fields in Lindeberg (\citeyear{Lin13-BICY,Lin21-Heliyon}), 
one may ask if biological vision has also developed corresponding
variabilities in receptive field shapes as would be predicted from the
presented theory? Specifically, one may ask if the shapes of the
simple cells in the primary visual cortex would have the ability to span the variabilities
corresponding to 2-D spatial affine transformations?

As previously stated, a highly useful property of affine covariant receptive fields in
computer vision is that they allow for substantially more accurate
inference of shape from monocular or binocular cues as opposed to
non-covariant receptive fields, see Tables~1--4 in Lindeberg and
G{\aa}rding (\citeyear{LG96-IVC}) for numerical results of computing
estimates of local surface orientation with using a procedure that
successively updates the shapes of the receptive fields to previous
orientation estimates, which then leads to substantially lower errors
in the surface orientation estimates after just a few iterations.

As previously argued in Lindeberg (\citeyear{Lin21-Heliyon}) Section~6, it could
specifically also constitute an evolutionary advantage for
higher species, that rely on visual information as a critical
source of information about the environment, to adapt their vision
systems to the geometrical properties of the image formation process.
Specifically, an expansion of the image data over the degrees of
freedom over the parameters of natural image transformations would be
consistent with the substantial number of receptive fields in the
early visual pathway, with
\begin{itemize}
\item
  about 100~M photoreceptors and
  1~M output channels in and from the retina
  to the lateral geniculate nucleus (LGN),
\item
  about 1~M neurons in the
  LGN and about 1~M output channels to the primary visual cortex (V1), then
\item
  with about 190~M neurons in V1 and about
  37~M output channels from V1,
\end{itemize}
see Figure 3 in
DiCarlo {\em et al.\/} (\citeyear{DiCZocRus12-Neuron}).

The subject of this section is to use the theoretical treatment developed in the
previous sections to consider whether we
could from such a view regard the receptive fields in the primary visual cortex to
exhibit variabilities in their shapes, that could be interpreted as
if the receptive fields would have the ability to span the variabilities generated
by the family of 2-D spatial affine transformations.

\subsection{Variability under uniform scaling transformations}

The issue of a possible variability of the receptive fields with respect
spatial scaling transformations is special, in the sense that,
according to the theory for affine Gaussian receptive fields, the receptive
fields at coarser levels of scale can, in principle, be computed from the
receptive fields at finer levels of scales.

Consider the semi-group property of the affine Gaussian kernel
(see Lindeberg (\citeyear{Lin93-Dis}) Equation~(15.35))
\begin{equation}
  g(\cdot;\; \Sigma_1) * g(\cdot;\; \Sigma_2) = g(\cdot;\; \Sigma_1 + \Sigma_2),
\end{equation} from which it follows that
by combining {\em e.g\/}\ the output from the receptive fields
corresponding to first-order spatial derivatives
\begin{equation}
  (\nabla_x \, L)(\cdot;\; \Sigma_1) = (\nabla_x \, g)(\cdot;\; \Sigma_1) * f(\cdot),
\end{equation}
where $\nabla_x = (\partial_{x_1}, \partial_{x_2})^T$,
over different positions in image space. Due to this semi-group
property, we can from the output of such
a first layer of visual processing compute the responses of the first-order
derivatives at any coarser level of scale
\begin{equation}
  (\nabla_x \, L)(\cdot;\; \Sigma_2) = (\nabla_x \, g(\cdot;\; \Sigma_2) ) * f(\cdot),
\end{equation}
by combining the outputs from the first layer with suitable weights,
such that the resulting computations implement a convolution
operation of the following form
\begin{equation}
  (\nabla_x \, L)(\cdot;\; \Sigma_2)
  = g(\cdot;\; \Sigma_2 - \Sigma_1)  * (\nabla_x \, L)(\cdot;\; \Sigma_1),
\end{equation}
provided that the difference between the spatial covariance matrices
\begin{equation}
  \Delta \Sigma = \Sigma_2 - \Sigma_1
\end{equation}
is a symmetric positive definite matrix.

In fact, concerning receptive fields in the retina, an interpretation of
results concerning measurements of receptive fields at different distances
from the center of the fovea, in combination with a theoretically
principled model of a foveal scale space, with the complementary
essential requirement of a limited processing capacity in terms of
a finite number of neurons, are consistent with the interpretation
that the minimum size of the receptive fields, at any distance
from the center of the fovea, should increase linearly with the distance
from the center of the fovea
(see Lindeberg (\citeyear{Lin13-BICY}) Section~7).

Thus, irrespective of whether the primary visual cortex would perform
an explicit expansion over multiple receptive field sizes over some
scale range or not,
one could, because of the cascade smoothing property of
spatial Gaussian derivative operators,
also conceive a possible design strategy for a vision
system to only implement a first layer of visual receptive fields
at a finest level of scale, and then computing the representations
at coarser scale in an implicit manner, from the finer-scale
receptive field responses.

Explicit suggestions for neurophysiological experiments to map the possible
variability of receptive field shapes over multiple sizes in the image
domain are given in Section~3.2.2 in Lindeberg  (\citeyear{Lin23-FrontCompNeuroSci}).

%Nevertheless, there are biological results by Howard and his co-workers
%(personal communication) that indicate that the receptive fields in
%the primary visual cortex of ANIMAL SPECIES that implement a
%variability of the size of the receptive fields by up to a factor of 4.

\subsection{Variability under rotations in the image plane}

\begin{figure}[hbtp]
   \begin{center}
    \begin{tabular}{c}
     \includegraphics[width=0.35\textwidth]{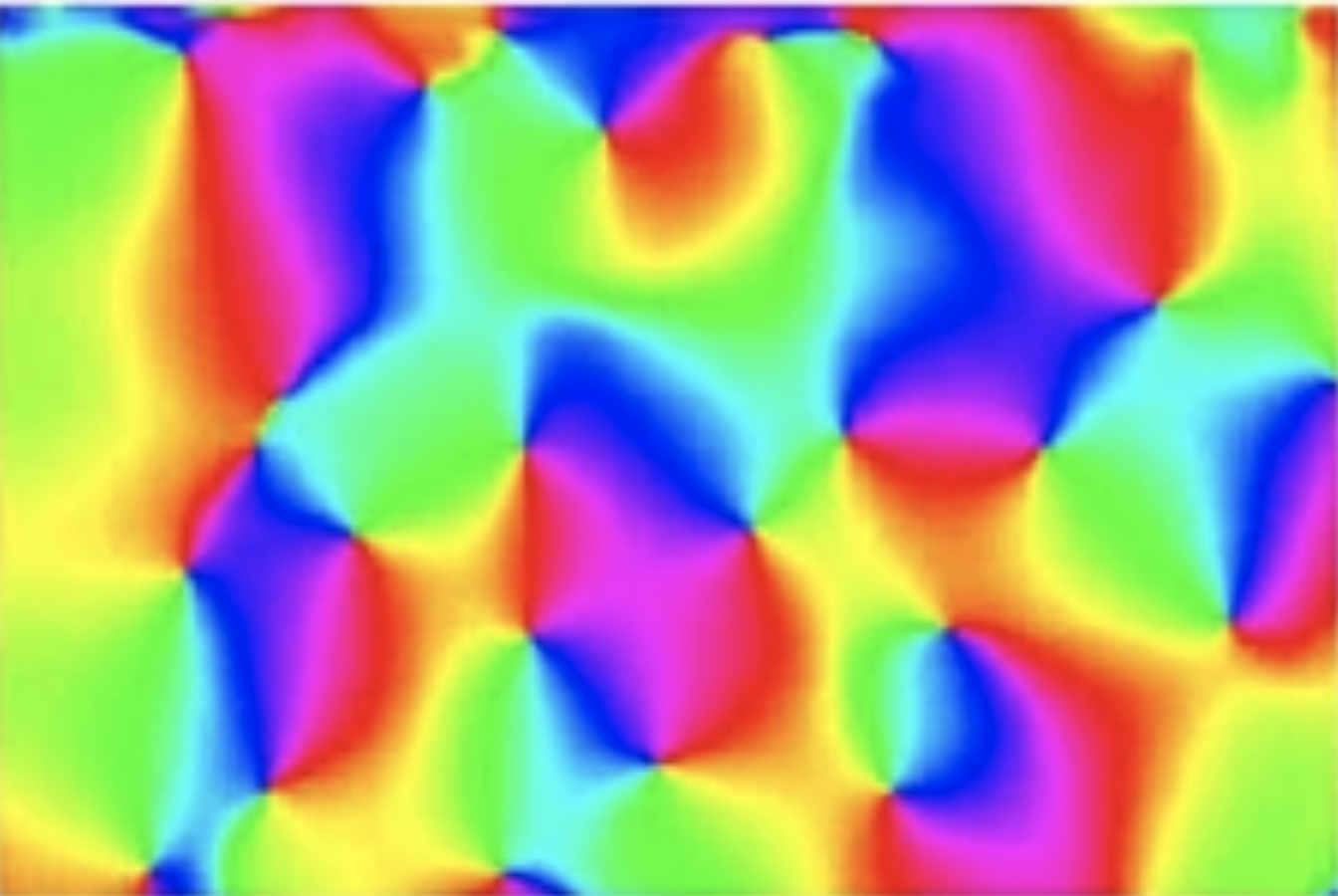}
    \end{tabular}
  \end{center}
  \caption{Orientation map in the primary visual cortex of cat, 
     as recorded by Koch {\em et al.\/} (\citeyear{KocJinAloZai16-NatComm})
     (OpenAccess), and and demonstrating that the
     visual cortex performs an explicit
     expansion of the receptive field shapes over spatial image
     orientations, as would be the result of combining the notion of
     covariance over the subgroup of pure image rotations, and
     corresponding to a variability over the orientation parameter
     $\varphi$ in the proposed decomposition  (\ref{eq-aff-decomp})
     of 2-D spatial affine transformations.}
   \label{fig-koch-ori-map-rot}
 \end{figure}

From the structure of the orientation maps around pinwheels,
as pioneered by Bonhoeffer and Grinvald (\citeyear{BonGri91-Nature})
and Blasdel (\citeyear{Bla92-JNeuroSci})
(see Figure~\ref{fig-koch-ori-map-rot}),
we can from the above theoretical
treatment interpret these results as the receptive fields in the
visual cortex could be regarded as spanning a variability over the pure image rotation
component $\varphi$ in (\ref{eq-aff-decomp}) in the affine group.

Furthermore, using a stimulus manifold analysis,
Beshkov and Einevoll (\citeyear{BesEin23-bioRxiv}) show that
rotating stimuli in the image domain leads to the generation of
circles in the primary visual cortex, also
in very good agreement with the assumption of a covariance property in
the primary visual cortex over the degree of freedom in affine image
transformations corresponding to pure rotations in the image domain.

\subsection{Variability under non-uniform scaling transformations?}
\label{sec-variab-non-uni-sc-transf}

In Lindeberg (\citeyear{Lin24-arXiv-HypoElongVarRF}), an in-depth treatment is given concerning whether the receptive fields in the primary visual cortex could be regarded as spanning a variability over the degree of elongation of the receptive fields.

Based on the relationships between the degrees of freedom in 2-D spatial affine transformations and the degrees of freedom in the affine Gaussian derivative model
established in Section~\ref{sec-rels-var-aff-transf-aff-gauss-ders},
the possible existence of such a variability, for the
receptive fields in the primary visual cortex of higher mammals with a
pinwheel structure, would then correspond to the family of the receptive fields
spanning an expansion over the
non-isotropic scaling component
${\cal D} = \diag(\sqrt{\frac{\rho_1}{\rho_2}}, \sqrt{\frac{\rho_2}{\rho_1}})$
in (\ref{eq-aff-decomp}) of the affine group, that is over the parameter
$\lambda = \rho_1/\rho_2$ according to (\ref{eq-def-lambda-ratio-sing-vals})
in the proposed decomposition of 2-D spatial affine transformations.

Unfortunately, there does, however, not appear to exist any
sufficiently extensive direct measurements of the eccentricity or the degree of elongation for sufficiently large populations of visual neurons to firmly answer this question.
In Lindeberg (\citeyear{Lin24-arXiv-HypoElongVarRF}), we have
therefore taken and alternative indirect approach to this topic, by
making use of existing biological measurements of orientation
selectivity by Nauhaus {\em et al.\/}\ (\citeyear{NauBenCarRin09-Neuron}).
They have reported that there is a substantial variability in the orientation
selectivity of the visual receptive fields in relation to the pinwheel
structure for monkeys and cats.
Goris {\em et al.\/}\ (\citeyear{GorSimMov15-Neuron}) have also accumulated
histograms of the resultant of the orientation selectivity curves for
for simple cells and complex cells.

In Lindeberg (\citeyear{Lin25-JCompNeurSci-orisel}), we have
performed an in-depth theoretical analysis of the orientation
selectivity properties of idealized models of the receptive fields of
simple and complex cells in terms of affine Gaussian derivatives, and
shown that there is a direct connection between the orientation
selectivity and the degree of elongation of the receptive fields
according to the affine Gaussian derivative model.
In Lindeberg (\citeyear{Lin24-arXiv-HypoElongVarRF}) we have
furthermore computed closed-form expressions for the resultant of the
orientation selectivity curves for the idealized model for visual
receptive fields based on affine Gaussian derivatives, and
demonstrated that such a closed-form theoretical analysis leads to
predictions about histograms of the resultants of orientation
selectivity curves for the corresponding idealized models of simple
cells that are in very good qualitative agreement with experimentally
obtained histograms from biological neurons by
Goris {\em et al.\/}\ (\citeyear{GorSimMov15-Neuron}).

If we could assume that the idealized generalized affine Gaussian
derivative model would constitute a sufficiently accurate model for
the receptive fields of the visual neurons in the primary visual
cortex, then we would be able to logically infer that there ought to
be a variability over the degree of elongation of the receptive fields
in the biological neurons. Such logically based modelling arguments
are, however, not necessarily guaranteed to hold, if there would be
other factors, not incorporated into the idealized models of the
receptive fields, that could also affect the orientation selectivity
of the receptive fields.

To more firmly determine whether the hypothesis about an expansion
over the degree of elongation of the receptive fields would hold for
actual biological neurons, a set of more explicit biological
hypotheses, with suggestions for complementary quantitative measurements,
have therefore been formulated in Sections~3.2--3.3 in
Lindeberg (\citeyear{Lin24-arXiv-HypoElongVarRF}) to answer this question.
If the working hypothesis would hold, the proposed types of
neurophysiological measurements could then also characterize how such a possible
variability in the degree of elongation of the receptive fields would relate to the pinwheel structures in the orientation maps of the primary visual cortex.

\subsection{Possible variability over a fourth degree of freedom?}
\label{sec-4th-degree-of-freedom}

Concerning a possible additional variability over the remaining fourth
degree of freedom of affine transformations, we did in
Section~\ref{sec-ext-4D-variabil} discuss two ways by which affine
Gaussian derivative model with a genuine 3-D variability could be
extended to four degrees of freedom.

Notably, in relation to his Gabor modelling of visual receptive
fields, Ringach {\em et al.\/} (\citeyear{Rin01-JNeuroPhys})
found that for a majority of the
receptive fields that he modelled in that study, he did not find
it necessary to add a parameter to vary the orientation of the
non-isotropic Gaussian kernel in relation to the orientation
of the cosine wave; such a parameter
``helped only in a small number of cases''
(see Ringach {\em et al.\/} \citeyear{Rin01-JNeuroPhys}, page~457).

Given that a few receptive fields, hence, obviously have been recorded,
for which a Gabor-based modelling of receptive fields would be helped
by having the principal axes of an affine Gaussian kernel being not
oriented in a similar orientation as the principal axis of the cosine
function in the Gabor model, one may ask the following:
If instead modelling
those receptive fields by directional derivative of affine Gaussian
kernels, would such modelling be helped by not having the
orientation of the directional derivative operator in the affine
Gaussian derivative model being aligned with a
principal direction of the affine Gaussian kernel used in that affine
Gaussian derivative model?

Given this observation, one may then raise the question of whether it
would be possible to accumulate support from neurophysiological
experiments for such a variability, which would then constitute
support for an expansion of the receptive field shapes over this
remaining fourth degree of freedom.

In this context, it is also interesting to note that for some of the receptive field
profiles reconstructed by
Yazdanbakhsh and Livingstone (\citeyear{YazLiv06-NatNeuroSci})
(see Figure~6 in that paper)
appear to be more similar to first- or second-order directional
derivatives of Gaussian kernels in directions different from
the principal directions of an affine Gaussian kernel compared to
directional derivatives of such kernels in directions that coincide
with the principal directions of affine Gaussian kernels.
Thus, those reconstructed receptive fields appear to be
more similar to members of the receptive field family illustrated
in Figure~\ref{fig-ori-variability-fixed}
than to the members of the receptive field families illustrated in
Figures~\ref{fig-size-variability}--\ref{fig-ecc-variability}.

\section{Summary and discussion}
\label{sec-summ-disc}

In this paper, we have given an overall treatment of the variabilities
in image structures on the retina or the image plane
that are generated by viewing 3-D objects
in the environment from different distances and viewing directions,
and how the resulting geometric image transformations will interact
with the receptive fields in the earliest layers in the visual
hierarchy. For convenience of mathematical analysis, we have throughout
modelled these geometric image transformations in terms of
spatial affine transformations, corresponding to the derivative
of the perspective mapping from smooth local surface patches
in the world to the retinal surface, or corresponding to the
derivative of the projective mappings between pairwise views
of the same surface patch from different viewing directions and
distances.

For the purpose of establishing an identity relation between receptive
field responses computed from different views of the same local
surface patch, we have investigated the consequences of the
assumption that receptive field family should
be covariant under the locally linearized perspective or projective
transformations, thereby assuming that the receptive fields should be
covariant under spatial affine transformations. As a main working
hypothesis, we have thus investigated the consequences of this assumption
in the respect of the simple cells in the primary visual cortex being
covariant to spatial affine transformations. This covariance property
would then mean that the visual receptive fields in the primary visual
cortex would meet the variabilities in spatial image structures as
generated by the geometric image transformations by a corresponding
variability in the receptive field shapes, in order to enable matching
of receptive field responses under different viewing conditions of the
same object. A manifest implication 
of that covariance property would then be that the receptive field shapes in the
primary visual cortex would span a similar variability as generated by
the variability of spatial affine transformations, and thereby span
the degrees of freedom of spatial affine transformations.

The main subject of this paper has been to develop a theoretical foundation for
investigating the consequences of this way of reasoning, by first
relating the degrees of freedom of spatial affine transformations to
the degrees of freedom in our highly idealized affine Gaussian
derivative model for visual receptive fields, and then considering for
which matches of degrees of freedom in the geometric image
transformations to the corresponding degrees of freedom in the
theoretical receptive field model, we could identify corresponding
variabilities in the receptive field shapes of simple cells in the
primary visual cortex.

A longer-term underlying motivation to this work is that the presented
theoretical foundation could then be used as a basis for more detailed
neurophysiological or psychophysical studies, concerning to what extent biological vision for different
types species may have developed receptive field structures that have
similar properties as the predictions obtained from the highly idealized theory.

More technically, we have, after describing contextual relations to previous work in
Section~\ref{sec-rel-work}, first in Section~\ref{sec-dofs-2d-affine}
performed an in-depth characterization of
the degrees of freedom in 2-D spatial affine transformations, based on a
decomposition of the affine transformation matrix to a product form very
closely related to a singular value decomposition, although on closed
form, and with the pre-multiplication and post-multiplication matrices
$U$ and $V$ in a regular singular value decomposition here required to be
pure rotation matrices.

For our target application, where we consider
affine transformation matrices reasonably close to a unit matrix
multiplied by a positive scaling factor, such a product form does
indeed guarantee positive diagonal entries in the diagonal matrix of
the proposed matrix decomposition.
In Appendix~\ref{sec-app-deriv-2d-decomp}, we have specifically shown
how the parameters $(\rho_1, \rho_2, \varphi, \psi)$ of the proposed matrix
decomposition (\ref{eq-aff-decomp}) can be determined from the
parameters $a_{ij}$ of the affine transformation matrix ${\cal A}$,
which in these ways extend the earlier treatment of this concept
in (Lindeberg \citeyear{Lin95-ICCV}).

Then, we have, after an overview of the covariance properties of the
affine Gaussian derivative model in Section~\ref{sec-cov-aff-rfs},
in Section~\ref{sec-dofs-aff-gauss-der-rfs} analyzed the
degrees of freedom in the affine Gaussian derivative model, to in
Section~\ref{sec-rels-var-aff-transf-aff-gauss-ders} relate the
degrees of freedom of the affine Gaussian derivative model to the
degrees of freedom in 2-D spatial affine transformations:
\begin{itemize}
\item
  For the uniform spatial scaling factor $S = \sqrt{\rho_1 \, \rho_2}$
  in a spatial affine transformation decomposed according
  to (\ref{eq-def-overall-sc-factor-aff}), there is a direct
  mapping to the the square root of the product of the standard
  deviations $\bar{\sigma} = \sqrt{\sigma_1 \, \sigma_2}$ according
  to (\ref{eq-def-sigma-bar}) of the Gaussian kernel in the principal
  directions of the spatial covariance matrix $\Sigma$.
\item
  For the ratio between the singular values $\lambda = \rho_1/\rho_2$
  in the affine decomposition according to
  (\ref{eq-def-lambda-ratio-sing-vals}), there is a
  direct (inverse) relationship to the eccentricity
  $\epsilon = \sigma_2/\sigma_1$ according to
  (\ref{eq-def-eccentricity-rf}) in the receptive field model
  based on affine Gaussian derivatives.
\item
  For the total amount of rotation
  ${\cal R}_{\varphi} = {\cal R}_{\frac{\varphi}{2}} {\cal  R}_{\frac{\varphi}{2}}$
  according to
  (\ref{eq-def-tot-amount-rot}) in the proposed matrix
  decomposition (\ref{eq-aff-decomp}) , there is a direct mapping to
  variabilities in the orientation $\varphi$ of the affine Gaussian
  receptive fields according to (\ref{eq-dir-der-aff-gauss-kern}),
  provided that the spatial covariance matrix covaries with the angle
  $\varphi$ according to (\ref{eq-expl-par-Cxx})--(\ref{eq-expl-par-Cyy}).
\end{itemize}
Due to the fact that the original formulation of idealized models of
simple cells $T_{\simple}(x_1, x_2;\; \sigma_1, \sigma_2, \varphi, m)$
according to the most strict formulation according to
(\ref{eq-dir-der-aff-gauss-kern}), with the directional derivative
operator parallel to either of the eigendirections of the spatial
covariance matrix $\Sigma$, only
represents a three-dimensional variability, while the variability of
affine transformation matrices ${\cal A}$ is four-dimensional,
the restriction of idealized simple cells to the form
(\ref{eq-dir-der-aff-gauss-kern})  cannot span the full variability of
2-D spatial affine transformations.

To enable matching of receptive field responses under the full
variability of 4-D spatial transformations, we have therefore in
Section~\ref{sec-ext-4D-variabil} considered two types of
extensions of the original model, to enable prefect matching of receptive field
responses under general 2-D spatial affine transformations: 
\begin{itemize}
\item
  either by extending the variability of the directions of the
  directional derivatives to a full variability over image
  orientations in relation to the eigendirections of the spatial
  covariance matrix $\Sigma$, which, however, then would lead to a highly
  redundant representation, since the directional derivatives in
  different directions are related in terms of linear combinations,
\item
  or indirectly representing just the subspace of such a full variability
  with a set of at least $m + 1$ sampled and sufficiently different
  image orientations $\varphi_k$ in relation to the eigendirections of
  the spatial covariance matrix, and then performing the matching to
  linear combinations of receptive field responses within that
  subspace, as opposed to perfect pairwise matching of receptive field
  responses between the two image domains that are related by a 2-D
  spatial affine transformations.
\end{itemize}
In these ways, we have described how the idealized affine Gaussian
derivative model allows for explicit matching of first of all spatial receptive
field responses under general 2-D spatial affine transformations,
based on the covariance properties of the spatial affine Gaussian
derivative model according to the treatment in
Section~\ref{sec-aff-cov-pure-spat-rfs}. With extension of the spatial
affine covariance properties to spatio-temporal affine covariance
according to Section~\ref{sec-aff-cov-spat-temp-rfs}, the proposed
model does additionally allow for matching of spatio-temporal
receptive field responses computed based on the idealized model for
spatio-temporal receptive fields according to
(\ref{eq-def-spat-temp-rf-model}), when complemented by spatial and
temporal differentiation according to the theory presented in
Lindeberg
(\citeyear{Lin23-FrontCompNeuroSci,Lin25-JMIV}).

Finally, we have in Section~\ref{sec-dofs-spanned-by-V1-RFs}
considered if we could from the theory predict if the receptive fields
of the simple cells and complex cells would have the ability to span similar types of
variabilities in receptive field shapes as inferred from the presented
theory:
\begin{itemize}
\item
  From existing results, by combining the spans over the degrees
  of freedom corresponding to uniform scaling transformations and
  rotations, it appears clear that we could regard the primary visual
  cortex to be able to be covariant under similarity transformations, that is
  to combinations of uniform scaling transformations and rotations. 
\item
  If we additionally would interpret the potential support in
  Lindeberg (\citeyear{Lin24-arXiv-HypoElongVarRF}) for the
  hypothesis that the receptive field shapes would additionally be
  expanded over different degrees of elongations, as predicted from
  the variabilities of the degree of orientation selectivity
  established from neurophysiological recordings by
  Nauhaus {\em et al.\/}\ (\citeyear{NauBenCarRin09-Neuron})
  and Goris {\em et al.\/}\ (\citeyear{GorSimMov15-Neuron}). Then, it seems
  natural to also predict that the receptive fields in the primary
  visual cortex could exhibit properties as would be obtained if the
  primary visual cortex would have the ability to be covariant also over a significant span
  over the group of 2-D spatial affine transformations, that is also
  to image transformations involving
  non-uniform scaling transformations, as naturally arise from
  variations of the slant angle, when observing smooth local surface
  patches in the world from different viewing directions.
\end{itemize}
To firmly establish if these predictions, based on our highly
idealized theory for visual receptive fields, would hold in reality in the
primary visual cortex of higher species,
further biological experiments would, however, be extremely valuable,
for which explicitly testable experimental hypotheses concerning
specific subgroups of 2-D spatial affine transformations have been
outlined in Section~3.2.1 in
Lindeberg (\citeyear{Lin23-FrontCompNeuroSci})
and Sections~3.2--3.3 in
Lindeberg (\citeyear{Lin24-arXiv-HypoElongVarRF}).

\subsection{Outlook}

If it could be established that the receptive field shapes in the
primary visual cortex would exhibit a variability corresponding to the
degrees of freedom in 2-D spatial affine transformations, then such a
result would constitute partial support for the working hypothesis
that the receptive fields in the primary visual cortex of higher
mammals could be regarded as an affine covariant family of basis
functions, that performs an expansion of the incoming image data over
the degrees of freedom of 2-D spatial image transformations.
A topic of particular interest could also be to investigate if the presence
of such covariance properties would be different for different
species, and, if so, concerning what species.

We thus propose the presented theory and predictions to have
immediate implications for electrophysiological characterization
of receptive fields as well as for psychophysical studies of vision.

One main line of possible future research would concern reconstructing
a large set of neurophysiologically recorded receptive field shapes in the primary visual cortex of
higher mammals, to map the variabilities in the parameters of the
receptive fields (their spatial extent in the principal directions,
their orientation, as well as the possible deviation between
the direction of a directional derivative operator from the principal
directions of the underlying spatial smoothing kernel).
In relation to the more detailed proposals concerning such
neurophysiological measurements outlined in some additional respects%
\footnote{Note, however, that the treatment in this paper goes
  conceptually further compared to the earlier treatments in
  Lindeberg (\citeyear{Lin23-FrontCompNeuroSci},
  \citeyear{Lin24-arXiv-HypoElongVarRF}),
  by explicitly addressing a variability in receptive field shapes
  along the fourth degree of freedom
  in Section~\ref{sec-4th-degree-of-freedom}, which was not addressed
  in Lindeberg (\citeyear{Lin23-FrontCompNeuroSci},
  \citeyear{Lin24-arXiv-HypoElongVarRF}).}
in Lindeberg (\citeyear{Lin23-FrontCompNeuroSci},
\citeyear{Lin24-arXiv-HypoElongVarRF}),
one may specifically investigate if there would be fundamental
differences between lower and higher species, as well as regarding
species that either have or not have a pinwheel structure.
The developed notion of affine image transformations
in relation to receptive field measurements
could possibly also be combined with the approach in
Yazdanbakhsh and Livingstone (\citeyear{YazLiv06-NatNeuroSci})
to extract second-order receptive fields, to
electrophysiologically characterize the aperture problem,
and infer the connectivities in the pathway from the retina,
upward to the lateral geniculate nucleus and the primary visual cortex.

Concerning psychophysical experiments, another interesting direction
could be to try to explore the influence of evolutionarily- or learning-based
priors in the visual system that enable biological vision to very
quickly solve visual tasks that from a purely mathematical modelling
perspective could be regarded as ambiguous or ill-posed, given that the
visual system faces the task of inferring 3-D cues about the
environment from lower-dimensional 2-D image data.
Could it be established that a pre-wiring of the visual system to
achieve constancy of object properties in the environment under
variations in the viewing conditions could be characterized in terms
of psychophysical measurements, to in turn relate to local affine
transformations of image patterns and covariance or
invariance properties of the underlying receptive fields under such
geometric image transformations, as addressed in this treatment?

For example, regarding visual illusions, the accordion grating
illusion, where a non-uniform deformation field perpendicular
to the lines is perceived when
viewing a set of parallel lines under variations of the distance to
the observer, would according to its theoretical explanation
(Yazdanbakhsh and Gori \citeyear{YazGor11-NeurNetw}) not exist
without the existence and influence of local affine image
transformations. Additionally, in the rotating tilted lines illusion,
where a non-zero radial motion is perceived when viewing a static
stimulus consisting of a set of gradually rotated line segments that
form a circle at a larger scale, with a geometric interpretation
corresponding to the interior of a 3-D tube, can according to its theoretical
explanation (Yazdanbakhsh and Gori \citeyear{YazGor08-NeuroSciLett})
be used to estimate the minimum receptive field size as function
of the distance from the centre of the visual field.

Irrespective of such possible biological implications, the theoretical
analysis of the degrees of freedom of 2-D spatial affine
transformations, with its relations to the degrees of freedom in the
affine Gaussian derivative model, is also important for
modelling and understanding the computational functions in computer
vision systems, that derive information about the environment from
dense measurements of image structures or surface patterns from smooth
surfaces in the environment.

\appendix

\section{Appendix}

\subsection{Derivation of the decomposition and the parameterization of
  2-D spatial affine transformation matrix}
\label{sec-app-deriv-2d-decomp}

\subsubsection{Basic definitions}

Consider a 2-D spatial affine transformation of the form
\begin{equation}
  \label{eq-def-A-app}
  {\cal A}
  =
  \left(
    \begin{array}{cc}
      a_{11} & a_{12} \\
      a_{21} & a_{22}
    \end{array}
  \right)
\end{equation}
and introduce the following
descriptors from the elements $a_{ij}$ of ${\cal A}$:
\begin{align}
  \begin{split}
    \label{eq-def-T-descr-app}
     T & = \frac{a_{11} + a_{22}}{2},
  \end{split}\\
  \begin{split}
    \label{eq-def-A-descr-app}    
    A & = \frac{a_{21} - a_{12}}{2},
  \end{split}\\
  \begin{split}
    \label{eq-def-C-descr-app}    
    C & = \frac{a_{11} - a_{22}}{2},
  \end{split}\\
  \begin{split}
    \label{eq-def-S-descr-app}    
    S & = \frac{a_{12} + a_{21}}{2}.
  \end{split}
\end{align}

\subsubsection{Product decompositions of the affine transformation
  matrix}

Let us next decompose the affine transformation matrix ${\cal A}$
according to a singular value decomposition
\begin{equation}
  {\cal A} = U \, \Sigma \, V^T,
\end{equation}
where $U$ and $V$ are unitary matrices and $\Sigma$ a diagonal matrix.
Let us, however, next rewrite this decomposition as
\begin{equation}
  \label{eq-decomp-A-app}
   {\cal A} = {\cal R}_{\alpha} \, \diag(\rho_1, \rho_2) \, {\cal R}_{\beta}^T,
\end{equation}
where
\begin{equation}
  {\cal R}_{\alpha}
  =
  \left(
    \begin{array}{cc}
       \cos \alpha & - \sin \alpha \\
       \sin \alpha & \cos \alpha 
    \end{array}
  \right)
\end{equation}
and 
\begin{equation}
  {\cal R}_{\beta}
 =
 \left(
    \begin{array}{cc}
       \cos \beta & - \sin \beta \\
       \sin \beta & \cos \beta
    \end{array}
  \right)
\end{equation}
are rotation matrices, and
$\diag(\rho_1, \rho_2)$ is a diagonal matrix with the elements
$\rho_1$ and $\rho_2$.

\subsubsection{Relations between the parameters of the product
  decomposition and the TACS descriptors}

By expanding the decomposition (\ref{eq-decomp-A-app})
\begin{equation}
  \label{eq-decomp-A-app-again}
  {\cal A}
  =
 \left(
    \begin{array}{cc}
       \cos \alpha & - \sin \alpha \\
       \sin \alpha & \cos \alpha 
    \end{array}
  \right)
 \left(
    \begin{array}{cc}
       \rho_1 & 0 \\
       0 & \rho_2 
    \end{array}
  \right)
 \left(
    \begin{array}{cc}
       \cos \beta & - \sin \beta \\
       \sin \beta & \cos \beta
    \end{array}
  \right)
\end{equation}
and identifying with the elements $a_{ij}$ in (\ref{eq-def-A-app}) we
then obtain
\begin{align}
  \begin{split}
    a_{11}
    = \rho_1 \, \cos \alpha \, \cos \beta
    + \rho_2 \, \sin \alpha \, \sin \beta,
  \end{split}\\
  \begin{split}
    a_{12}
    = \rho_1 \, \cos \alpha \, \sin \beta
    - \rho_2 \, \sin \alpha \, \cos \beta,
  \end{split}\\
  \begin{split}
    a_{21}
    = \rho_1 \, \sin \alpha \, \cos \beta
    - \rho_2 \, \cos \alpha \, \sin \beta,
  \end{split}\\
  \begin{split}
    a_{22}
    = \rho_1 \, \sin \alpha \, \sin \beta
    + \rho_2 \, \cos \alpha \, \cos \beta.
  \end{split}  
\end{align}
By next using the following trigonometric relationships
\begin{align}
  \begin{split}
    \sin (x + y) = \sin x \, \cos y + \cos x \, \sin y,
  \end{split}\\
  \begin{split}
    \sin(x - y) = \sin x \, \cos y - \cos x \, \sin y,
  \end{split}\\
  \begin{split}
    \cos(x + y) = \cos x \, \cos y - \sin x \, \sin y,
  \end{split}\\
  \begin{split}
    \cos(x - y) = \cos x \, \cos y + \sin x \, \sin y,
  \end{split}  
\end{align}
we then obtain that the explicit expressions for the TACS descriptors
according to (\ref{eq-def-T-descr-app})--(\ref{eq-def-S-descr-app})
are given by
\begin{align}
  \begin{split}
    \label{eq-def-T-descr-trig-app}
     T & = \frac{1}{2} \, (\rho_1 + \rho_2) \, \cos (\alpha - \beta),
  \end{split}\\
  \begin{split}
    \label{eq-def-A-descr-trig-app}    
    A & = \frac{1}{2} \, (\rho_1 + \rho_2) \, \sin (\alpha - \beta),
  \end{split}\\
  \begin{split}
    \label{eq-def-C-descr-trig-app}    
    C & = \frac{1}{2} \, (\rho_1 - \rho_2) \, \cos (\alpha + \beta),
  \end{split}\\
  \begin{split}
    \label{eq-def-S-descr-trig-app}    
    S & = \frac{1}{2} \, (\rho_1 - \rho_2) \, \sin (\alpha + \beta).
  \end{split}
\end{align}

\subsubsection{Relationships between the singular values $\rho_1$ and
  $\rho_2$ and the derived PQ descriptors}

By further introducing the derived PQ descriptors according to
\begin{align}
 \begin{split}
    P  & = \sqrt{T^2 + A^2},
  \end{split}\\
  \begin{split}
    Q & = \sqrt{C^2 + S^2},
  \end{split}
\end{align}
we then obtain that
\begin{align}
 \begin{split}
    P  & = \frac{1}{2} \, | \rho_1 + \rho_2 |,
  \end{split}\\
  \begin{split}
    Q & = \frac{1}{2} \, | \rho_1 - \rho_2 |.
  \end{split}
\end{align}

\subsubsection{Explicit expressions for the singular values $\rho_1$ and
  $\rho_2$}

For the geometric application we are interested in, we are primarily
interested in affine transformation matrices ${\cal A}$ that are
reasonably close to the unit matrix multiplied by some scalar scaling
factor. Therefore, we will assume
that both $\rho_1 > 0$ and $\rho_2 > 0$.
Additionally, we will for the following parameterizations use the
convention to order the singular values such that $\rho_1 > \rho_2$.
Then, we obtain
\begin{align}
 \begin{split}
    P  & = \frac{1}{2} \, (\rho_1 + \rho_2),
  \end{split}\\
  \begin{split}
    Q & = \frac{1}{2} \, (\rho_1 - \rho_2),
  \end{split}
\end{align}
which in turn gives
\begin{align}
  \begin{split}
    \label{eq-expr-rho1}
    \rho_1 & = P + Q,
  \end{split}\\
  \begin{split}
    \label{eq-expr-rho2}    
    \rho_2 & =  P - Q.
  \end{split}
\end{align}

\subsubsection{Relationships for the angular parameters $\alpha$, 
  $\beta$, $\varphi$ and $\psi$}

By combining (\ref{eq-def-T-descr-trig-app}) and
(\ref{eq-def-A-descr-trig-app}), we additionally get
\begin{equation}
  \frac{A}{T} = \tan(\alpha - \beta),
\end{equation}
and by combining  (\ref{eq-def-C-descr-trig-app}) and
(\ref{eq-def-S-descr-trig-app}), we get
\begin{equation}
  \frac{S}{C} = \tan(\alpha + \beta).
\end{equation}
Let us next introduce the new variables
\begin{align}
 \begin{split}
    \varphi & = \alpha - \beta,
  \end{split}\\
  \begin{split}
    \psi & =  \alpha + \beta,
  \end{split}
\end{align}
which correspond to the relationships
\begin{align}
 \begin{split}
    \alpha & = \frac{\psi + \varphi}{2},
  \end{split}\\
  \begin{split}
    \beta & =  \frac{\psi - \varphi}{2}.
  \end{split}
\end{align}
Then, we obtain the solutions
\begin{align}
 \begin{split}
    \varphi & = \arctan \left( \frac{A}{T} \right) + n_{\varphi} \, \pi,
  \end{split}\\
  \begin{split}
    \psi & =  \arctan \left( \frac{S}{C} \right) + n_{\psi} \, \pi,
  \end{split}
\end{align}
for some integers $n_{\varphi}$ and $n_{\psi}$,
where we, for the purpose of parameterizing affine transformation
matrices ${\cal A}$, %close to a matrix,
prefer  
to choose the
principal solutions according to 
\begin{align}
 \begin{split}
   \varphi & = \atantwo(A, T),
  \end{split}\\
  \begin{split}
    \psi & = \atantwo(S, C),
  \end{split}
\end{align}
where $\atantwo(y, x)$ denotes the function that returns the angle between
the positive $x$-axis and the vector from the origin to the point
$(x, y)$, with the restriction that this angle assumes values in the
interval $]-\pi, \pi]$.

\subsubsection{Resulting matrix decompositions}
\label{app-result-mat-decomp}

From the above explicit expressions for the angles $\alpha$ and
$\beta$, we then obtain that the decomposition (\ref{eq-decomp-A-app}) can
be written as
\begin{align}
  \begin{split}
    A
    & = {\cal R}_{\alpha} \, \diag(\rho_1, \rho_2) \, {\cal R}_{\beta}^T
  \end{split}\nonumber\\
  \begin{split}
    & = {\cal R}_{\frac{\psi + \varphi}{2}} \, \diag(\rho_1, \rho_2)
    \, {\cal R}_{\frac{\psi - \varphi}{2}}^T  
  \end{split}\nonumber\\
  \begin{split}
    \label{eq-aff-decomp-raw-app}
    & = {\cal R}_{\frac{\psi}{2}} \, {\cal R}_{\frac{\varphi}{2}} \, \diag(\rho_1, \rho_2) \,
    {\cal R}_{\frac{\varphi}{2}} \, {\cal R}_{\frac{\psi}{2}}^{T},
  \end{split}
\end{align}
which can also be expressed on the more symmetric form
\begin{equation}
  \label{eq-aff-decomp-app}
  {\cal A}
  = \sqrt{\rho_1 \, \rho_2} \,\,
         {\cal R}_{\frac{\psi}{2}} \, {\cal R}_{\frac{\varphi}{2}}
         \diag(\sqrt{\frac{\rho_1}{\rho_2}}, \sqrt{\frac{\rho_2}{\rho_1}}) \, 
         {\cal R}_{\frac{\varphi}{2}} \, {\cal R}_{-\frac{\psi}{2}}.
\end{equation}

\subsection{Analysis of the proposed decomposition of 2-D affine
  transformations for special subgroups of affine transformations}
\label{app-aff-decomp-spec-cases}

In the following, we will compute the parameters in the proposed
decomposition of affine transformations according to
(\ref{eq-aff-decomp-app}) for four basic classes of image
transformations, in terms of uniform scaling transformations,
pure rotations and non-uniform scaling transformation.

\subsubsection{Uniform scaling transformations}

For a uniform scaling transformation with spatial scaling factor
$S_x > 0$ and with the corresponding spatial scaling transformation matrix
\begin{equation}
  \label{eq-def-scale-mat-app}
  {\cal S}
  =
  \left(
    \begin{array}{cc}
      S_x & 0 \\
      0 & S_x
    \end{array}
  \right),
\end{equation}
we have
\begin{align}
  \begin{split}
     T & = \frac{a_{11} + a_{22}}{2} = S_x
  \end{split}\\
  \begin{split}
    A & = \frac{a_{21} - a_{12}}{2} = 0,
  \end{split}\\
  \begin{split}
    C & = \frac{a_{11} - a_{22}}{2} = 0,
  \end{split}\\
  \begin{split}
    S & = \frac{a_{12} + a_{21}}{2} = 0,
  \end{split}
\end{align}
which then gives
\begin{align}
  \begin{split}
     P & = \sqrt{T^2 + A^2} = S_x,
  \end{split}\\
  \begin{split}
    Q & = \sqrt{C^2 + S^2} = 0,
  \end{split}\\
  \begin{split}
    \rho_1 & = P + Q = S_x,
  \end{split}\\
  \begin{split}
    \rho_2 & = P - Q = S_x,
  \end{split}\\
    \begin{split}
    \varphi & = \atantwo \left( A, T \right) = 0,
  \end{split}\\
  \begin{split}
    \psi & = \atantwo \left( S, C \right) = \mbox{undefined}.
  \end{split}
\end{align}
That the angle $\psi$ is undefined here, means that this angle is
formally underdetermined, and that any value of $\psi$ will be
consistent with the decomposition (\ref{eq-aff-decomp-app}).
Without loss of generality, we can therefore make the simplest
choice of $\psi = 0$.

\subsubsection{Pure rotations}

For a pure rotation with rotation angle $\gamma$ and the
corresponding rotation matrix
\begin{equation}
  \label{eq-def-rot-mat-app}  
  {\cal R}
  =
  \left(
    \begin{array}{cc}
      \cos \gamma & -\sin \gamma \\
      \sin \gamma & \cos \gamma
    \end{array}
  \right),
\end{equation}
we have
\begin{align}
  \begin{split}
     T & = \frac{a_{11} + a_{22}}{2} = \cos \gamma,
  \end{split}\\
  \begin{split}
    A & = \frac{a_{21} - a_{12}}{2} = \sin \gamma,
  \end{split}\\
  \begin{split}
    C & = \frac{a_{11} - a_{22}}{2} = 0,
  \end{split}\\
  \begin{split}
    S & = \frac{a_{12} + a_{21}}{2} = 0,
  \end{split}
\end{align}
which then gives
\begin{align}
  \begin{split}
     P & = \sqrt{T^2 + A^2} = 1,
  \end{split}\\
  \begin{split}
    Q & = \sqrt{C^2 + S^2} = 0,
  \end{split}\\
  \begin{split}
    \rho_1 & = P + Q = 1,
  \end{split}\\
  \begin{split}
    \rho_2 & = P - Q = 1,
  \end{split}\\
  \begin{split}
      \varphi & = \atantwo \left( A, T\right)
      = \atantwo \left( \sin \gamma, \cos \gamma \right)
      = \gamma
  \end{split}\\
  \begin{split}
    \psi & = \atantwo \left( S, C \right) = \mbox{undefined}.
  \end{split}
\end{align}
Again, since the angle $\psi$ is not formally restricted in the
decomposition (\ref{eq-aff-decomp-app}),
we make the simplest choice of choosing $\psi = 0$.

\subsubsection{Non-uniform scaling transformations}

For a non-uniform scaling transformation of the form
\begin{align}
  \begin{split}
    A
    & = {\cal R}_{\frac{\gamma}{2}} \diag(S_1, S_2) \, {\cal R}_{\frac{\gamma}{2}}^T
  \end{split}\nonumber\\
  \begin{split}
    \label{eq-non-uni-scale-transf-app}
    & =
    \left(
      \begin{array}{cc}
        \cos \frac{\gamma}{2} & -\sin \frac{\gamma}{2} \\
        \sin \frac{\gamma}{2} & \cos \frac{\gamma}{2}
      \end{array}
    \right)
    \left(
      \begin{array}{cc}
        S_1 & 0 \\
        0 & S_2
      \end{array}
    \right)
   \left(
      \begin{array}{cc}
        \cos \frac{\gamma}{2} & \sin \frac{\gamma}{2} \\
        -\sin \frac{\gamma}{2} & \cos \frac{\gamma}{2}
      \end{array}
    \right)
  \end{split}
\end{align}
with the spatial scaling factors $S_1 > S_2 > 0$ and the orientation 
$\frac{\gamma}{2}$ of the preferred symmetry axis,
we do after a straightforward expansion of the matrix product
followed by simplifications get that
\begin{align}
  \begin{split}
     T & = \frac{a_{11} + a_{22}}{2} = \frac{S_1 + S_2}{2},
  \end{split}\\
  \begin{split}
    A & = \frac{a_{21} - a_{12}}{2} = 0,
  \end{split}\\
  \begin{split}
    C & = \frac{a_{11} - a_{22}}{2} = \frac{(S_1 - S_2)}{2} \, \cos \gamma,
  \end{split}\\
  \begin{split}
    S & = \frac{a_{12} + a_{21}}{2} = \frac{(S_1 - S_2)}{2} \, \sin \gamma,
  \end{split}
\end{align}
which then gives
\begin{align}
  \begin{split}
     P & = \sqrt{T^2 + A^2} = \frac{S_1 + S_2}{2},
  \end{split}\\
  \begin{split}
    Q & = \sqrt{C^2 + S^2} = \frac{|S_1 - S_2|}{2} = \frac{S_1 - S_2}{2},
  \end{split}\\
  \begin{split}
    \rho_1 & = P + Q = S_1,
  \end{split}\\
  \begin{split}
    \rho_2 & = P - Q = S_2,
  \end{split}\\
    \begin{split}
      \varphi & = \atantwo(A, T) = 0,
  \end{split}\\
  \begin{split}
    \psi & = \atantwo \left( S, C \right)
    = \atantwo \left( \sin \gamma, \cos \gamma \right) = \gamma.
  \end{split}
\end{align}
In this respect, the proposed sign conventions for the singular values
$\rho_1$ and $\rho_2$, as well as the choices for determining the angle
$\psi$, correctly recover the parameters in the original non-uniform
transformation of the form (\ref{eq-non-uni-scale-transf-app}).

\subsubsection{Pure skewing transformations}

Consider a pure skewing transformation with skewing angle $\gamma$ of the form
\begin{align}
  \begin{split}
    x' & = x + y \, \tan \gamma,
  \end{split}\\
  \begin{split}
    y' & = y,
  \end{split}
\end{align}
that is with an affine transformation matrix of the form
\begin{equation}
  \label{eq-def-skew-mat-app} 
  {\cal A}
  =
  \left(
    \begin{array}{cc}
      1 & \tan \gamma \\
      0 & 1
    \end{array}
  \right),
\end{equation}
we have
\begin{align}
  \begin{split}
     T & = \frac{a_{11} + a_{22}}{2} = 1,
  \end{split}\\
  \begin{split}
    A & = \frac{a_{21} - a_{12}}{2} = - \frac{\tan \gamma}{2},
  \end{split}\\
  \begin{split}
    C & = \frac{a_{11} - a_{22}}{2} = 0,
  \end{split}\\
  \begin{split}
    S & = \frac{a_{12} + a_{21}}{2} = - \frac{\tan \gamma}{2},
  \end{split}
\end{align}
which then gives
\begin{align}
  \begin{split}
     P & = \sqrt{T^2 + A^2} = \sqrt{1 + \frac{\tan^2 \gamma}{4}},
  \end{split}\\
  \begin{split}
    Q & = \sqrt{C^2 + S^2} = \frac{| \tan \gamma |}{2},
  \end{split}\\
  \begin{split}
    \rho_1 & = P + Q = \sqrt{1 + \frac{\tan^2 \gamma}{4}} + \frac{| \tan \gamma |}{2},
  \end{split}\\
  \begin{split}
    \rho_2 & = P - Q = \sqrt{1 + \frac{\tan^2 \gamma}{4}} - \frac{| \tan \gamma |}{2},
  \end{split}\\
    \begin{split}
      \varphi & = \atantwo \left( A, T \right)
      = \atantwo \left( - \frac{\tan \gamma}{2}, 1 \right) 
      = \arctan \left( - \frac{\tan \gamma}{2} \right),
  \end{split}\\
  \begin{split}
    \psi & = \atantwo \left( S, C \right) 
    = \atantwo \left( - \frac{\tan \gamma}{2}, 0 \right)
    = - \frac{\pi}{2} \, \sign \gamma.
  \end{split}
\end{align}
In other words, according to the proposed decomposition
(\ref{eq-aff-decomp-app}) of affine
image transformations, a pure skewing transformation is regarded as a
more complex image transformation, that corresponds to
(i)~first rotating the input image to a preferred symmetry orientation in the image
plane by the angle $-\frac{\psi}{2}$, then (ii)~rotating the image by the
angle $\frac{\varphi}{2}$ before (iii)~performing a pure non-uniform scaling
transformation with the spatial scaling factors $\rho_1$ and $\rho_2$,
after which the image is again (iv]~rotated by the same
angle $\frac{\varphi}{2}$, before finally being (vi)~rotated back from the
preferred symmetry orientation.

\section*{Statements and declarations}

\subsection*{Competing interests}

The author declares no competing interests.

\bibliographystyle{abbrvnat}

{\footnotesize
\bibliography{defs,tlmac}
}

\end{document}